\documentclass{egpubl}
\usepackage{eg2022}

\usepackage{amssymb}
\usepackage{amsmath}
\usepackage{appendix}
\usepackage{float}
\usepackage{booktabs} 
\usepackage{wrapfig}
\usepackage{color}
\usepackage{diagbox}
\usepackage{makecell}
\usepackage{bbm}
\usepackage{threeparttable}
\usepackage{dsfont}
\usepackage{arydshln}
\usepackage[percent]{overpic}
\usepackage{multirow}
\usepackage{subfigure}

\DeclareRobustCommand*{\ora}{\overleftrightarrow}
\DeclareRobustCommand*{\oraright}{\overrightarrow}

\ConferencePaper        
\usepackage[T1]{fontenc}
\usepackage{dfadobe}  

\biberVersion
\BibtexOrBiblatex
\usepackage[backend=biber,bibstyle=EG,citestyle=alphabetic,backref=true]{biblatex} 
\electronicVersion
\PrintedOrElectronic

\usepackage{egweblnk} 
\definecolor{darkspringgreen}{rgb}{0.09, 0.45, 0.27}

\title[Coverage Axis: Inner Point Selection for 3D Shape Skeletonization] {Coverage Axis: Inner Point Selection for 3D Shape Skeletonization}

\author[Dou et al.]
{\parbox{\textwidth}{\centering Zhiyang Dou$^{1}$\orcid{0000-0003-0186-8269}, Cheng Lin$^{2}$\orcid{0000-0002-3335-6623}, Rui Xu$^{3}$\orcid{0000-0001-8273-1808}, Lei Yang$^{1}$\orcid{0000-0002-3284-4019}, Shiqing Xin$^{3}$\orcid{0000-0001-8452-8723}, Taku Komura$^{1}$\orcid{0000-0002-2729-5860}, Wenping Wang$^{4}$\orcid{0000-0002-2284-3952}
        }
        \\
{\parbox{\textwidth}{\centering $^1$ The University of Hong Kong \\
         $^2$Digital Content Technology Center, CROS, Tencent Games\\ 
         $^3$Shandong University \\
         $^4$Texas A\&M University       }
}
}

\begin{document}

\teaser{
\vspace{-9mm}
 \includegraphics[width=\linewidth]{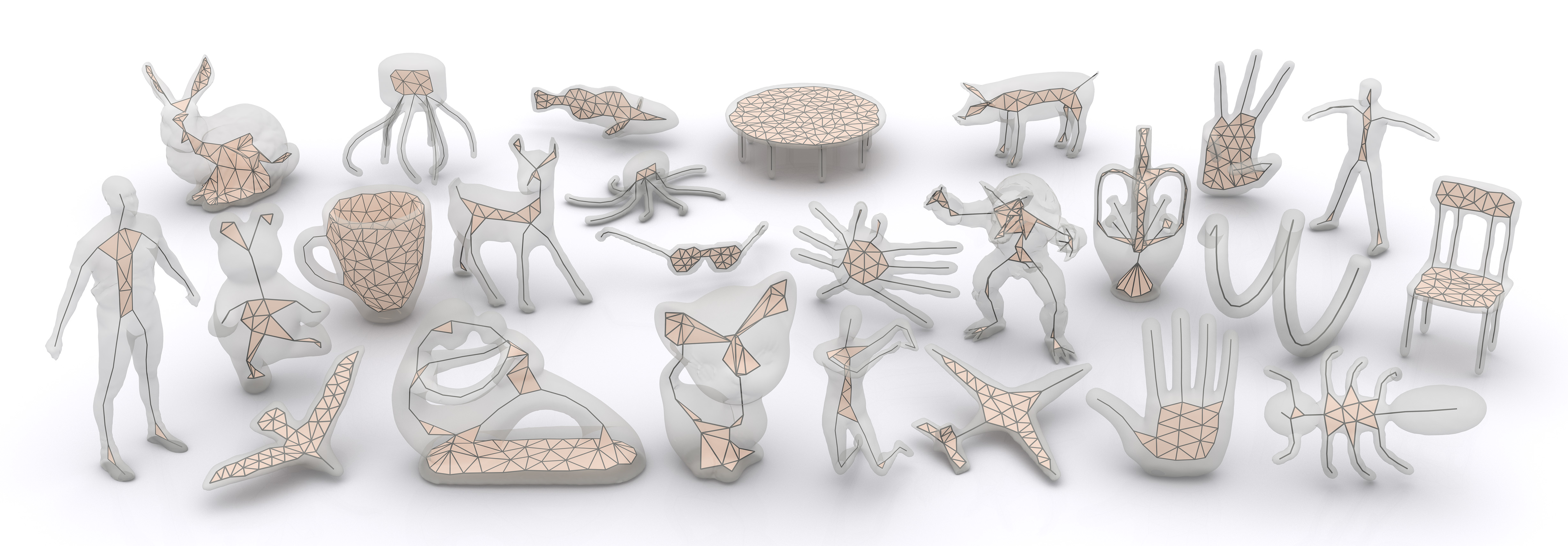}
 \centering
 \vspace{-8mm}
\caption{A gallery of 3D shape skeletonization results by Coverage Axis from surface mesh input.}
\label{fig:mesh_result}
\vspace{2mm}
}

\maketitle

\begin{abstract}
  In this paper, we present a simple yet effective formulation called Coverage Axis for 3D shape skeletonization. Inspired by the set cover problem, our key idea is to cover all the surface points using as few inside medial balls as possible. This formulation inherently induces a compact and expressive approximation of the Medial Axis Transform (MAT) of a given shape. Different from previous methods that rely on local approximation error, our method allows a global consideration of the overall shape structure, leading to an efficient high-level abstraction and superior robustness to noise. Another appealing aspect of our method is its capability to handle more generalized input such as point clouds and poor-quality meshes. Extensive comparisons and evaluations demonstrate the remarkable effectiveness of our method for generating compact and expressive skeletal representation to approximate the MAT.


\end{abstract}  
\section{Introduction}

With the capability of effectively capturing the underlying structures of 3D shapes, skeletal representations have been an important tool in various applications of shape analysis and geometric processing, such as 3D reconstruction \cite{wu2015deep, tang2019skeleton, amenta2001power}, volume approximation~\cite{stolpner2011medial, sun2013medial}, shape segmentation~\cite{lin2020seg}, shape abstraction ~\cite{dou2020top}, pose estimation~\cite{shotton2011real,yang2021learning} and  animation~\cite{baran2007automatic, yang2018dmat}, etc. 

Curve skeletons, consisting of only 1D curves, have been extensively researched~\cite{au2008skeleton, ma2003skeleton, tagliasacchi2012mean, xu2019predicting} due to their simplicity and intuitiveness. Curve skeletons are usually empirically understood, while Dey and Sun~\cite{dey2006defining} give a mathematical definition based on Medial Geodesic Function. Generally, the curve representation only applies to tubular components instead of arbitrary shapes, which thus cannot be considered as a generalized tool for shape analysis.

Another prominent example of skeletal representations is called the Medial Axis Transform (MAT)~\cite{blum1967transformation}. The MAT is defined by a union of the maximally inscribed balls inside the shape with the associated radius functions. Different from the curve skeleton, the MAT has a consistent definition for arbitrary shapes. In addition to curves, the MAT consists of both curve-like and surface-like structures thus leading to significantly better representational ability. That being said, the MAT is difficult to use which mainly manifests in two aspects. First, the MAT is notoriously sensitive to boundary noise, i.e., small perturbations on the boundary surface will result in dramatic changes on the medial axis. Second, the computation of the MAT usually relies on stringent requirements of the input geometry, such as the watertightness and manifoldness of the surface.

In this paper, we propose Coverage Axis, a novel and simple formulation to generate skeletal representations for 3D shapes. Our goal is to give a compact approximation of the MAT, while this approximation should inherit good geometric and topological properties of the MAT but overcome its aforementioned drawbacks.

We observe a medial sphere can be considered as an abstraction of local geometry, while the union of all the local geometries forms the entire shape. With this insight, our key idea is to formulate a Set Cover Problem, of which goal is to identify the smallest sub-collection of whose union equals the universe. Specifically, we aim to find the minimum number of dilated inner balls that cover sub-regions on the surface (i.e., sub-collection) to approximate the whole shape (i.e., universe).

This formulation inherently induces a compact and expressive representation that approximates the MAT. First, the coverage constraint enforces consistency with the shape structure. Second, selecting the fewest possible candidates for shape approximation not only gives a simplified representation, but also favors the interior points that dominate larger area, which corresponds to the definition of MAT, i.e., maximally inscribed spheres. 

Compared to the existing methods that mainly focus on MAT simplification~\cite{amenta2001power, li2015q, giesen2009scale, foskey2003efficient, dey2002approximate, sud2007homotopy,miklos2010discrete}, our method has several appealing aspects. First, its formulation is simple yet effective, which does not require any complex geometric processing nor a computationally-costly pipeline to derive skeletal points. Compared to metrics like QEM~\cite{garland1997surface} and Spherical QEM~\cite{thiery2013sphere} which are used to construct the medial axes by only considering the local geometry, our set coverage formulation jointly leverages all the surface points and inner points for computing MAT approximation, which results in considering the global shape structure, and thus leads to a better abstraction of the overall shape as well as a shape-aware point distribution.  More importantly, we can handle the input of poor-quality meshes or point clouds since our method is based on the coverage of point sets, while other algorithms usually rely on watertight surfaces with decent quality. Finally, our method takes as input a set of overfilled inner point candidates and selects the most expressive ones as the skeletal points; thus it does not necessarily need to compute a rigorously defined MAT, but can work with randomly generated points inside the input.

We demonstrate the effectiveness and robustness of our method on a variety of 3D shapes. The extensive evaluation results reveal that our novel and simple formulation for inner point selection effectively captures the global structure and the fundamental geometry of input shapes, thus providing informative skeletal representations to approximate the structure of the MAT.

\section{Related Work}
\paragraph{Curve skeletonization}
Curve skeletonization has been extensively researched in computer vision and computer graphics. Traditional methods rely on hand-crafted rules to utilize geometric information for curved skeleton computation \cite{attali1996modeling, ma2003skeleton, sharf2007fly, au2008skeleton, natali2011graph, tagliasacchi2012mean, xu2019predicting, cheng2020skeletonization}. Specifically, Ma et al.~\cite{ma2003skeleton} applies  using radial basis functions (RBFs) for skeleton extraction. Sharf et al. \cite{sharf2007fly} adopt a deformable model evolution that captures the object’s volumetric shape and then generates the approximation for the curve skeleton.  Au et al.~\cite{au2008skeleton} compute the skeleton via mesh extraction. 
Livesu et al.~\cite{livesu2012reconstructing} reconstruct the curve skeletons of 3D shapes using the visual hull.Mean Curvature Skeleton~\cite{tagliasacchi2012mean} formulates the skeletonization problem via mean curvature flow, which drives the curvature flow towards the extreme so as to collapse the input mesh. The mesh collapse also induces an intermediate result called meso-skeleton that consists of both surface-like and curve-like structures. Recently, Cheng et al.~\cite{cheng2020skeletonization} propose a method for skeletonization using the dual of shape segmentation.

At the same time, learning-based methods~\cite{xu2019predicting} have been proposed for predicting curve skeletons. Although curve skeletons are able to represent tubular shapes, it has issues representing arbitrary shapes, e.g., shapes with flat components.

\paragraph{Medial axis transform}  As a more general skeletal representation, medial axis transform (MAT)~\cite{blum1967transformation} is able to encode arbitrary shapes with curve-like and mesh-like structures. The maximal balls defined in the volume together with their locii complete the representation of the shape.

The most commonly used technique for extracting MAT is to initialize the medial surface using the Voronoi diagrams. Then a simplified medial surface is achieved by applying optimization on spikes pruning or mesh tessellations based on various rules. Following this way, there are many existing approaches. Angle-based filtering methods~~\cite{amenta2001power, foskey2003efficient, dey2002approximate, sud2007homotopy} reach a simplification based on the angle formed by the point of MA and its two closest points on the boundary. These methods usually produce a simplified result with local features well preserved; however, they suffer from preserving the original topology of the objects. $\lambda$-medial axis methods ~\cite{chazal2005lambda, chaussard2011robust} 
are another series of methods to simplify the computation of MAT,
which adopt cumradius of the closest points of a medial point as a pruning criterion. The main drawback of these methods is poor feature preserving ability at different scales~\cite{attali2009stability}.

Compared with the aforementioned methods, Scale Axis Transform (SAT)~\cite{miklos2010discrete} prunes spikes more effectively. Unstable medial axis points are identified if the corresponding medial balls are covered by the neighboring medial balls during the multiplicative growing.
However, SAT typically has a significant computational cost and tends to destroy the topology by introducing new topological structures at large scales. To overcome these issues, progressive medial axis filtration (PMAF)~\cite{faraj2013progressive} proposes to perform successive edge collapse based on sphere absorption to preserve the topology as well as improve time efficiency.
Our idea also uses a dilated medial balls; we pay special attention to the relationship between inner balls and surface samples. Different from SAT that favors a dense representation with a large number of vertices on the medial axis, our method is able to yield a compact skeletal representation without destroying the shape structure.

Meanwhile, other strategies have been proposed to achieve simplified MAT, e.g., Delta Medial Axis (DMA)~\cite{marie2016delta}, Bending Potential Ratio (BPR) pruning ~\cite{shen2011skeleton}, Erosion Thickness (ET) measure~\cite{yan2016erosion}, voxelization based $\lambda$-pruning~\cite{yan2018voxel}. The simplified MAT, without doubt, cannot reconstruct the original shape exactly, and thus Sun et al.~\cite{sun2013medial} propose to control the approximation error by computing the Hausdorff distance. Rebain et al.~\cite{rebain2021deep}~introduce medial fields derived from the MAT to represent the original shape using learning techniques. 

To achieve high accurate approximation of the original shape, Li et al.~\cite{li2015q} propose Q-MAT that adopts quadratic error metric (QEM)~\cite{garland1997surface} for MAT simplification. As an edge collapse method, Q-MAT is fast and produces a piecewise linear approximation of the MAT.
Q-MAT collapses edges based solely on local information, which sometimes leads to unreasonable spatial distribution of vertices. Also, Q-MAT is limited to 2-manifold surface scope. 
Additionally, Q-MAT heavily relies on a good MAT initialization; when the initial MA is low quality, Q-MAT can not produce correct results by edge collapse. 
Instead, we tackle this problem from a global view, which can preserve representative inner points with relatively even spatial distribution. 
As our method is based on the coverage of point sets, we can handle inputs of poor-quality meshes or point clouds. Furthermore, our method, as we will demonstrate, has good adaptability to low-quality candidate points. We refer readers to \cite{tagliasacchi20163d} for a detailed survey covering various forms of skeletons.

\paragraph{Point cloud skeletonization} Nowadays, skeletonization of point cloud is drawing people's attention due to the easy availability of point cloud data~\cite{tagliasacchi_sig09, cao2010point, livesu2012reconstructing, huang2013l1, rebain2019lsmat, wu2015deep, yang2020p2mat}.
In particular, L1-medial skeleton~\cite{huang2013l1} contracts point clouds based on locally optimal projections (LOP)~\cite{lipman2007parameterization}. LSMAT~\cite{rebain2019lsmat} takes a densely sampled oriented point set as input and computes an MAT approximation on the basis of Signed Distance Function (SDF). 
Wu et al.~\cite{wu2015deep} associates the surface points to the inner point residing on the meso-skeleton~\cite{tagliasacchi2012mean}.
However, represented by unstructured points, their skeletonization lacks topological constraints. Additionally, the reconstruction results of \cite{wu2015deep} are still with large error.

With the success of deep neural networks, learning-based methods have been proposed for predicting skeletons of point clouds ~\cite{yang2020p2mat, lin2021point2skeleton}. P2MAT-NET~\cite{yang2020p2mat} transforms sparse point cloud to an output set of spheres to approximate the medial balls. 
Point2Skeleton (P2S)~\cite{lin2021point2skeleton} learns skeletal representations from generalized point clouds in an unsupervised manner.
Learning-based methods do not guarantee accurate computation of geometric features. Furthermore, they often suffer from generalization ability due to the dependency on the training data.
\section{Preliminaries}
\subsection{Medial Surface}

\begin{wrapfigure}{r}{2cm}
\vspace{0mm}
  \hspace*{-5mm}
  \centerline{
  \includegraphics[width=26mm]{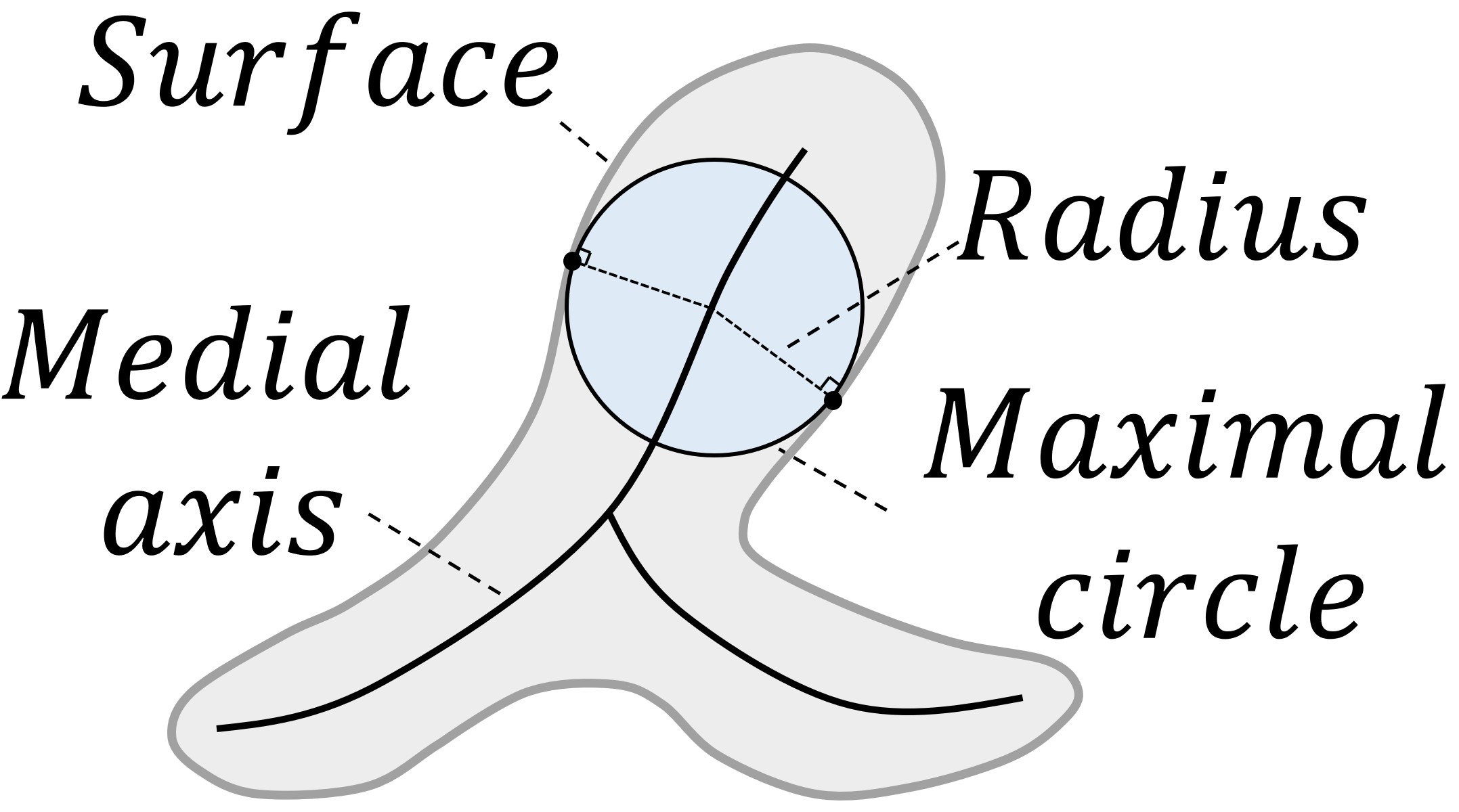}}
  \vspace*{-4mm}
\end{wrapfigure}
Given a closed, oriented and bounded two-manifold surface $\mathcal{S}$ in $\mathbb{R}^3$, \textit{medial axis} is defined as the locus of the center of maximally inscribed spheres that 
are tangent to
$\mathcal{S}$ at two or more points. The medial axis $\mathcal{M}$ of $\mathcal{S}$, together with its radius function $\mathcal{R}$, forms the \textit{medial axis transform} (MAT), denoted by a pair $\big(\mathcal{M}, \mathcal{R}\big)$.

\subsection{Voronoi Diagram \& Power Diagram} 
\begin{figure}[H]
    \centering
    \vspace{-1.5mm}
    \begin{overpic}
[width=0.86\linewidth]{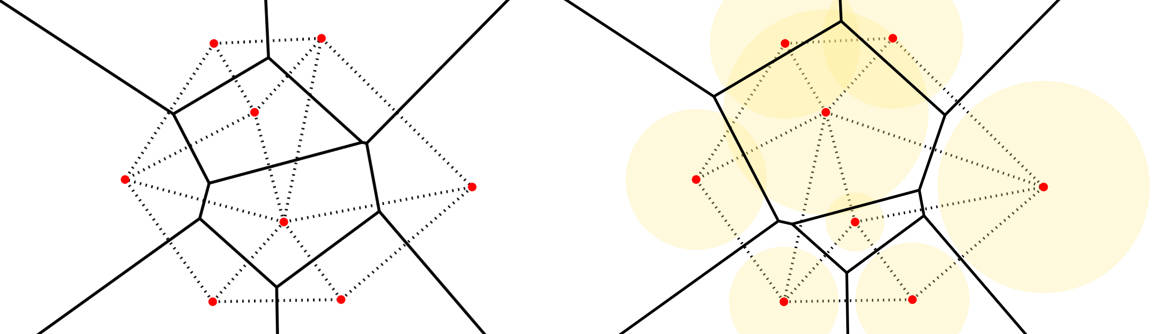}
\put(7,-4){(a) Voronoi diagram}
\put(56,-4){(b) Power diagram}
\end{overpic}
\vspace{1mm}
    \caption{2D Voronoi diagram and Power diagram~(solid lines) and their dual graphs: Delaunay triangulation and Regular triangulation~(dashed lines). The weight of each seed in the Power diagram is visualized by a disk.}\label{fig:Voronoi:power}
\end{figure}\vspace{-5mm}
Voronoi diagram is a partition of the domain $\Omega\subset \mathbb{R}^d$ into regions, close to a set of points called generators $\{\mathbf{x}_i\in\Omega\}_{i=1}^n$. 
Each region, also called cell, is defined as 
\begin{equation}
\Omega_i^{\text{vor}}:~\{\mathbf{x}\in\Omega\;\big|\;\|\mathbf{x}-\mathbf{x}_i\|\leq\|\mathbf{x}-\mathbf{x}_j\|,j\neq i\}.
\nonumber
\end{equation}
A well-known MAT initialization technique is to use the Voronoi diagram inside a model generated by sampled points on the surface.

Power diagrams \cite{aurenhammer1987power} can be viewed as an extension of Voronoi diagrams, where each generator $\mathbf{x}_i$ is equipped with a weight $w_i$ to control its influence. By defining the power distance $d^{\text{pow}}(x,x_i)$ between $x$ and the weighted generator $x_i$ to be $\|\mathbf{x}-\mathbf{x}_i\|^2-w_i$, the cell associated with $\mathbf{x}_i$ is defined by
\begin{equation}
\Omega_i^{\text{pow}}:~\{\mathbf{x}\in\Omega\;\big|\;d^{\text{pow}}(x,x_i)\leq d^{\text{pow}}(x,x_j),j\neq i\}.
\nonumber
\end{equation}
A generator with a larger weight is more dominant, and when all the weights are equal, the Power diagram is reduced to a Voronoi diagram. In~\cite{amenta2001power}, the approximate MAT structure is built based on regular triangulation, the dual of Power diagram, with the weights as squared radius of inside and outside poles (a subset of Voronoi vertices). Examples of a Voronoi diagram and a Power diagram are shown in Figure~\ref{fig:Voronoi:power} respectively.

\begin{figure*}
\centering
\begin{overpic}
[width=\linewidth]{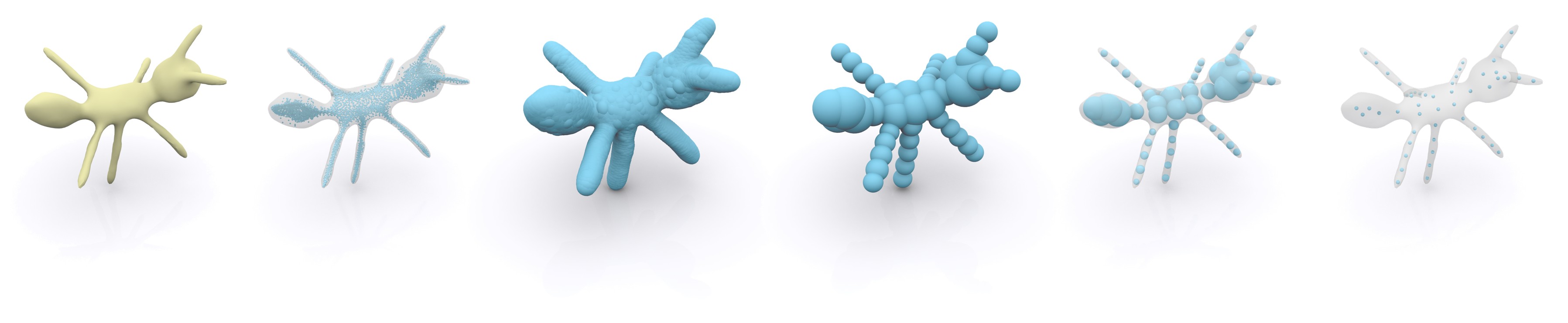}
\put(5.5,5){(a)}
\put(21,5){(b)}
\put(38,5){(c)}
\put(56.5,5){(d)}
\put(73.5,5){(e)}
\put(90,5){(f)}
\end{overpic}
\vspace{-13mm}
\caption{Pipeline of inner point selection. (a) Input 3D model. (b) Inner point candidates. (c) Dilated inner balls. (d) Selected dilated balls based on surface point coverage. (e) Selected inner balls with original radius. (f) Selected inner points. }
\vspace{-8mm}
\label{fig:3D_demo_pipeline}
\end{figure*}

\section{Method}
Our goal is to generate a compact skeletal representation of a given 3D shape that approximates its MAT, for capturing the fundamental structure of the shape. 

Our method consists of three main steps. We first generate a set of inner points as the candidate skeletal points. Then, we introduce a novel formulation called coverage axis, which is based on the set cover problem, to find a minimal sub-collection of candidates to recover the input shape. Finally, we analyze the connectivity of the selected points to form a connected structure with edges and triangles. 

\subsection{Inner Points Generation} 
\label{sec:inner_point_gen}
Our method starts with the generation of candidate points located inside a given 3D shape. The input to our system can be either surface meshes or point clouds with normals.

For a mesh input, following a common initialization technique, we compute the Voronoi diagram of a set of points sampled on the mesh and extract the inner points, which become the candidate points of the 
skeleton.
The inside Voronoi diagram becomes the initial coverage axis structure.

 For a point cloud input (with normals), similar to a mesh input, we first compute the Voronoi diagram w.r.t. surface samples.~Then, we extract the inner points as follows. We first compute its Delaunay triangulation which is the dual of the Voronoi diagram w.r.t. the input point cloud. Consider a Voronoi vertex $p$ and the vertices of its dual tetrahedron $p'_0, p'_1, p'_2$ and $p'_3$. The candidate $p$ is considered as an inner point only if we have  
 $\overrightarrow{pp'_i}\cdot \overrightarrow{n}(p'_i) >0$, $\forall i = 0,1,2,3$. Here $\overrightarrow{n}(p'_i)$ is the input normal of  $p'_i$ and $\cdot$ is dot product. See more details in Appendix~A. Note other inside-outside query methods such as~\cite{Barill:FW:2018} are also applicable.
 
The Voronoi-based point generation is based on principled geometric transform which can generate candidates with good quality. Nevertheless, note our method does not rely on strictly defined MAT for initialization. We can also randomly sample a set of points inside the shape and use them as skeleton candidates (see Sec.~\ref{sec:exp4_center} for a detailed discussion). This property significantly enhances the flexibility of our method for handling various inputs in different applications.
\begin{figure*}
\centering
  \includegraphics[width=\linewidth]{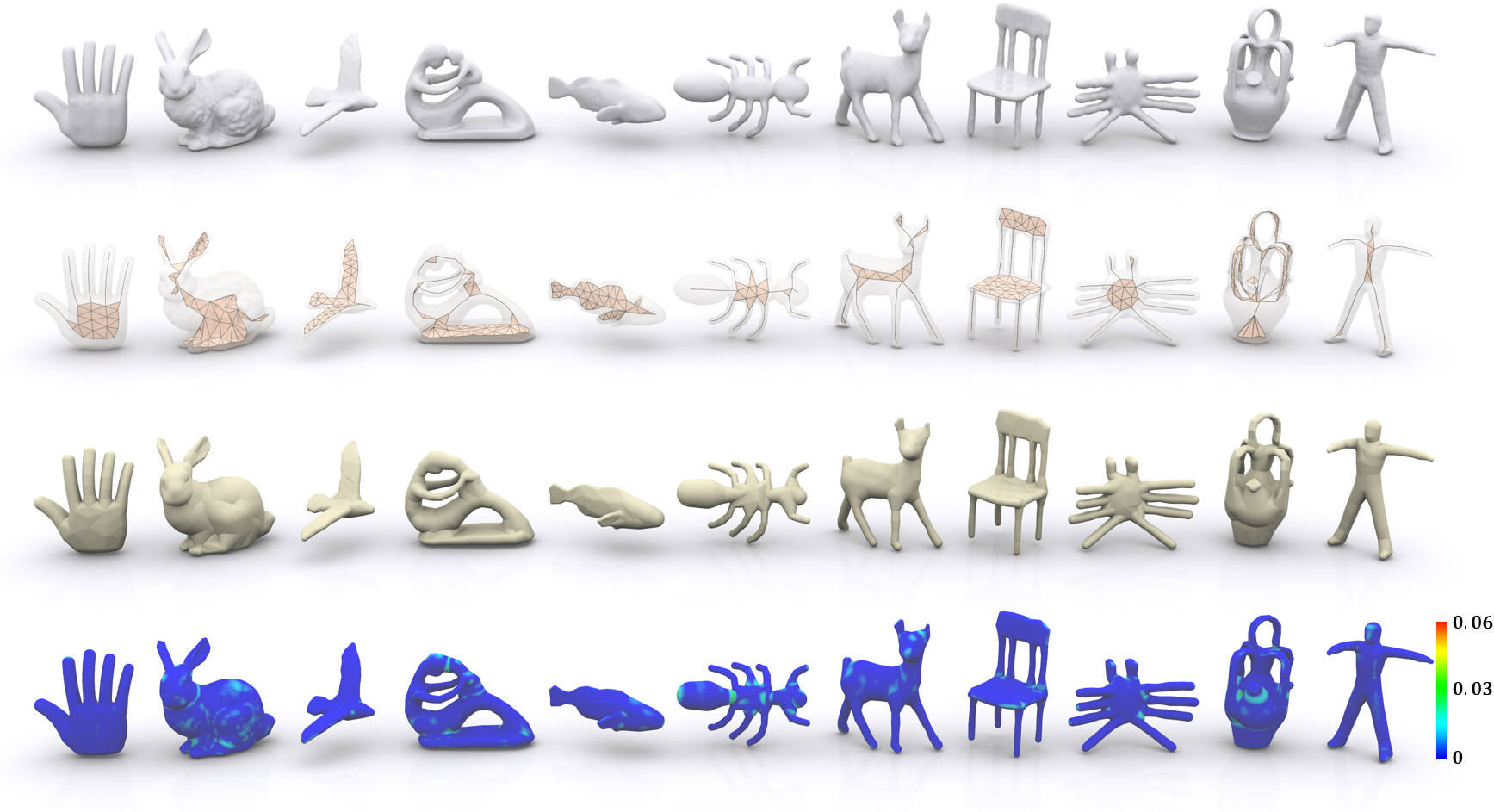}
    \vspace{-9mm}
\caption{Shape approximation results of Coverage Axis. First row: the input shapes. Second row: the output skeletal representations for MAT approximation. Third row: the reconstructed shapes by the skeletal representations. Fourth row: color coding of the reconstruction error between each point in the reconstruction and the input shape, defined relative to the diagonal length of the bounding box.} 
\vspace{-6mm}
\label{fig:mesh_recon}
\end{figure*}

\subsection{Point Selection Based on Set Coverage}
\label{sec:point_selection}

Given a set of elements (called universe) and its subsets whose union equals the universe, the set cover problem (SCP) is to identify the minimum sub-collection of these subsets whose union equals the universe. Inspired by this formulation, we now introduce our Coverage Axis. In the last step, we obtain a set of inner point candidates $P=\{p_i\}$. We estimate the radii of these points $R=\{r_i\}$ as their closest distances to the boundary surface, resulting in candidate balls. Now we slightly dilate all the balls by adding a small value to their radii, i.e., $r_i'=r_i+\delta_r$, leading to a set of dilated balls $B=\{(p_i, r_i')\}$. The dilation makes our algorithm robust to insignificant details and noise, for which we give detailed discussions later in Sec.~\ref{sec:noise}.

Our goal is to find the minimum number of dilated balls that cover all the sampled points $S=\{s_j\}$ on the surface. For this purpose, we introduce a coverage matrix $\mathbf{D}\in \{0,1\}^{m\times n}$, where $m$ and $n$ are total numbers of sampled surface points and candidate skeletal points, respectively. Each element $d_{ji}\in\{0,1\}$ of $\mathbf{D}$ indicates if a surface point $s_j$ is covered by the dilated ball $(p_i, r_i')$:
\begin{equation}
d_{ji}=\left\{
\begin{aligned}
1, & \ \text{if} \ \left\|p_i - s_j\right\|_2 \le r_i', j = 1,...,m, \ i = 1, ... , n\\
0, & \ \text{if} \ \left\|p_i - s_j\right\|_2 > r_i', j = 1,...,m, \ i = 1, ... , n.  \\
\end{aligned}
\right.
\end{equation}

Let $\mathbf{v} \in \{0,1\}^{n\times1}$ be a decision vector, where the $i$-th element $v_i\in\{0,1\}$ indicates if a candidate skeletal point $p_i$ is selected. Now we can derive the formulation of our 0-1 integral optimization problem:

\begin{equation}  
\begin{split}  
&\min \left\|\mathbf{v}\right\|_2 \\  
&s.t.\  \begin{array}{c}  
\mathbf{Dv} \ge \mathds{1},\\
\end{array}  
\end{split}  
\label{eq:core_op}
\end{equation}  
where $\mathds{1}$ is a vector of ones and $\ge$ is applied element-wise to the vector entries.
Here we minimize the norm of the decision vector $\mathbf{v}$ under a certain constraint. The constraint enforces each sampled point to be covered by the selected skeletal balls, which ensures consistency between the selected skeleton and the input shape. With this simple formulation, our method allows a global consideration of the overall shape structure and effectively selects the most predominant inner candidates, giving an expressive and compact skeletal representation to faithfully capture the fundamental geometry and topology. We solve Eq.~\ref{eq:core_op} using Mixed-integer linear programming (MILP) solver in Matlab (MathWorks 2021). The pipeline of inner point selection is demonstrated in Figure~\ref{fig:3D_demo_pipeline}. We further discuss a series of properties of our method, e.g., centrality, robustness to noise, etc., in Sec.~\ref{sec:_discussion}.

It is notable that the constraint in Eq.~\ref{eq:core_op} does not always lead to a solution. For example, if there are points not covered by any dilated balls (assume the dilation is tiny), no solution can be found. Nevertheless, this barely happens and the fixed parameters always lead to a solution for the extensive experiments as we will show later. This is owing to the proper initialization for generating inside candidates: Voronoi Diagram or densely sampled points. Voronoi Diagram produces balls maximally inscribed to the surface, thus already giving tight-fitting candidates. The randomly generated inner points are considerably dense, of which a subset approximates the vertices of the Voronoi Diagram, and the excessive balls provide abundant candidates for solving SCP.

\subsection{Connection Establishment}
\label{subsec:connection}

Once the skeletal points are determined, in this step, we analyze their connectivity to form a structured mesh. Generally, if the input is a surface mesh, the topological connections of the skeletal points are easier to be derived from the surface. However, this is difficult when the input is a point cloud that only consists of unorganized points.

Considering our method flexibly allows different input types (i.e., surface mesh and point cloud) with different candidate generation strategies (i.e., Voronoi-based and random sampling), we propose to use different ways to build the connections of skeletal points. More details can be found in Appendix~B.

\begin{wrapfigure}{r}{3cm}
\vspace{-4.5mm}
  \hspace*{-4mm}
  \centerline{
  \includegraphics[width=39mm]{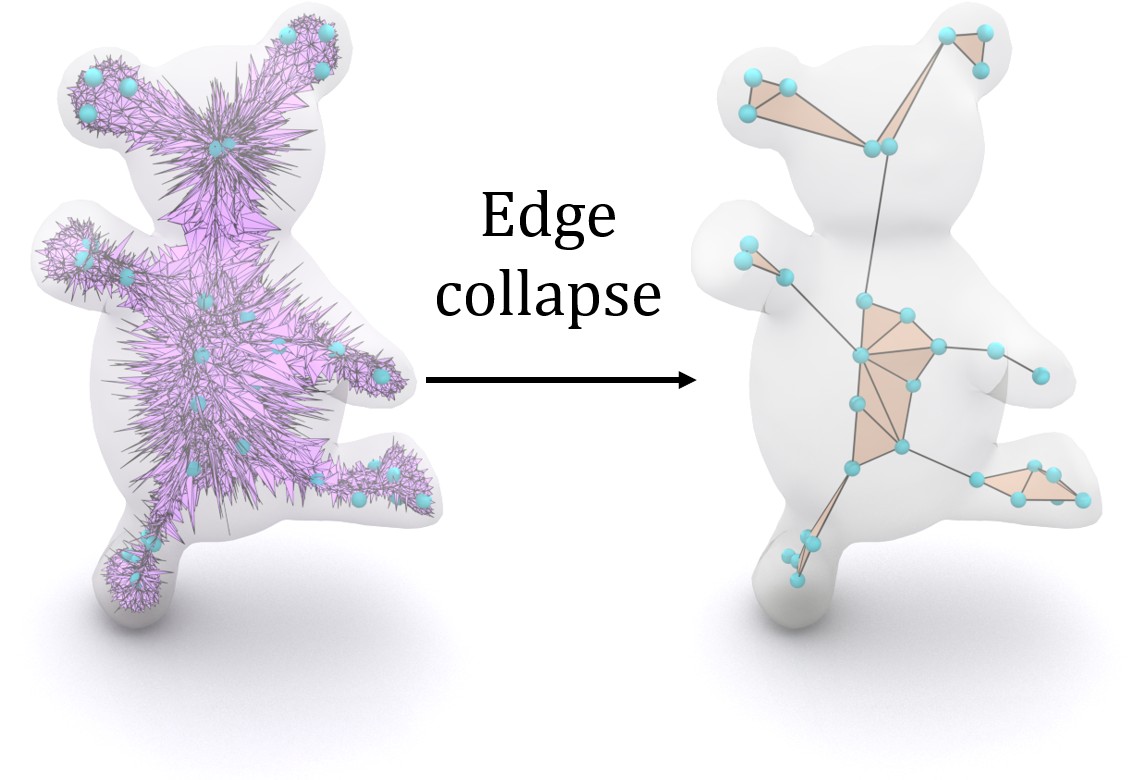}
  }
  \vspace*{-8mm}
\end{wrapfigure}
\paragraph{Mesh with candidates from Voronoi diagram} This configuration usually gives the best skeletonization quality since not only the boundary surface is structured, but the inner points are derived from principled geometric transform, thus also having a good structure. Accordingly, we can obtain selected points embedded on the Voronoi diagram (the part locating inside the model). Note that such a Voronoi diagram provides a correct topological connection w.r.t. the input 3D model. We then can simplify this structure using off-the-shelf method with selected points. In this paper, we follow the edge collapse strategy of Q-MAT \cite{li2015q}. 
But the key difference is that the result selected by set coverage serves as a series of anchors; these points, like nails, are guiding the direction of further simplification so as to ensure a more reasonable distribution w.r.t. original geometry. 
In particular, for an edge to be collapsed, if one endpoint is not the selected point but another is, we merge the point to the selected point; if two endpoints are both selected points, we preserve this edge and skip to the next edge; if both two endpoints are not selected, we merge them to the optimal contraction target same as~\cite{li2015q}. Meanwhile, topology preservation \cite{dey1998topology} and mesh inversion avoidance algorithm ~\cite{garland1997surface} are applied during the collapse.

\paragraph{More generalized configurations}  This case takes the input of either mesh or point cloud, while the candidate skeletal points can be generated by Voronoi diagrams or random sampling. Different from the first case, where both inner and outside structures are well defined, it is challenging to derive precise topological connections of the skeletal points. We seek for an approximation of the MAT connection similar to \cite{amenta2001power} as follows. \\
Suppose $P^+$ is the set of all selected points by set coverage. Let $S$ denotes the set of the surface. We then compute the Power diagram $\textbf{PD}$ on the set of $\{B'(p_i,r_i'), p_i \in P^+\} \cup \{B(s_j, \delta_r), s_j \in S\} $ and establish the connection by extracting edges among the selected inner points on the dual of $\textbf{PD}$, regular triangulation $\textbf{RT}$. In order to suppress over-connection, we up-sample on the input point cloud following~\cite{huang2013edge}, leading to $30000$ surface samples. These points are treated as surface points in $\textbf{RT}$ during connection establishment. Note that $r'$ can be adjusted accordingly during the computation of $\textbf{RT}$ to achieve flexible control over the connection.

\section{Experimental Results}
\label{sec:exp_res}

In this section, we conduct qualitative and quantitative evaluations of the proposed method to comprehensively demonstrate its effectiveness. We implement and experiment on a computer with a 4.00 GHz Intel(R) Core(TM) i7-6700K CPU and 16 GB memory. The size of all the models are normalized to the $[0,1]$ range. 

All the experimental results are generated using a consistent parameter setting, i.e., for all kinds of input, we always set $\delta_r=0.02$.
For a mesh input, we sample $|C| = 4000$ points from the surfaces. The resulting set is denoted by $C$ and is used to compute the Voronoi diagram, whose vertices inside the input mesh serve as candidate inner points set, denoted by $P$. Then we sample $|S| = 1500$ surface points as the set $S$ to be covered in the SCP. Here $|S|$ is set to be less than $|C|$ to reduce the computational cost (see Sec.~\ref{sec:running_time} for a detailed evaluation). The input point clouds all have $2000$ points, and thus we have $|S| = |C| = 2000$. Detailed analyses on these parameters $\delta_r, S, C$ and $P$ are given in Sec.~\ref{sec:parameter_analysis}.

As for evaluation metrics, besides one-sided Hausdorff distance (HD) from the input surface to the surface reconstructed by the skeleton that is employed by Q-MAT~\cite{li2015q}, we adopt two-sided Hausdorff distance since the former is not informative to fully reveal the consistency between two surfaces.
We use $\overrightarrow{\epsilon}$ to denote the HD from surface to reconstruction, $\overleftarrow{\epsilon}$ to denote the HD from reconstruction to surface, and $\overleftrightarrow{\epsilon}$ to represent two-sided HD.

We show a set of qualitative results on various 3D shapes in Figure~\ref{fig:mesh_result}. The shape approximation results are shown in Figure~\ref{fig:mesh_recon}. The reconstruction results are obtained by interpolating the medial balls based on the skeleton connectivity similar to~\cite{li2015q}.
Please refer to the Appendix for more implementation details.

\subsection{Coverage Axis from Mesh Input}
We conduct comparisons with representative MAT simplification methods including Q-MAT~\cite{li2015q} and SAT~\cite{miklos2010discrete}. 
The errors are scaled by the length of the diagonal of the input model bounding box. 

\paragraph{Comparison with Q-MAT~\cite{li2015q}} 

For a fair comparison, the number of vertices in the medial axis are set the same for both methods. The quantitative results are reported in Table~\ref{table:QMAT}. It can be seen that our method is comparable with or better than Q-MAT in terms of reconstruction error. Additionally, we conduct comparisons with Q-MAT on the robustness against the input noise, whose results are given in Sec.~\ref{sec:noise}.

\begin{figure}
    \centering
   \includegraphics[width=\linewidth]{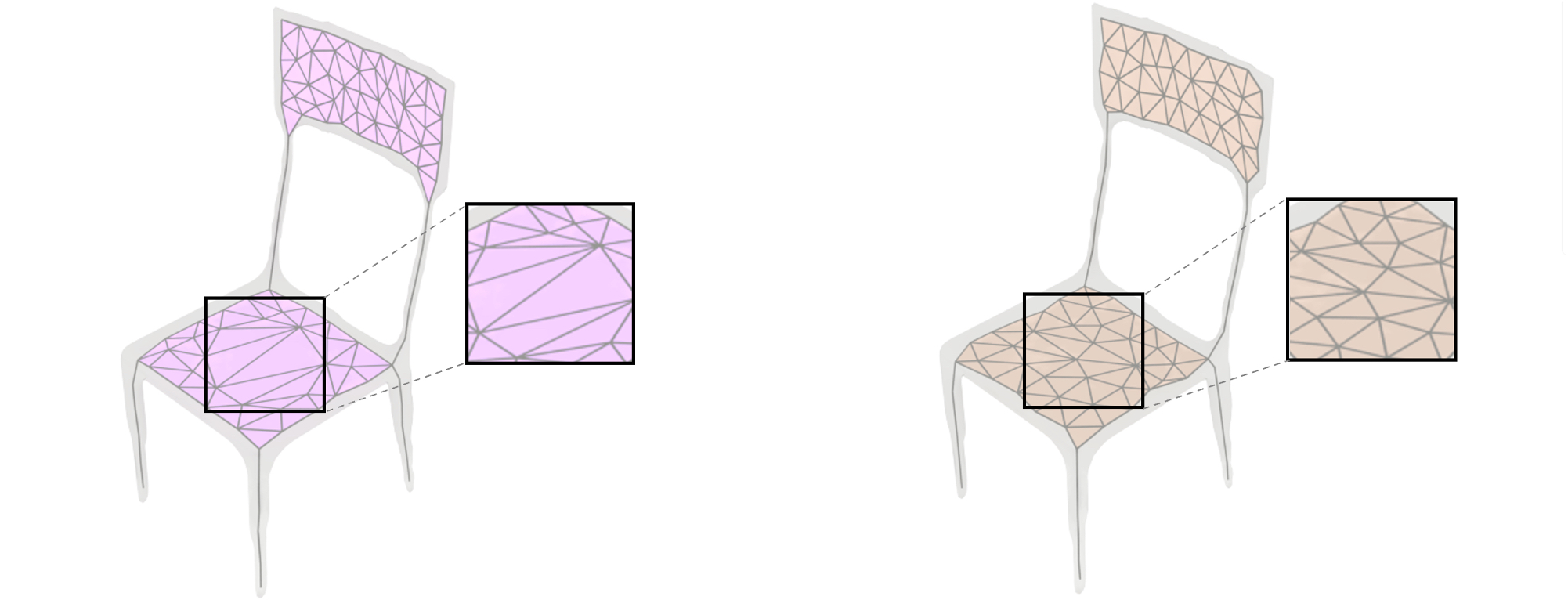}\\
      \vspace{-1mm}
  \includegraphics[width=\linewidth]{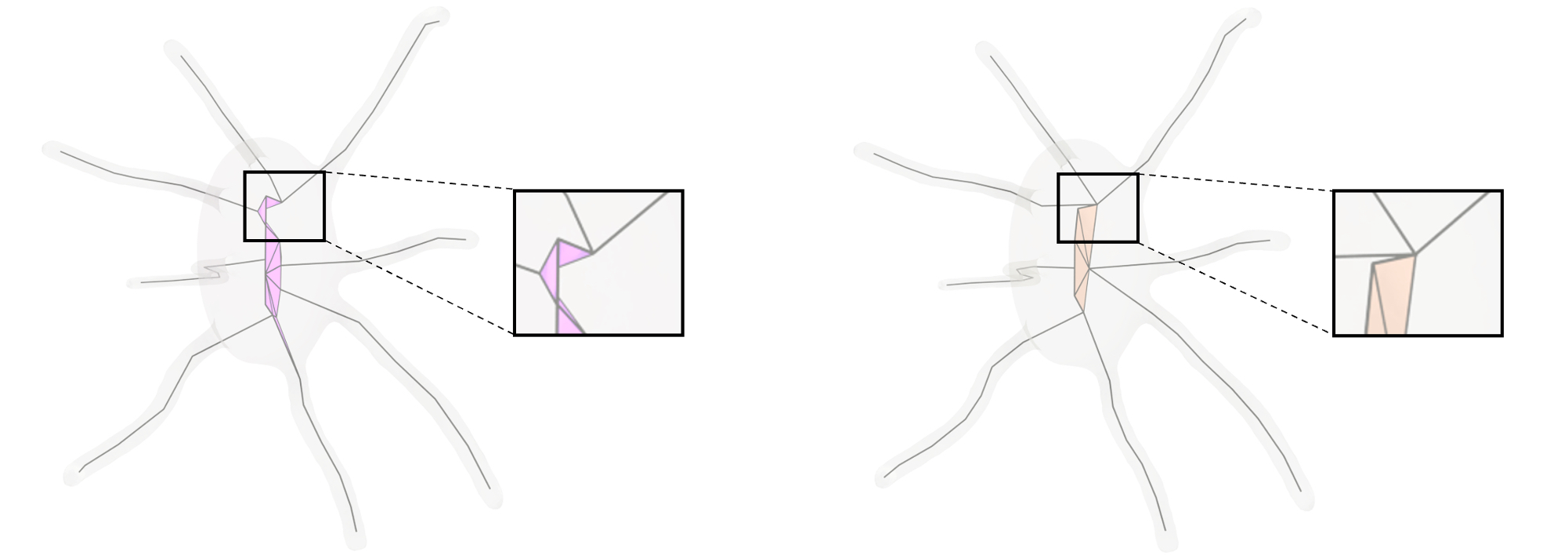}\\
    \vspace{1mm}
   \makebox[0.49\linewidth][c]{(a) Q-MAT}
        \makebox[0.49\linewidth][c]{(b) Coverage Axis}\\
        \vspace{-2mm}
    \caption{Comparison between Q-MAT and Coverage Axis on the quality of the generated medial surfaces. Note that the number of vertices is same.}
    \label{fig:qmat-failure1}
    \vspace{-2mm}
\end{figure}

From the experiments, it is revealed that the problem with Q-MAT is that it may yield an uneven distribution of inner points in parts where the geometry is similar  (Figure~\ref{fig:qmat-failure1}~(a) first and second row). This is because the edge collapse strategy of Q-MAT is solely based on local approximation errors, which lacks the global constraints of the whole geometry. This problem is even worse once a highly simplified representation is needed.
\begin{table}
\small
\begin{center}
\caption{Quantitative comparison on shape approximation error between Q-MAT and Coverage Axis.}
 \vspace{-1mm}
\label{table:QMAT}
\setlength{\tabcolsep}{0.45mm}{
\begin{tabular}{p{1.19cm}|p{0.8cm}<{\centering}|ccc|ccc}
\hline
\multirow{2}{*}{Model} & \multirow{2}{*}{ \makecell[c]{$|V|$} } & \multicolumn{3}{c|}{Q-MAT} & \multicolumn{3}{c}{Coverage Axis}\\
  &  &      $\overrightarrow{\epsilon}$ &  $\overleftarrow{\epsilon}$ &  $\overleftrightarrow{\epsilon}$ &   $\overrightarrow{\epsilon}$ &  $\overleftarrow{\epsilon}$ &  $\overleftrightarrow{\epsilon}$  \\
    \hline
    Ant-1 & $46$& $2.011\%$ & $2.236\%$ & $\textbf{2.236\%}$  & $ 2.894 \%$ & $ 2.601 \%$ & $ 2.894\%$ \\
    Armodillo &  $95$  & $4.178\%$ & $2.125\%$ & $ 4.178\%$ & $ 2.947 \%$ & $ 2.793 \%$ & $ \textbf{2.947\%}$ \\
    Bird & $54$ & $0.954\%$ & $1.539\%$ & $\textbf{ 1.539\%}$ & $ 1.682 \%$ & $ 1.816 \%$ & $ 1.816\%$ \\
    Bunny & $106$ & $2.127\%$ & $2.596\%$ & $ 2.596\%$ & $ 2.526 \%$ & $ 2.496 \%$ & $ \textbf{2.526\%}$ \\
    Chair-1 &  $79$ & $1.413\%$ & $1.534\%$ & $ \textbf{1.534\%}$ & $ 2.246 \%$ & $ 1.976 \%$ & $ 2.246\%$ \\ 
    Crab &$58$& $1.422\%$ & $1.794\%$ & $ 1.794\%$ & $ 1.458 \%$ & $ 1.759 \%$ & $ \textbf{1.759\%}$ \\
    Cup & $132$ &  $1.213\%$ & $2.481\%$ & $ \textbf{2.481\%}$ & $ 2.599 \%$ & $ 2.636 \%$ & $ 2.636\%$ \\
    Desk &$151$ & $1.068\%$ & $1.930\%$ & $ \textbf{1.930\%}$ & $ 2.273 \%$ & $ 2.164 \%$ & $ 2.273\%$ \\
    Dog & $49$&  $2.582\%$ & $3.266\%$ & $ 3.266\%$ & $ 2.077 \%$ & $ 1.850 \%$ & $ \textbf{2.077\%}$ \\
    Fertility & $79$ & $3.016\%$ & $4.259\%$ & $ 4.259\%$ & $ 2.703 \%$ & $ 2.377 \%$ & $ \textbf{2.703\%}$ \\
    Fish & $53$ & $2.336\%$ & $1.764\%$ & $ 2.336\%$ & $ 2.325 \%$ & $ 2.281 \%$ & $ \textbf{2.325\%}$ \\
Hand-1 & $42$ & $1.594\%$ & $1.780\%$ & $ 1.780\%$ & $ 1.334 \%$ & $ 1.750 \%$ & $ \textbf{1.750\%}$ \\
Hand-2 & $60$ & $1.203\%$ & $1.747\%$ & $ 1.747\%$ & $ 1.715 \%$ & $ 1.716 \%$ & $ \textbf{1.716\%}$ \\
Human-1 & $45$ & $3.231\%$ & $1.929\%$ & $ 3.231\%$ & $ 2.729 \%$ & $ 2.644 \%$ & $ \textbf{2.729\%}$ \\ 
Human-2 & $46$ & $1.614\%$ & $1.572\%$ & $ 1.614\%$ & $ 1.498 \%$ & $ 1.526 \%$ & $ \textbf{1.526\%}$ \\ 
Human-3 & $61$ & $2.265\%$ & $1.561\%$ & $ 2.265\%$ & $ 1.788 \%$ & $ 1.707 \%$ & $ \textbf{1.788\%}$ \\
Kitten & $45$ & $2.490\%$ & $2.309\%$ & $ \textbf{2.490\%}$ & $ 2.875 \%$ & $ 2.893 \%$ & $ 2.893\%$ \\
Octopus-1 & $49$ & $1.811\%$ & $1.845\%$ & $1.845\%$ & $1.825\%$ & $1.836\%$ & $\textbf{1.836\%}$ \\
Octopus-2 & $68$ &  $2.356\%$ & $1.765\%$ & $ 2.356\%$ & $ 2.256 \%$ & $ 2.316 \%$ & $ \textbf{2.316\%}$ \\
Pig & $56$ & $2.503\%$ & $2.420\%$ & $ 2.503\%$ & $ 2.107 \%$ & $ 2.183 \%$ & $ \textbf{2.183\%}$ \\ 
Plane & $53$ & $1.108\%$ & $1.184\%$ & $ \textbf{1.184\%}$ & $ 1.430 \%$ & $ 1.424 \%$ & $ 1.430\%$ \\
Snake & $39$ & $1.183\%$ & $1.639\%$ & $ 1.639\%$ & $ 1.335 \%$ & $ 1.623 \%$ & $ \textbf{ 1.623\%}$ \\
Spectacle & $45$ & $2.191\%$ & $1.061\%$ & $ 2.191\%$ & $ 2.133 \%$ & $ 1.542 \%$ & $ \textbf{2.133\%}$ \\
Bear & $32$& $2.348\%$ & $2.619\%$ & $ \textbf{2.619\%}$ & $ 3.185 \%$ & $ 3.229 \%$ & $3.229\%$ \\
Vase & $106$ & $2.906\%$ & $2.453\%$ & $ \textbf{2.906\%}$ & $ 3.183 \%$ & $ 2.954 \%$ & $ 3.183\%$ \\
\hline
{Average} & {-} & {$2.045\%$} & {$2.056\%$} & {$2.341\%$} & {$2.205 \%$} & {$ 2.164 \%$} & $\textbf{2.264\%}$ \\
\hline
\end{tabular}}
\end{center}
 \begin{tablenotes}
  \small
     \item[1] $|V|$ \textit{The number of skeletal points.}
     \item[2] $\overrightarrow{\epsilon}$ \textit{One-sided HD from surface to reconstruction.}
     \item[3] $\overleftarrow{\epsilon}$ \textit{One-sided HD from reconstruction to surface.}
     \item[4] $\overleftrightarrow{\epsilon}$  \textit{Two-sided HD between original surface and reconstruction.}
  \end{tablenotes}
\vspace{-2mm}
\end{table}

\begin{figure}
    \centering
    \begin{overpic}[width=\linewidth]{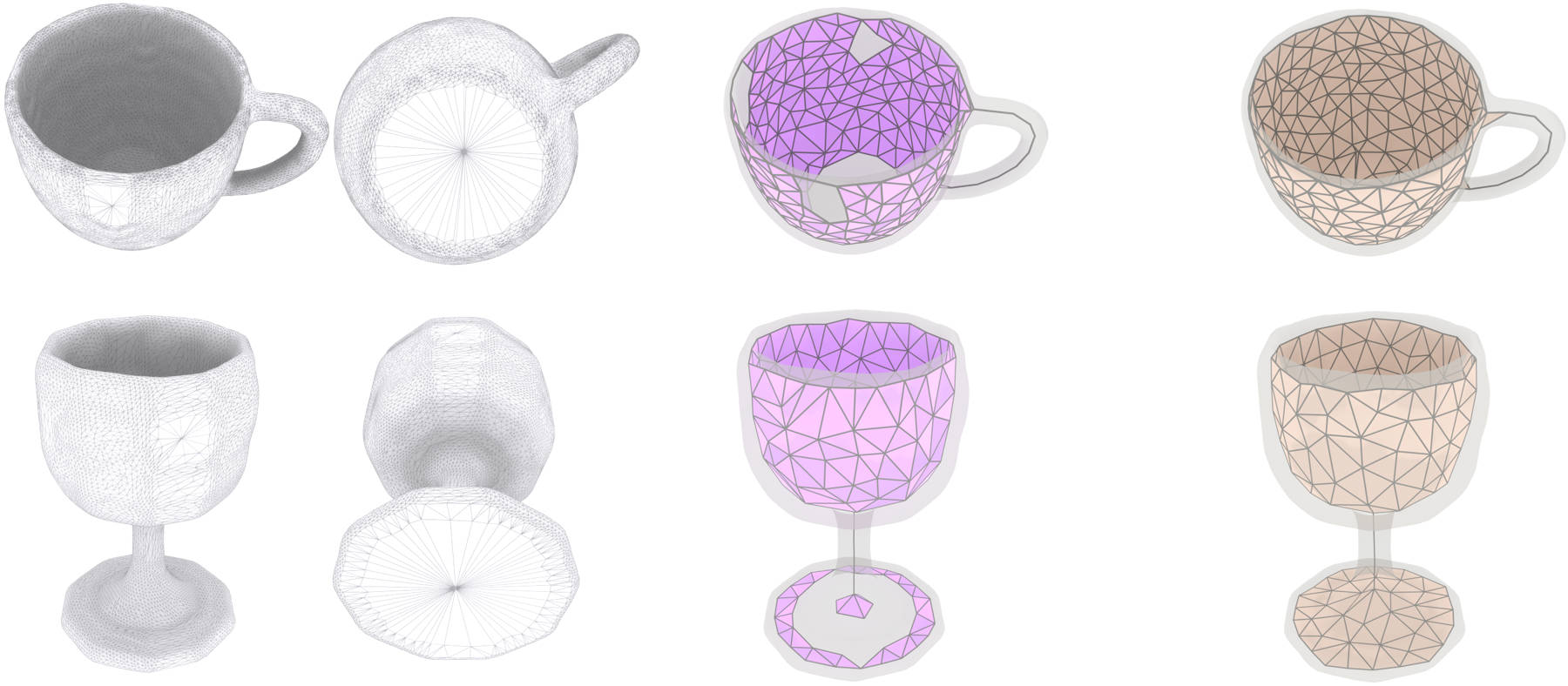}
\put(9,-5){(a) Input Mesh}
\put(44,-5){(b) Q-MAT}
\put(72,-5){(c) Coverage Axis}
\end{overpic}
\vspace{-1mm}
\caption{Comparison with Q-MAT for poor-quality mesh that can fail the construction of Voronoi Diagram.}
    \label{fig:qmat_failure2}
\end{figure}

In contrast, our method effectively leverages the global geometric and topological features of the input shape for skeleton generation. The point selection step jointly considers the overall input points and candidates, while the selected expressive points are used as fixed anchors to preserve the global topology. Hence, the skeletal representations generated by our method are equipped with shape-aware point distributions as well as connectivity. 

It is worth noting that Q-MAT heavily relies on the quality of the input mesh and the initial Voronoi diagram to compute the medial axis skeleton. In contrast, our method does not have this restriction. Figure~\ref{fig:qmat_failure2} shows a group of examples where Q-MAT yields incorrect structures due to the failure of initialization caused by the poor mesh quality, where our algorithm successfully gives the proper skeletal representation.

\begin{figure}[H]
\centering
\subfigure{
\small
    \begin{minipage}[t]{0.23\linewidth}
        \centering
        \makebox[\linewidth][c]{ {$\alpha = 1.1$}}\\                \makebox[\linewidth][c]{}\\
        \includegraphics[width=0.86\linewidth]{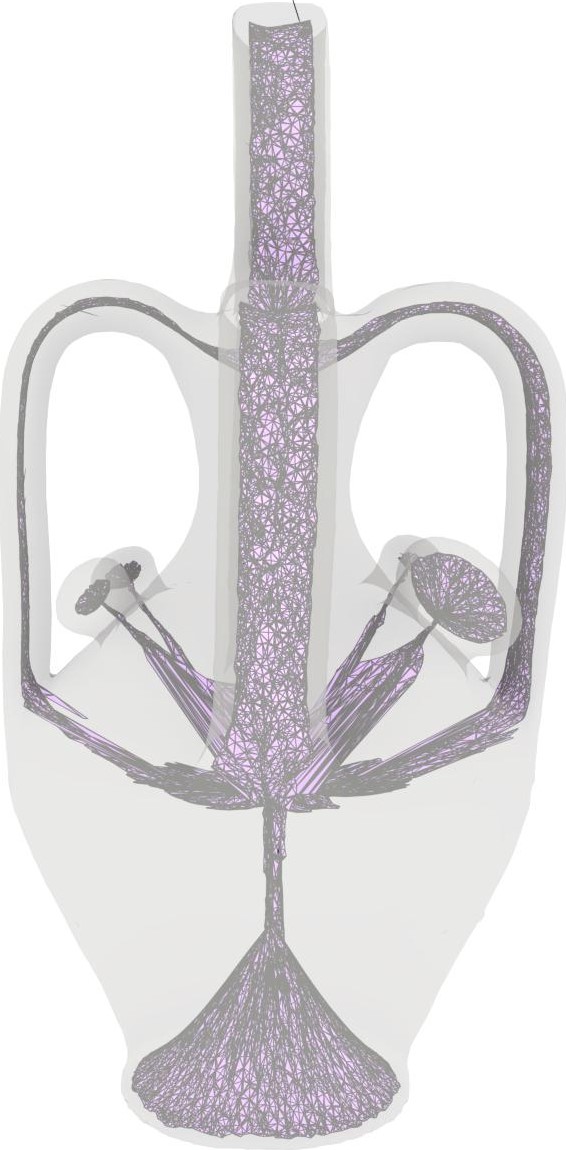}\\
    \makebox[\linewidth][c]{$|V|=91\text{k}$}\\
    \makebox[\linewidth][c]{$\epsilon=2.021\%$}\\
    \vspace{3mm}
        \includegraphics[width=1.05\linewidth]{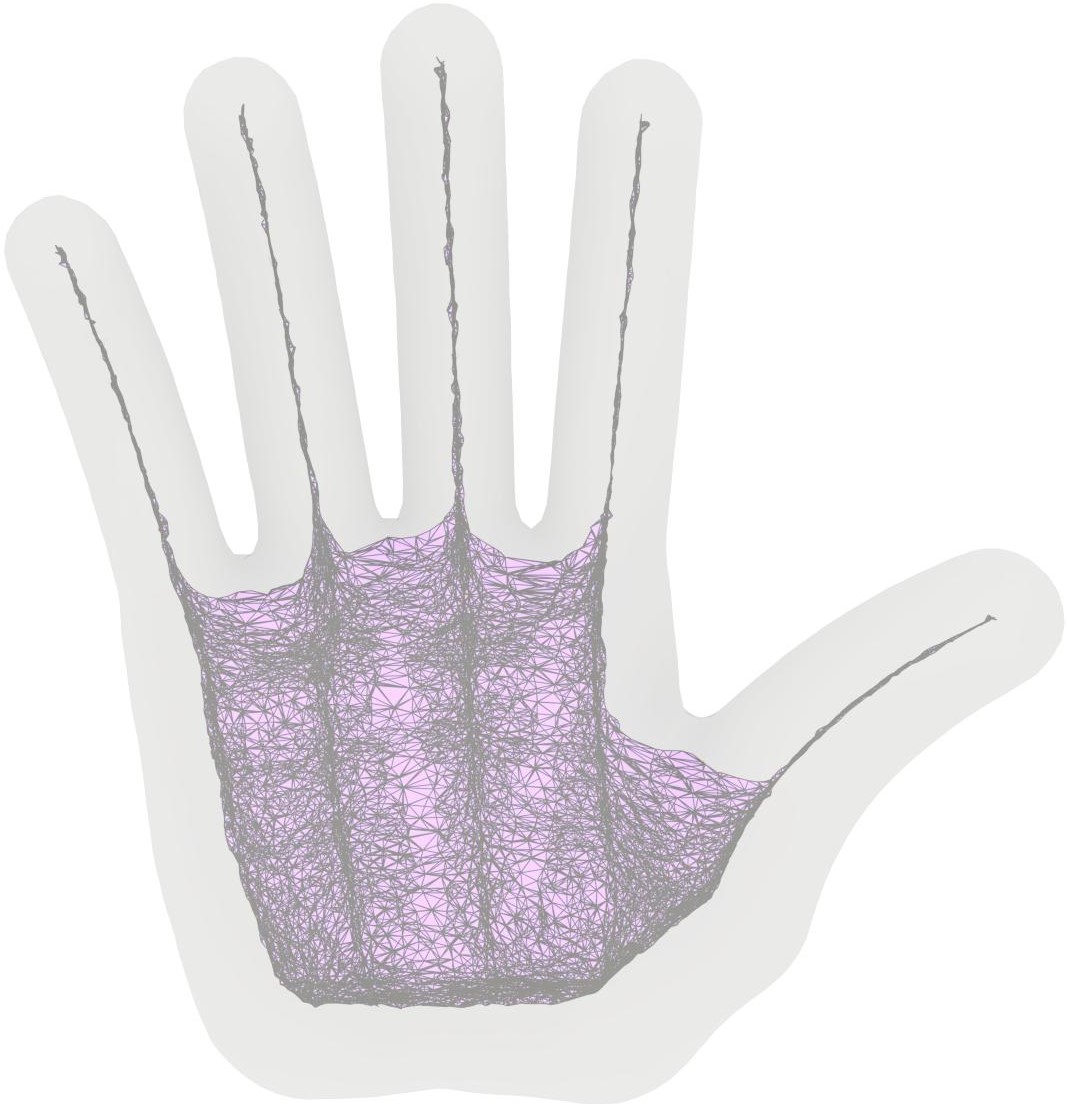}\\
    \makebox[\linewidth][c]{ $|V|=64\text{k}$}\\
    \makebox[\linewidth][c]{$\epsilon=1.688\%$}\\
        \vspace{4mm}
        \includegraphics[width=\linewidth]{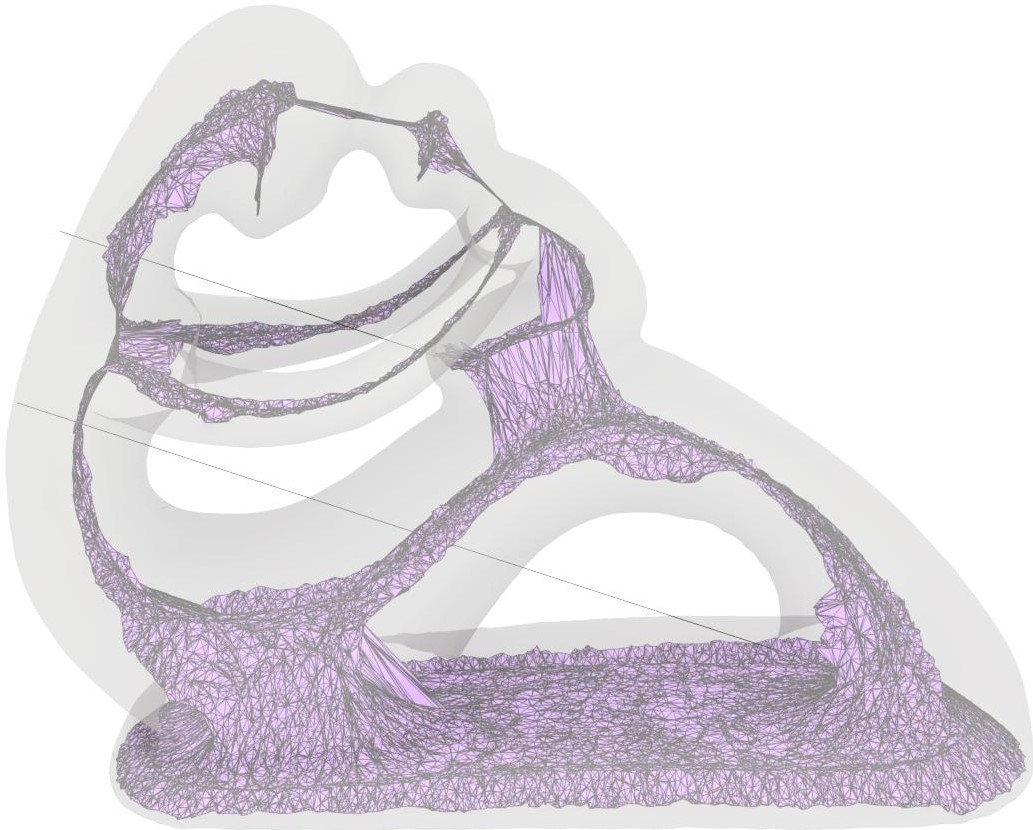}\\
    \makebox[\linewidth][c]{ $|V|=110\text{k}$}\\
    \makebox[\linewidth][c]{$\epsilon=1.883\%$}\\
        \vspace{3mm}
         \includegraphics[width=\linewidth]{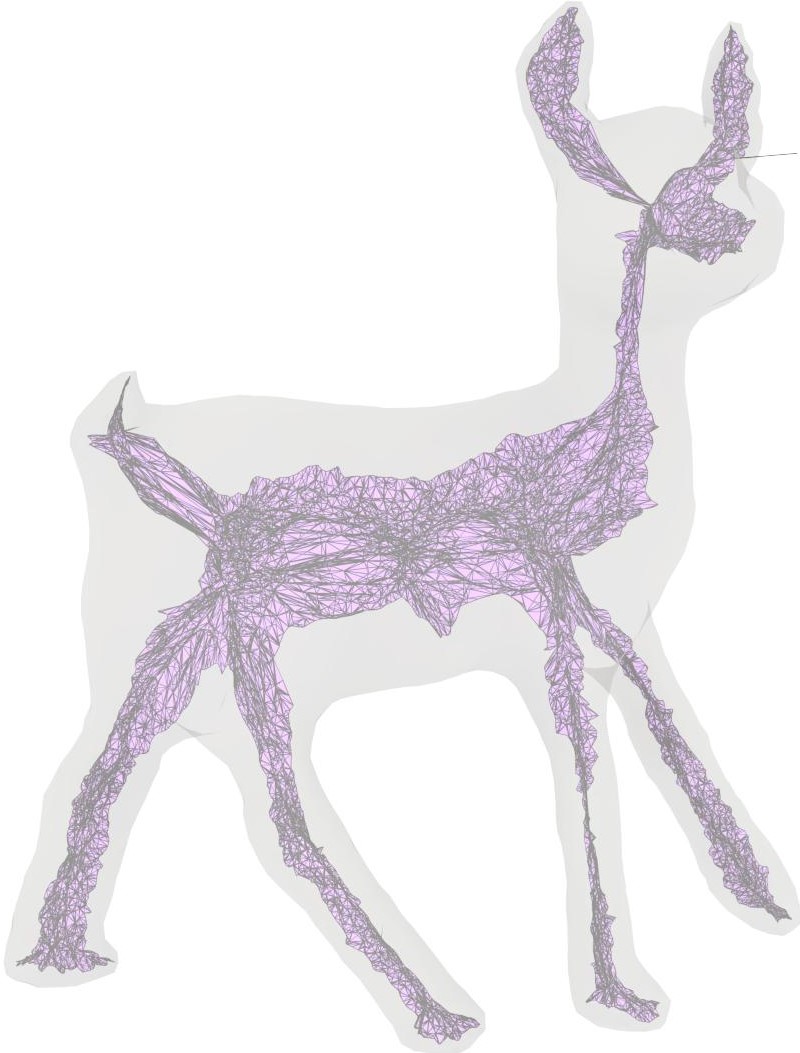}\\
    \makebox[\linewidth][c]{ $|V|=55\text{k}$}\\
    \makebox[\linewidth][c]{$\epsilon=1.743\%$}\\
        \vspace{3mm}
         \includegraphics[width=1.1\linewidth]{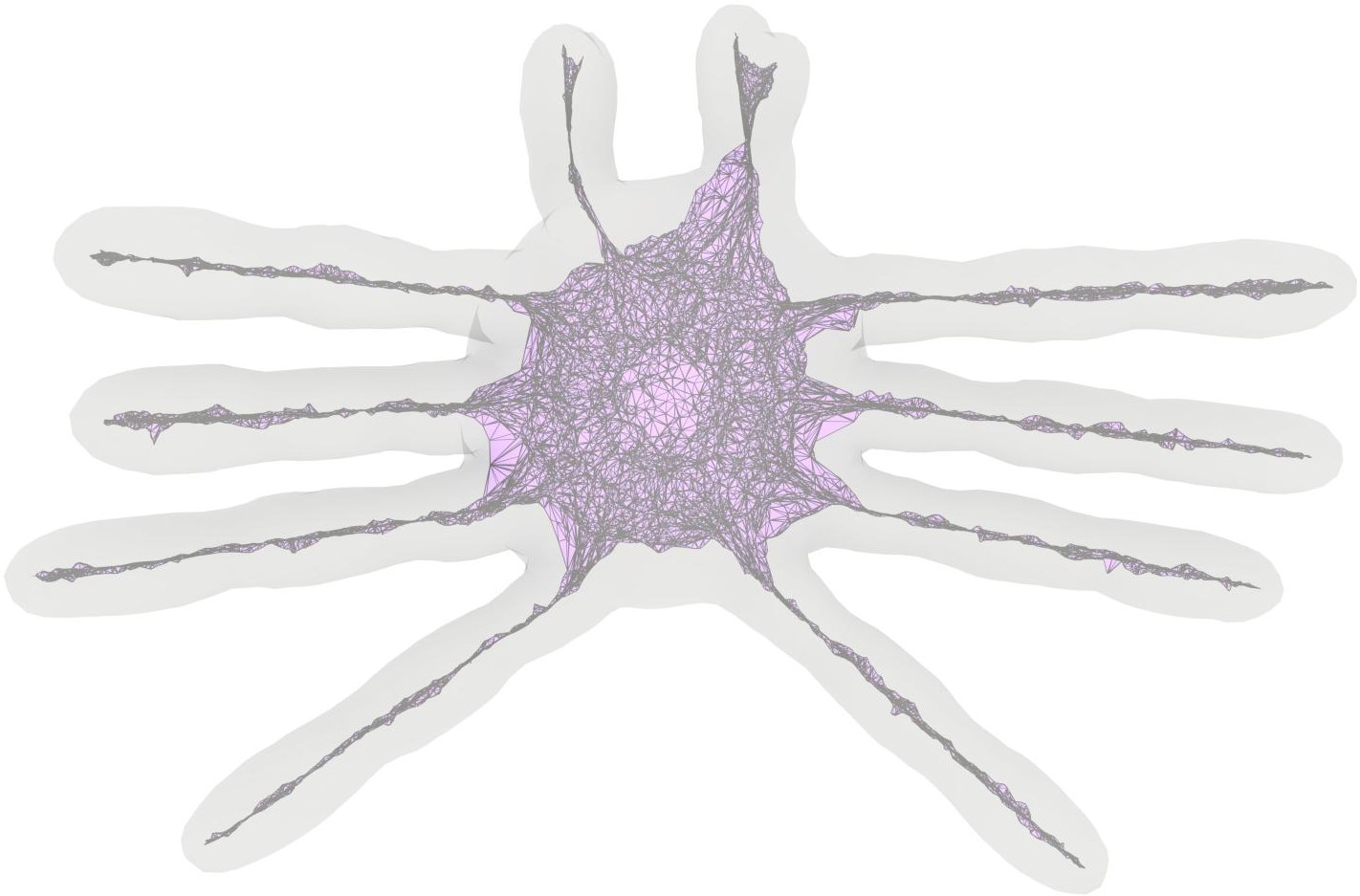}\\
    \makebox[\linewidth][c]{ $|V|=61\text{k}$}\\
    \makebox[\linewidth][c]{$\epsilon=1.336\%$}\\
    \end{minipage}%
}
\subfigure{
\small
    \begin{minipage}[t]{0.23\linewidth}
        \centering
        \makebox[\linewidth][c]{ $\alpha = 1.5$}\\                \makebox[\linewidth][c]{}\\
        \includegraphics[width=0.86\linewidth]{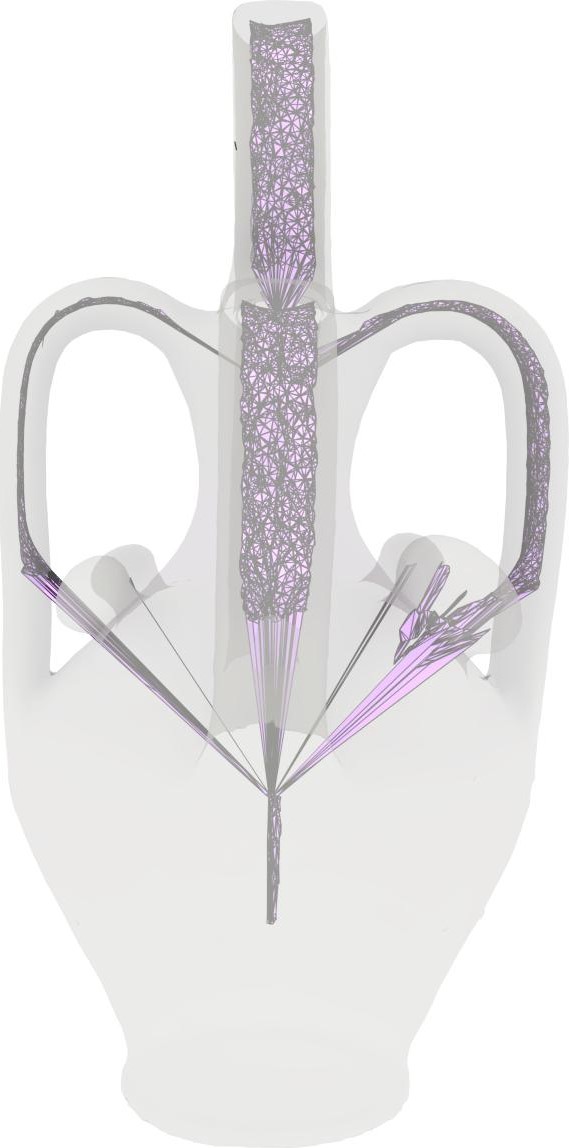}\\
    \makebox[\linewidth][c]{ $|V|=38\text{k}$}\\
    \makebox[\linewidth][c]{$\epsilon=6.858\%$}\\
        \vspace{3mm}
        \includegraphics[width=1.05\linewidth]{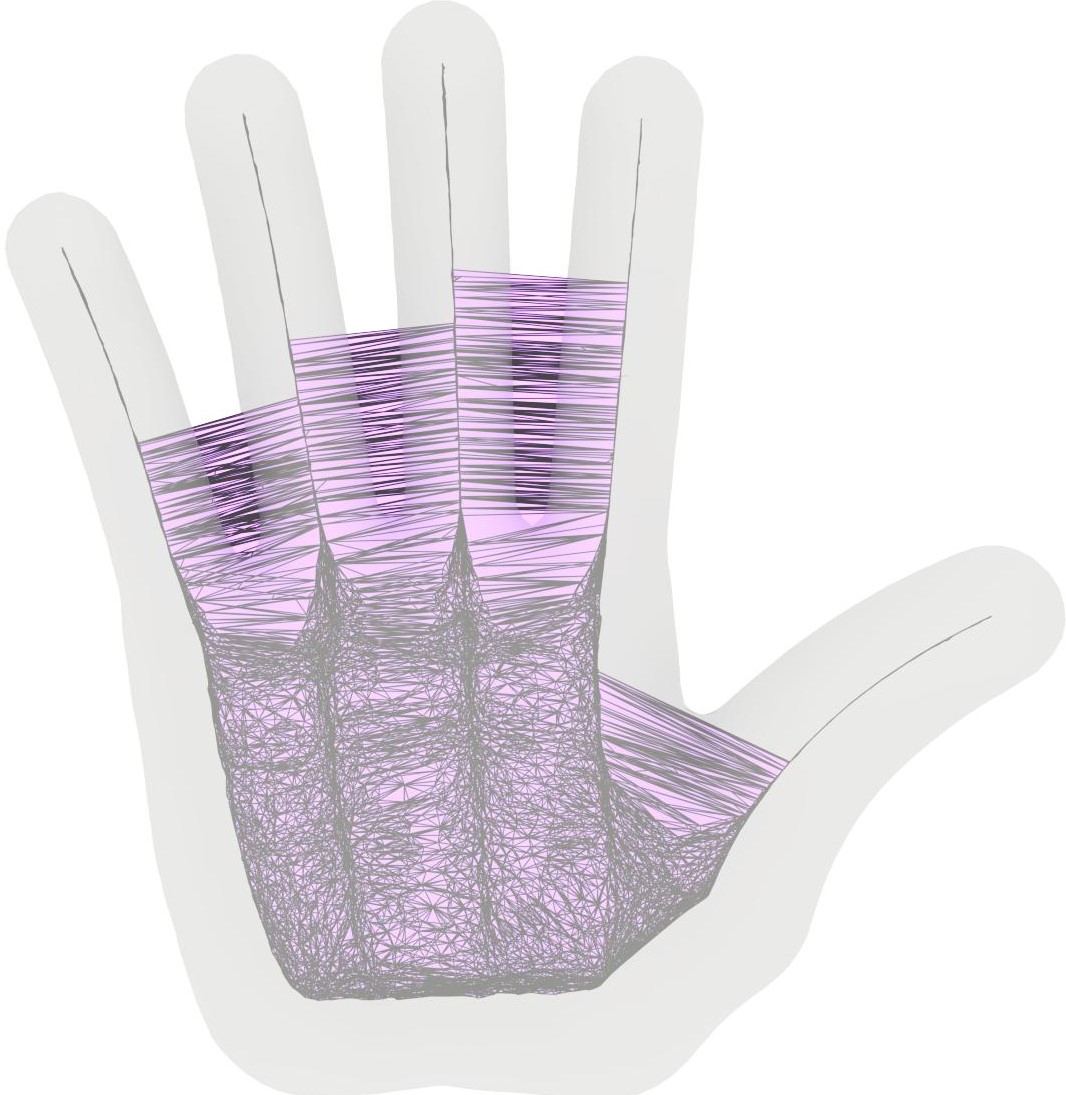}\\
    \makebox[\linewidth][c]{ $|V|=47\text{k}$}\\
    \makebox[\linewidth][c]{$\epsilon=3.556\%$}\\
            \vspace{4mm}
        \includegraphics[width=\linewidth]{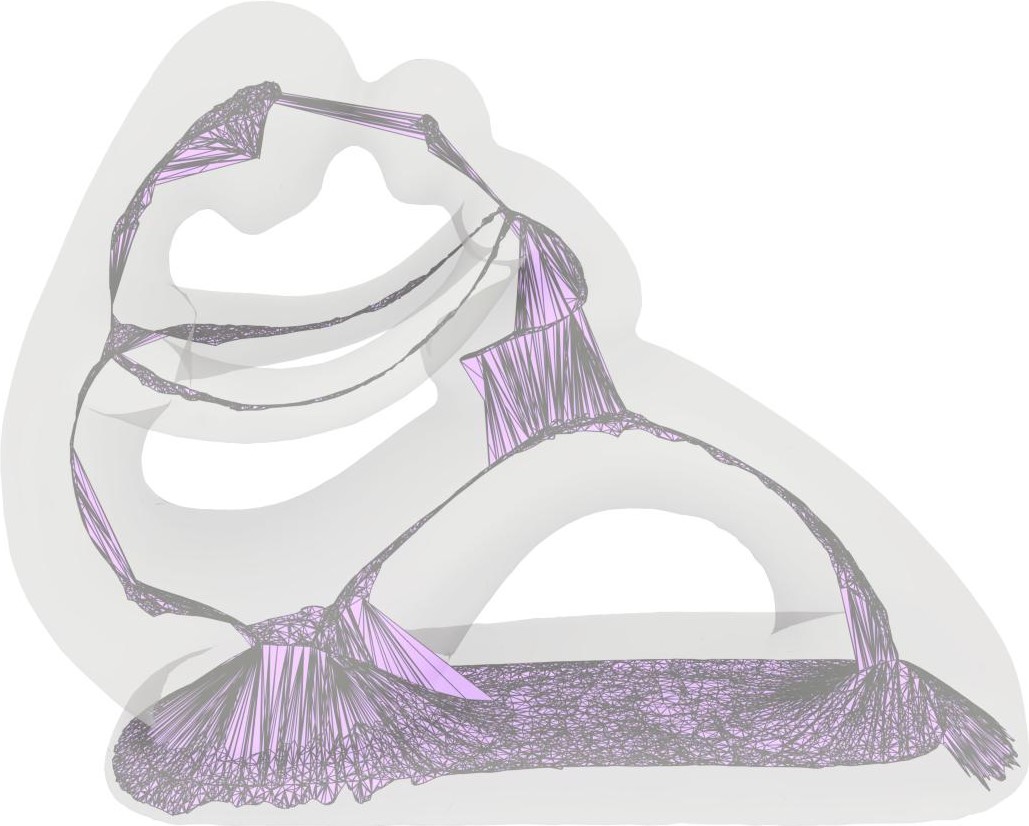}\\
    \makebox[\linewidth][c]{ $|V|=56\text{k}$}\\
    \makebox[\linewidth][c]{$\epsilon=2.735\%$}\\
            \vspace{3mm}
         \includegraphics[width=\linewidth]{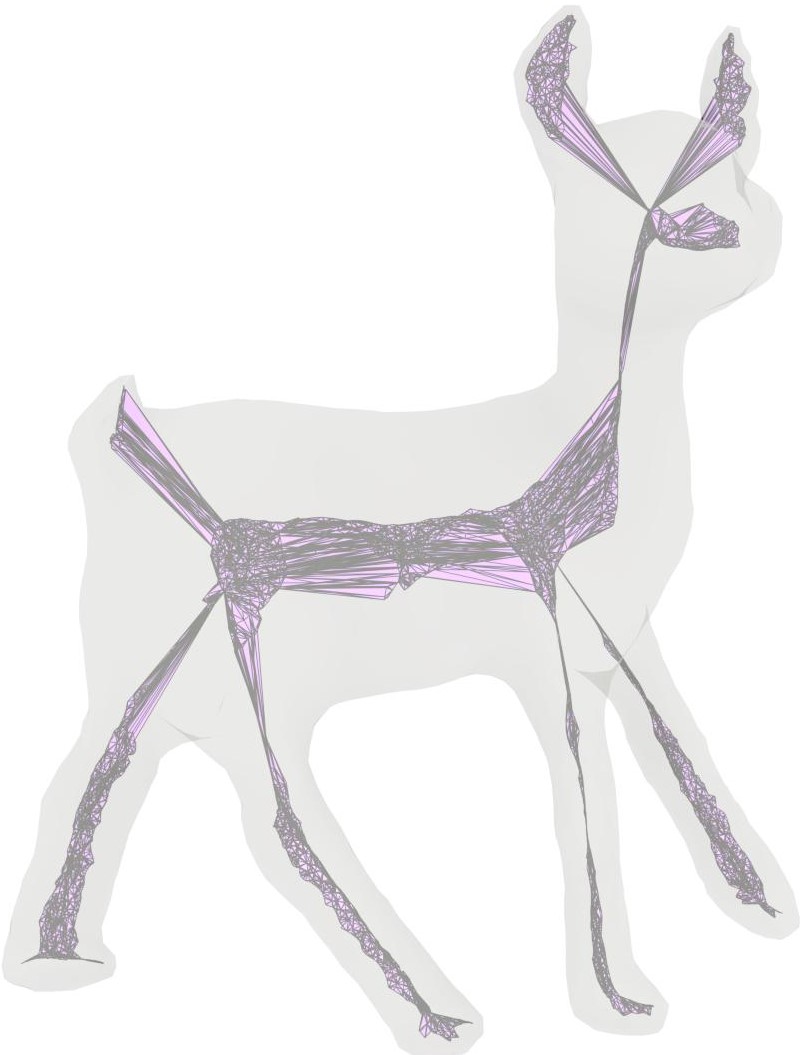}\\
    \makebox[\linewidth][c]{ $|V|=26\text{k}$}\\
    \makebox[\linewidth][c]{$\epsilon=2.507\%$}\\
            \vspace{3mm}
         \includegraphics[width=1.1\linewidth]{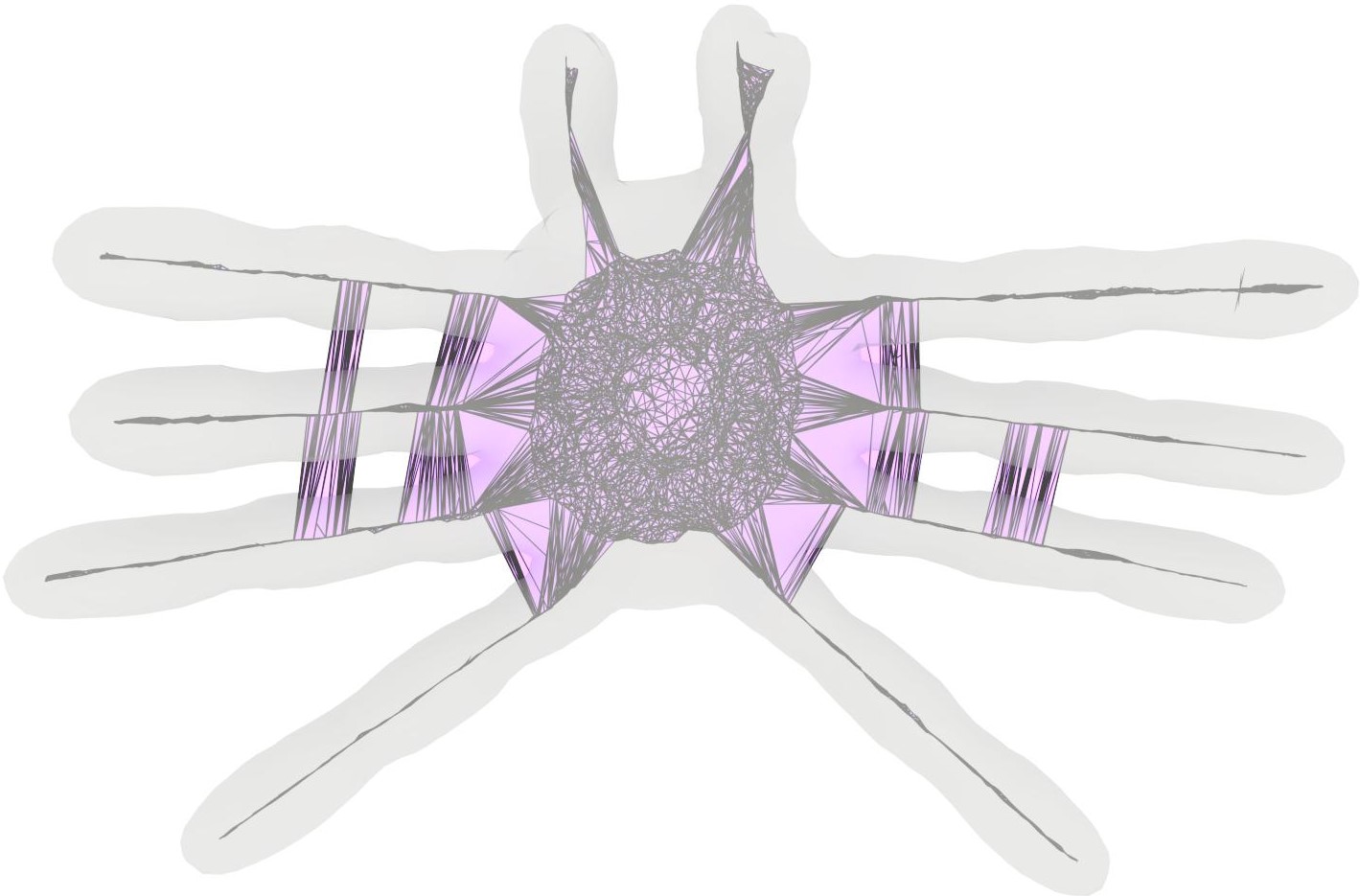}\\
    \makebox[\linewidth][c]{ $|V|=37\text{k}$}\\
    \makebox[\linewidth][c]{$\epsilon=2.953\%$}\\
    \end{minipage}%
}
\subfigure{
\small
    \begin{minipage}[t]{0.23\linewidth}
        \centering
        \makebox[\linewidth][c]{ $\alpha = 2.0$}\\                \makebox[\linewidth][c]{}\\
        \includegraphics[width=0.86\linewidth]{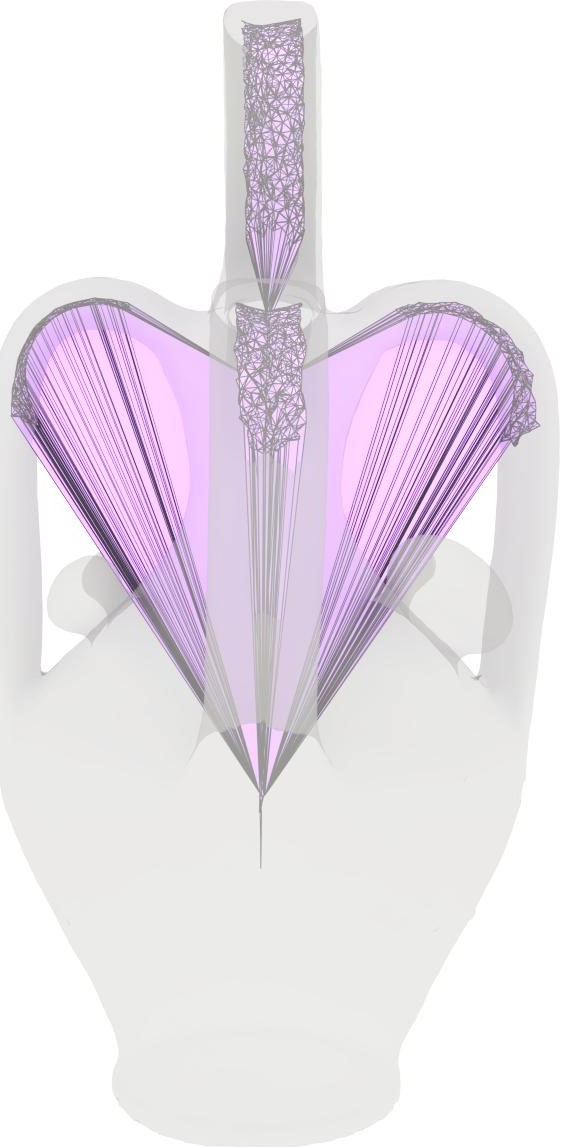}\\
        \makebox[\linewidth][c]{ $|V|=16\text{k}$}\\
        \makebox[\linewidth][c]{$\epsilon=8.581\%$}\\
                    \vspace{3mm}
        \includegraphics[width=1.05\linewidth]{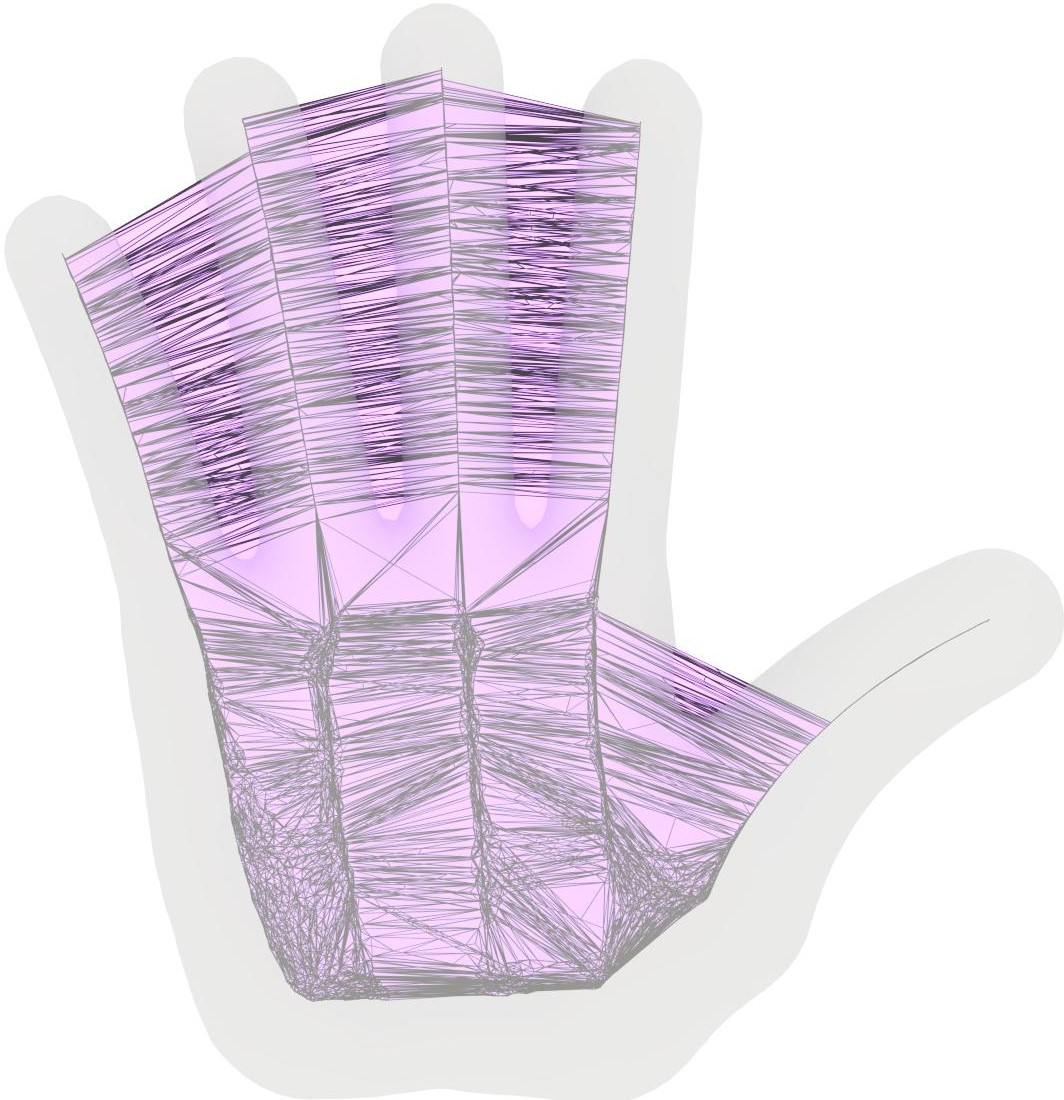}\\      
        \makebox[\linewidth][c]{ $|V|=58\text{k}$}\\
        \makebox[\linewidth][c]{$\epsilon=4.455\%$}\\
                    \vspace{4mm}
        \includegraphics[width=\linewidth]{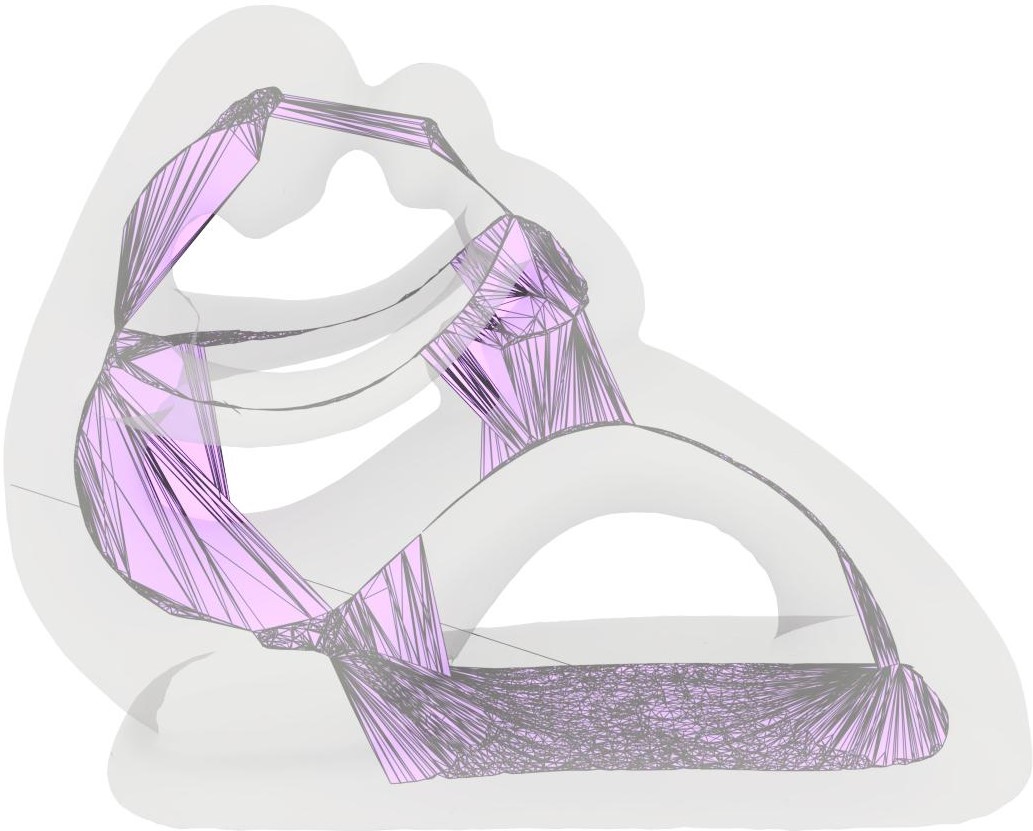}\\
        \makebox[\linewidth][c]{ $|V|=28\text{k}$}\\
        \makebox[\linewidth][c]{$\epsilon=6.690\%$}\\
                    \vspace{3mm}
         \includegraphics[width=\linewidth]{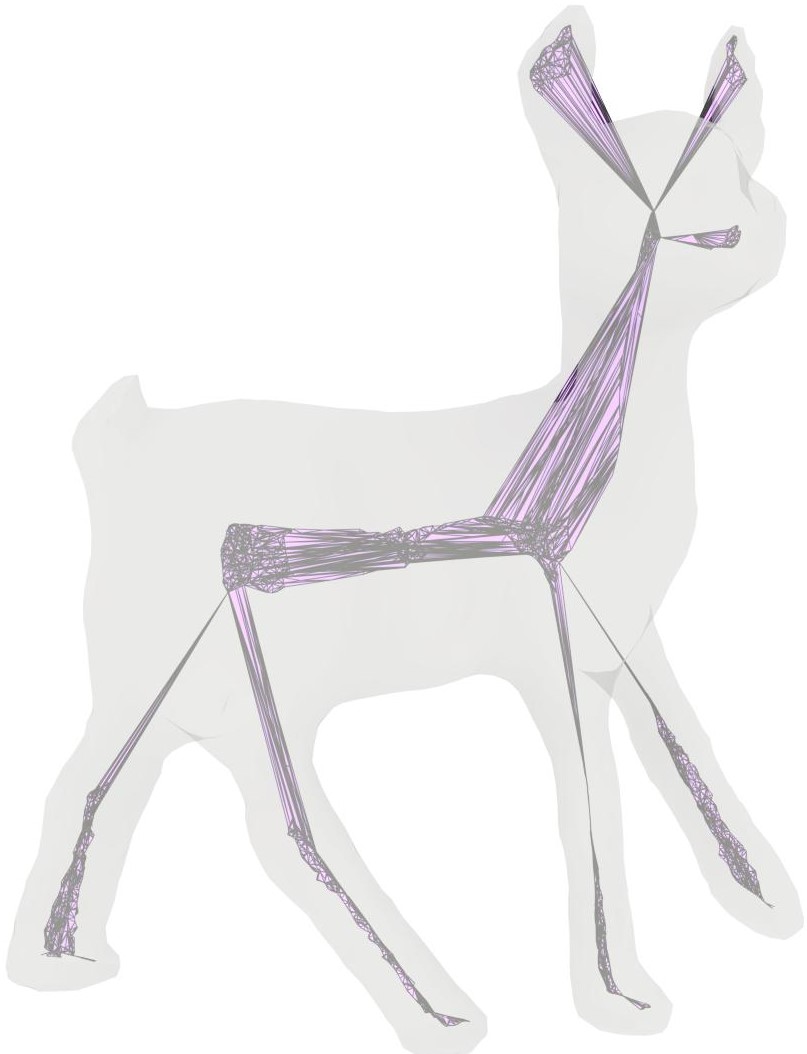}\\
        \makebox[\linewidth][c]{ $|V|=11\text{k}$}\\
        \makebox[\linewidth][c]{$\epsilon=4.662\%$}\\
                    \vspace{3mm}
         \includegraphics[width=1.1\linewidth]{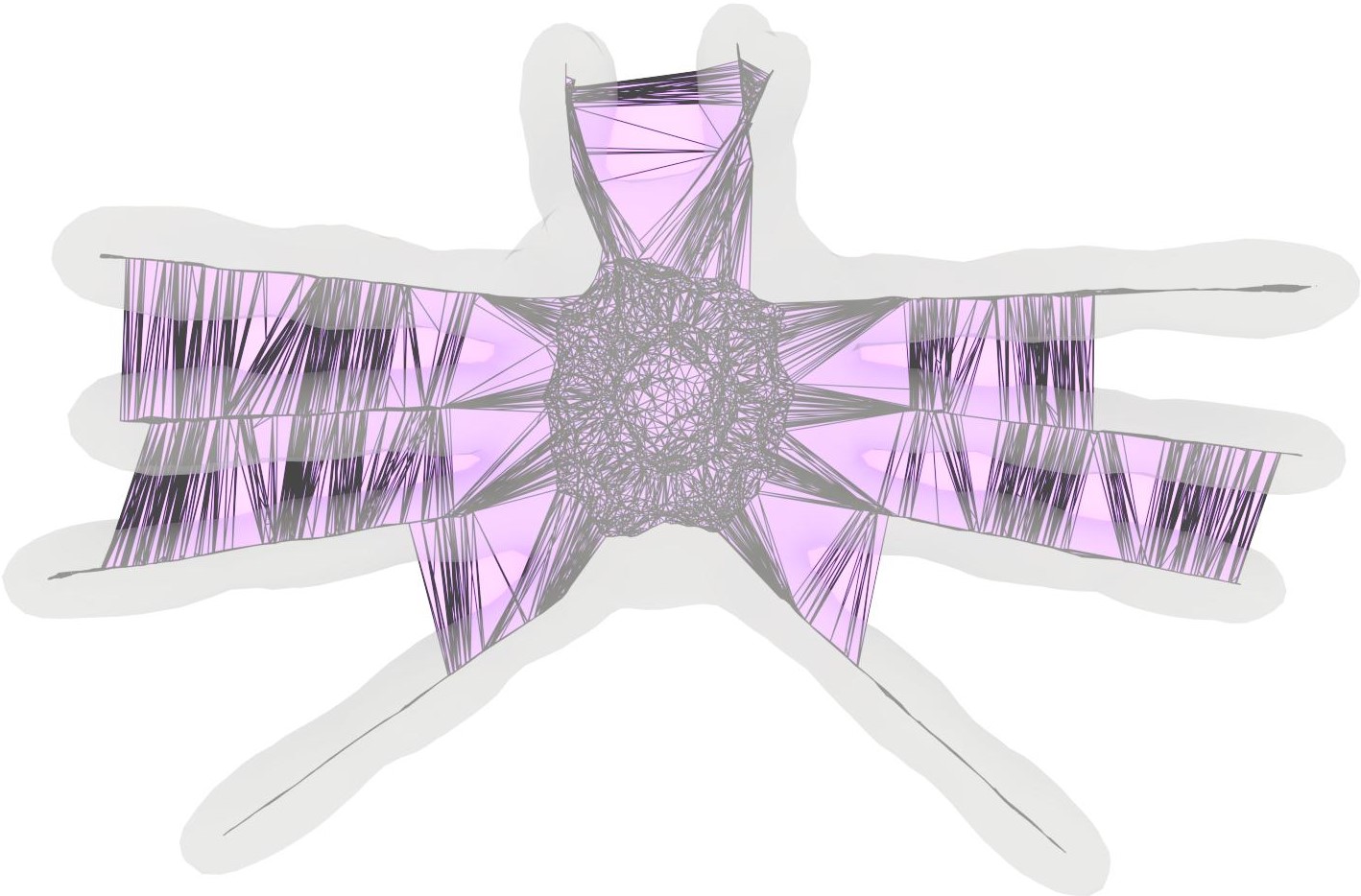}\\
        \makebox[\linewidth][c]{ $|V|=26\text{k}$}\\
        \makebox[\linewidth][c]{$\epsilon=4.290\%$}
    \end{minipage}}
\subfigure{
\small
    \begin{minipage}[t]{0.23\linewidth}
        \centering
        \makebox[\linewidth][c]{ offset=$0.02$}\\                \makebox[\linewidth][c]{}\\
        \includegraphics[width=0.86\linewidth]{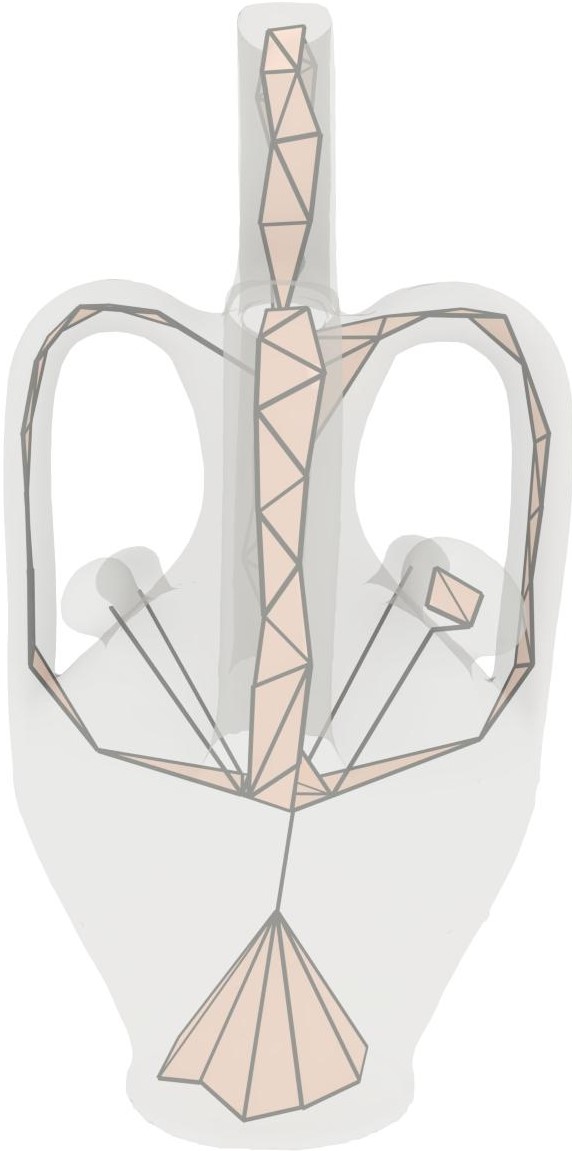}\\ 
        \makebox[\linewidth][c]{ $|V|=106$}\\
        \makebox[\linewidth][c]{$\epsilon=3.183\%$}\\
                        \vspace{3mm}
        \includegraphics[width=1.05\linewidth]{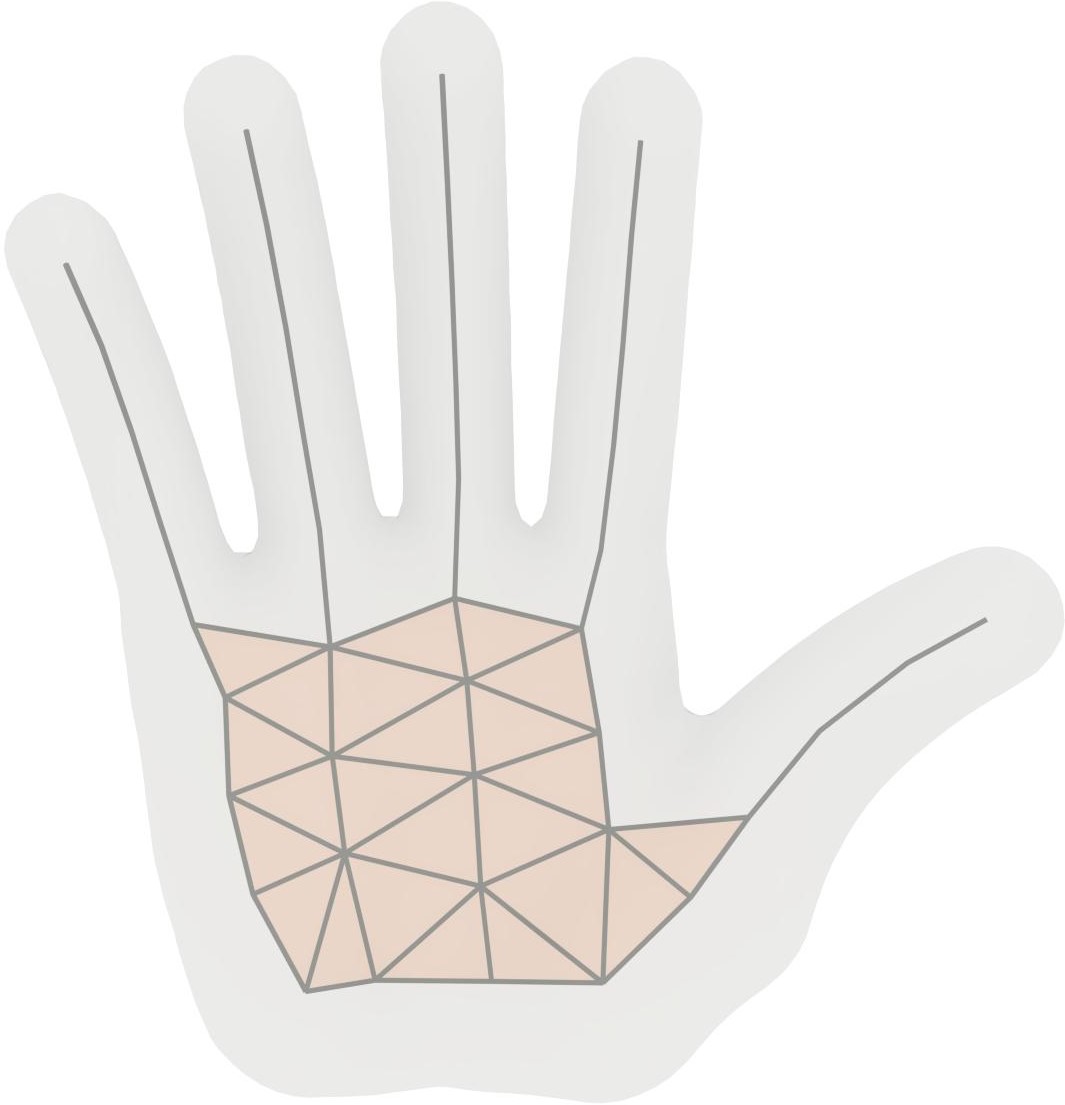}\\
        \makebox[\linewidth][c]{ $|V|=42$}\\
        \makebox[\linewidth][c]{$\epsilon=1.750\%$}\\
                        \vspace{4mm}
        \includegraphics[width=\linewidth]{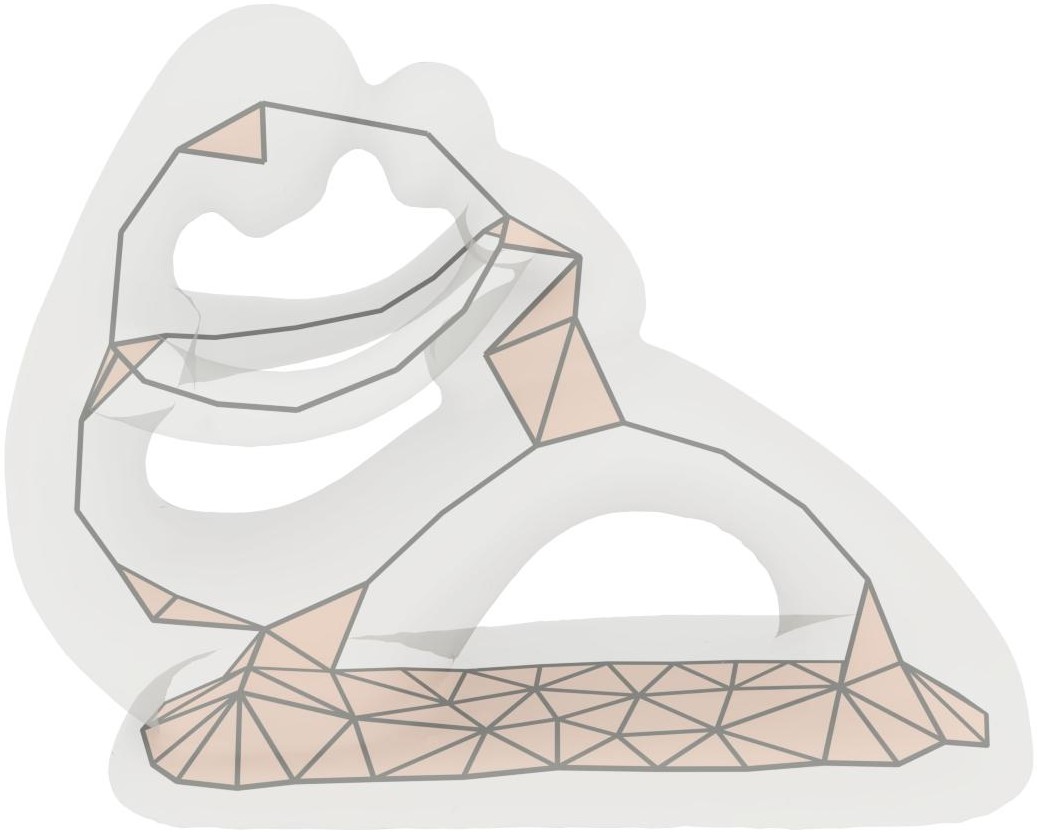}\\
        \makebox[\linewidth][c]{ $|V|=79$}\\
        \makebox[\linewidth][c]{$\epsilon=2.703\%$}\\
                        \vspace{3mm}
         \includegraphics[width=\linewidth]{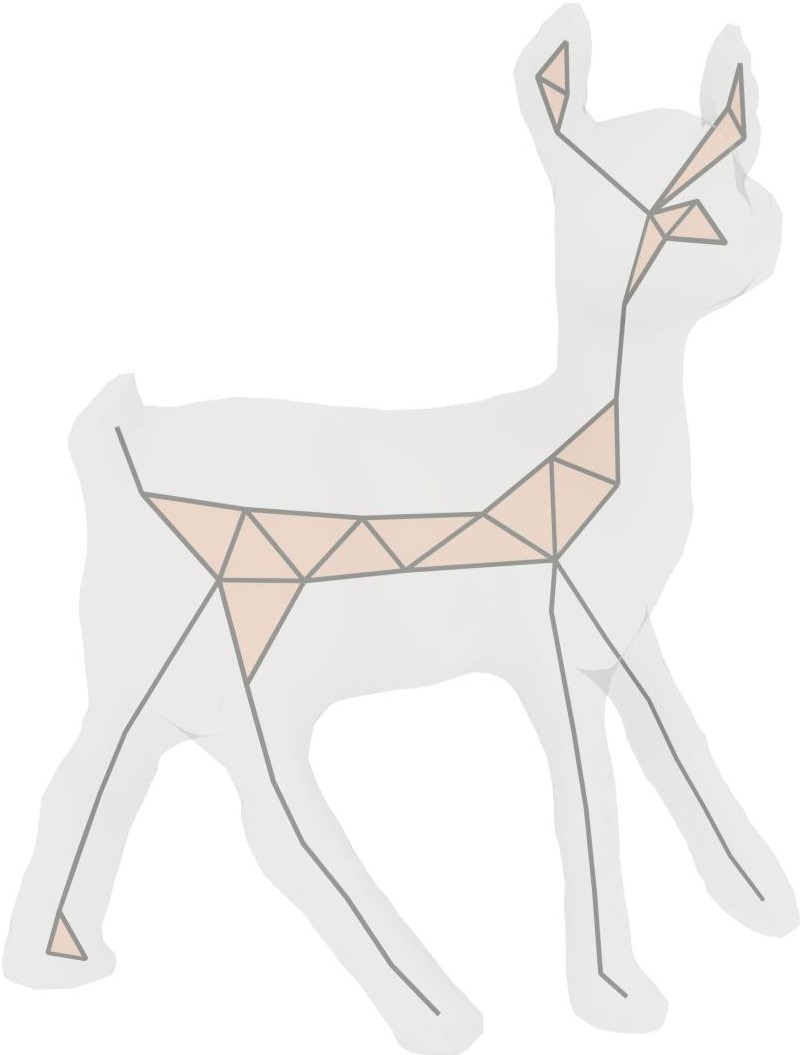}\\        
        \makebox[\linewidth][c]{ $|V|=49$}\\
        \makebox[\linewidth][c]{$\epsilon=2.077\%$}\\
                        \vspace{3mm}
         \includegraphics[width=1.1\linewidth]{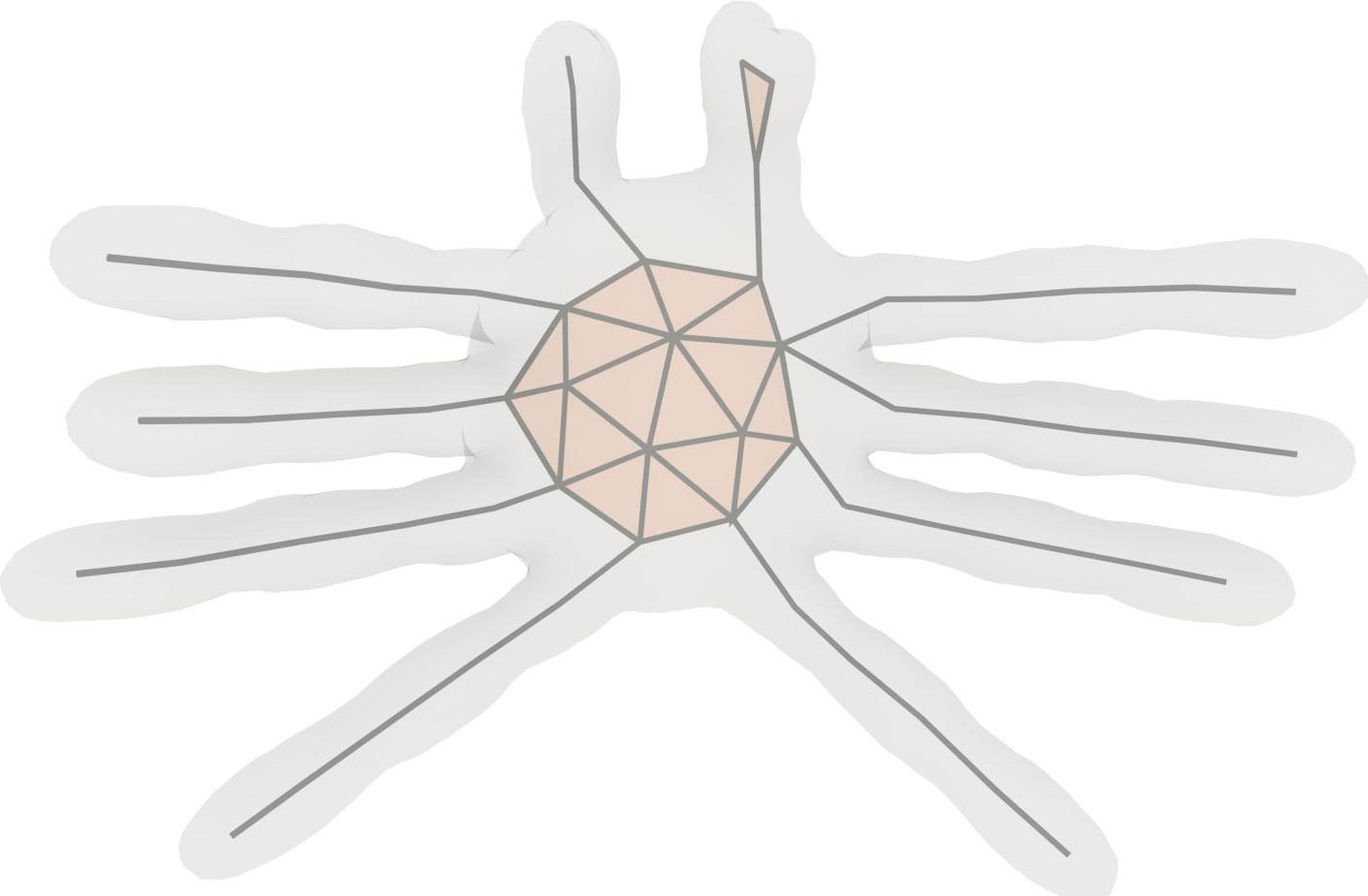}\\  
        \makebox[\linewidth][c]{ $|V|=58$}\\
        \makebox[\linewidth][c]{$\epsilon=1.759\%$}\\
    \end{minipage}%
}%
\\
 \makebox[0.01\linewidth][l]{ }
   \makebox[0.25\linewidth][c]{(a)}
   \makebox[0.23\linewidth][c]{(b)}
   \makebox[0.24\linewidth][c]{(c)}
        \makebox[0.23\linewidth][c]{(d)}\\
        \vspace{-2mm}
  \caption{Qualitative comparison with SAT. (a-c) SAT with mesh input. (d) Coverage Axis with mesh input. Note that Coverage Axis (ours) achieves trade-offs across fidelity, efficiency and compression abilities compared with SAT.}
    \label{fig:SAT}
\end{figure}

\paragraph{Comparison with SAT~\cite{miklos2010discrete}}
One of the main setbacks of this method is that SAT favors a dense representation with a large number of vertices, thus incapable of generating a simple and compact skeleton for shape abstraction. In this comparison with SAT, we test different values of its scaling factor, i.e., $\alpha = 1.1$, $1.5$ and $2.0$. The qualitative and quantitative comparison results are shown in Figure~\ref{fig:SAT}. On the one hand, despite high accuracy yielded by SAT using smaller $\alpha$, its representation has a large number of vertices (Figure~\ref{fig:SAT}~(a-c)). This redundant representation is not amenable for various applications that require skeleton to be simple and expressive, e.g., shape matching and retrieval~\cite{sundar2003skeleton}, skeletal animation~\cite{magnenat1988joint}, animated mesh approximation~\cite{yang2018dmat, thiery2016animated}, etc. On the other hand, using a larger $\alpha$ for a higher abstraction level destroys the shape structure and causes large approximation errors. Our method, on the contrary, provides trade-offs across fidelity, efficiency and compression abilities.

Both Q-MAT and SAT can only excel at handling watertight surface meshes. As aforementioned, the surface mesh input allows robust computation of the Voronoi diagram, which offers good structural information from both inside and outside. Instead, generating skeletal representations for point clouds is more challenging and beyond the capacity of these methods. Our method is able to take unstructured point sets as input and generate skeletal representations of which connected structures are recovered. We will evaluate this property of our method in the following section.

\begin{figure*}
\centering
\begin{overpic}[width=0.98\linewidth]{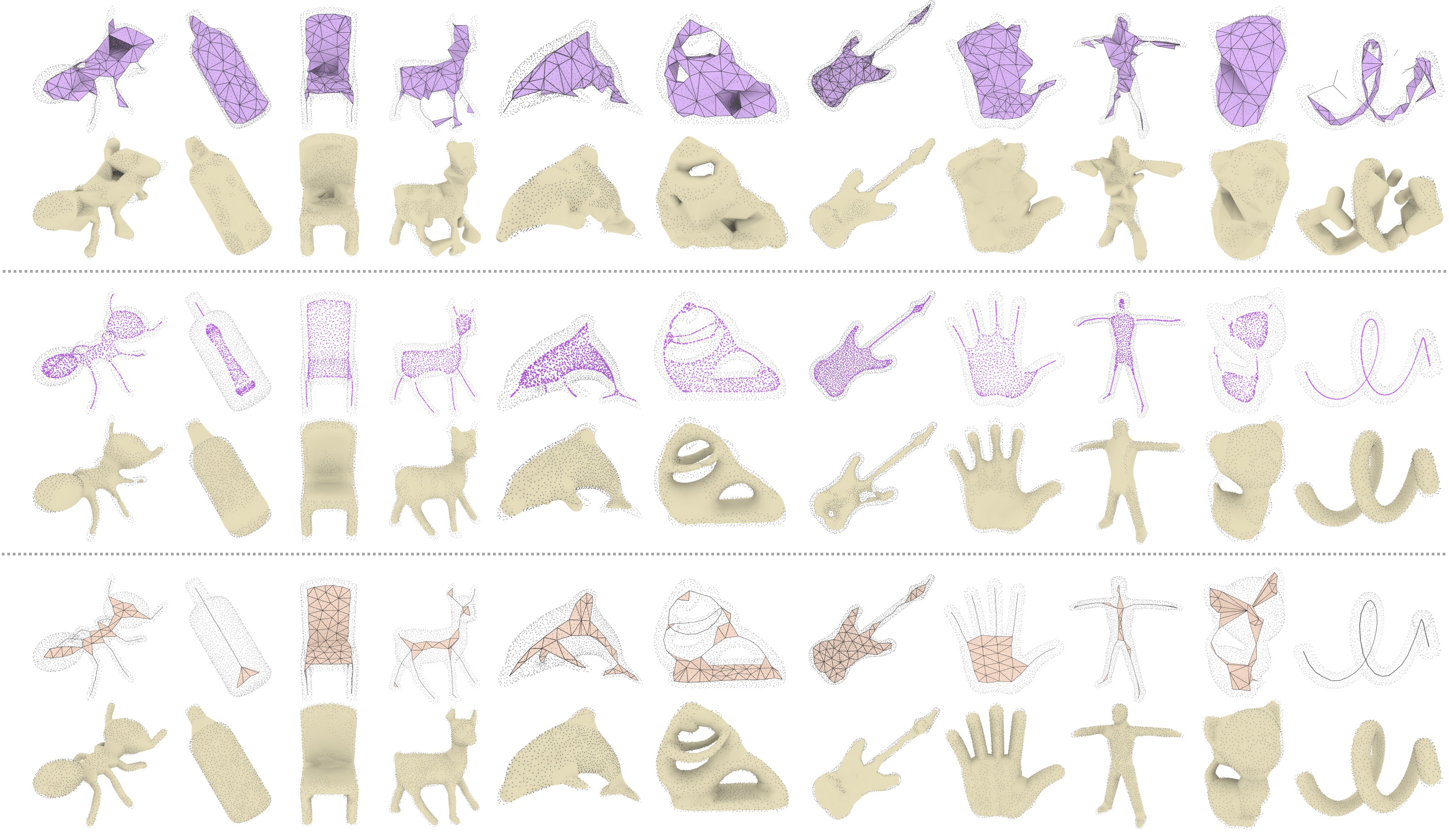} 
\put(-1,42){\rotatebox{90}{Point2Skeleton}}
\put(-1,26){\rotatebox{90}{DPC}}
\put(-1,3){\rotatebox{90}{Coverage Axis}}
\end{overpic}
\vspace{-2mm}
    \caption{Comparison with existing shape skeletonization methods for point clouds.}
        \vspace{-8mm}
\label{fig:pc}
\end{figure*}

\subsection{Coverage Axis from Point Cloud Input}
\label{sec:pc}
For point cloud input, we compare with two closely relevant methods:~Deep Point Consolidation (DPC)~\cite{wu2015deep} and Point2Skeleton~\cite{lin2021point2skeleton}, given that these two methods can also predict MAT-based skeletons directly from point clouds. Note Point2Skeleton is a deep learning-based method that requires training from a large amount of data.

\paragraph{Comparison with Deep Point Consolidation~\cite{wu2015deep}}  As shown in Figure~\ref{fig:pc} and Table~\ref{table:pc}, this method can only produce unstructured inner points without connections. Although the skeletons generated by DPC have denser points, the reconstruction results still exhibit large errors.  Even worse, the reconstructed topologies by deep point consolidation are usually inconsistent with the original input. In contrast, our reconstruction results reach better approximation accuracy with respect to the original geometry using fewer skeletal points. 
\paragraph{Comparison with Point2Skeleton~\cite{lin2021point2skeleton}}
More recently, using deep neural networks to predict skeletal representations from point clouds is beginning to be studied. Point2Skeleton~\cite{lin2021point2skeleton} is a representative approach that directly learns skeletal representations from point clouds based on the MAT in an unsupervised manner. We use their pre-trained network on a large quantity of data for comparison. 

\begin{table}
\small
\begin{center}
\caption{Quantitative comparison on shape approximation errors among Point2Skeleton (P2S), Deep Point Consolidation (DPC) and Coverage Axis.}
 \vspace{-1mm}
\label{table:pc}
\begin{tabular}{p{1.18cm}|cc|cc|cc}
\hline
\multirow{2}{*}{Model}  & \multicolumn{2}{c|}{P2S} & \multicolumn{2}{c|}{DPC} & \multicolumn{2}{c}{Coverage Axis}\\
  & $|V|$ &  $\overleftrightarrow{\epsilon}$ & $|V|$ &   $\overleftrightarrow{\epsilon}$ & $|V|$ &  $\overleftrightarrow{\epsilon}$\\
    \hline
Ant-2 & $100$ & $ 16.412\%$ & $ 1194$ & $ 8.863 \%$ & $58$ & $ \textbf{2.350\%}$ \\ 
Bottle & $100$& $2.955\%$ & $1194$ &$ \textbf{2.752\%}$ & $14$ & $  2.956\%$ \\
Chair-2& $100$ & $6.552\%$ & $1194$ & $ 4.807\%$ & $ 89$ & $ \textbf{2.890\%}$ \\
Dog & $100$ & $6.047\%$ & $ 1194$ & $ 5.237 \%$ & $49$ & $ \textbf{2.174\%}$ \\
Dolphin & $100$ & $5.925\%$ & $ 1194$ & $ 7.916 \%$ & $ 49$ & $ \textbf{1.971\%}$ \\
Fertility & $100$ & $8.162\%$ & $ 1194 $ & $ 4.226 \%$ & $ 79$ & $ \textbf{3.428\%}$ \\
Guitar & $100$ & $ 2.052\%$ & $1194$ & $3.231\%$& $ 60 $ & $ \textbf{2.032\%}$ \\
Hand-1 &  $100$ & $9.672\%$ & $ 1194$ & $ 4.110 \%$ & $ 44$ & $ \textbf{3.441\%}$ \\
Human-2 &  $100$ & $6.517\%$ & $ 1194$ & $ 4.758 \%$ & $ 44$ & $ \textbf{1.667\%}$ \\
Kitten &  $100$ & $8.724\%$ & $ 1194$ & $ 5.332 \%$ & $ 50$ & $ \textbf{3.450\%}$ \\
Snake &  $100$ & $15.021\%$  & $ 1194$ & $  1.736 \%$ & $ 40$ & $ \textbf{1.309\%}$ \\
\hline
Average &  - & $8.003\%$  & - & $  4.815 \%$ & - & $ \textbf{2.515\%}$ \\
\hline
\end{tabular}
\end{center}
 \begin{tablenotes}
  \small
     \item[1] $|V|$ \textit{The number of skeleton points.}
     \item[2] $\overleftrightarrow{\epsilon}$  \textit{Two-sided HD between original surface and reconstruction.
     }
  \end{tablenotes}
\end{table}

Figure~\ref{fig:pc} as well as Table~\ref{table:pc} show the comparison results. Note that Guitar and Chair-2 are from ShapeNet~\cite{chang2015shapenet} dataset which is used for training Point2Skeleton. First, an obvious problem with Point2Skeleton is their limited generalization abilities; despite being trained on numerous data, it still generates unsatisfactory results for unseen shapes. Furthermore, it performs poorly on models with higher topology complexity, e.g., input with more than genus one. Also, the neural network shows incapability for capturing the properties of geometric transforms, such as the centrality of skeletal points. Conversely, our approach is robust to various shapes and can produce high-quality skeletal representations that have accurate geometries and faithful structures.

\section{Discussions} 
\label{sec:_discussion}

In this section, in order to provide more insights and have better understandings of the proposed algorithm, we further give an in-depth analysis of some interesting properties of the Coverage Axis.

\subsection{Centrality and Random Initialization}
\label{sec:exp4_center}

\begin{figure}
    \centering
    \vspace{0mm}
    \begin{overpic}[width=\linewidth]{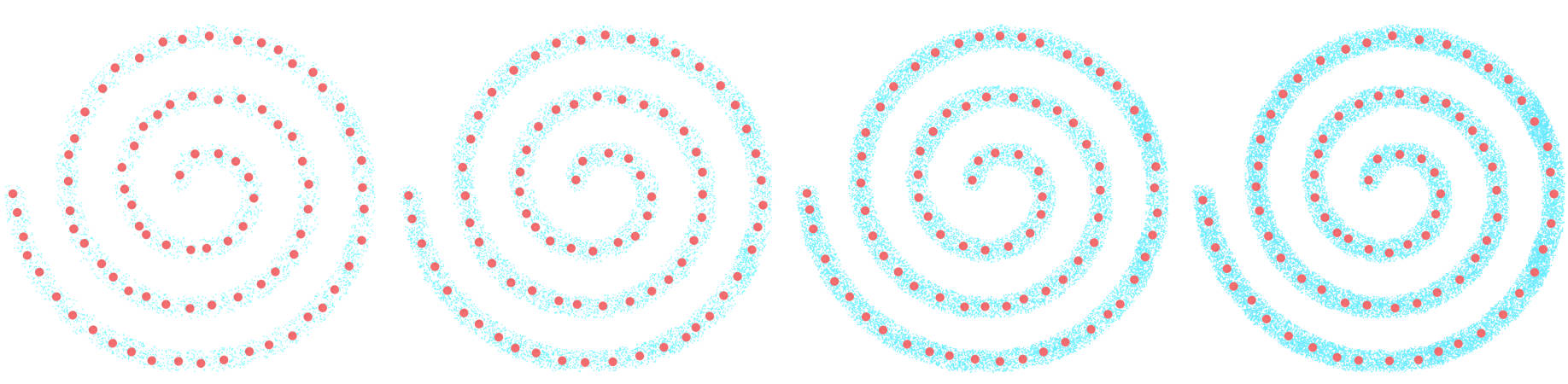}
\put(5,-4){$|P|=5000$}
\put(28,-4){$|P|=10000$}
\put(53,-4){$|P|=20000$}
\put(79,-4){$|P|=30000$}
\end{overpic}
\vspace{-1mm}
\caption{Effect of different numbers of randomly generated inner points on the centrality of the selected points. The blue points are randomly generated inner points while red points are selected skeletal points.} 
    \label{fig:centrality_ablation}
    \vspace{-4mm}
\end{figure}

A good skeleton requires each skeletal point to be located at the center of its corresponding local geometry. Our method is able to effectively select the local centers for generating the skeletal representation. This property also enables our algorithm to handle randomly generated inner points instead of relying on computing the strictly defined Voronoi diagram as an input. In this section, we elaborate on this centrality property of our point selection strategy. 
\begin{figure}
    \centering
       \includegraphics[width=0.95\linewidth]{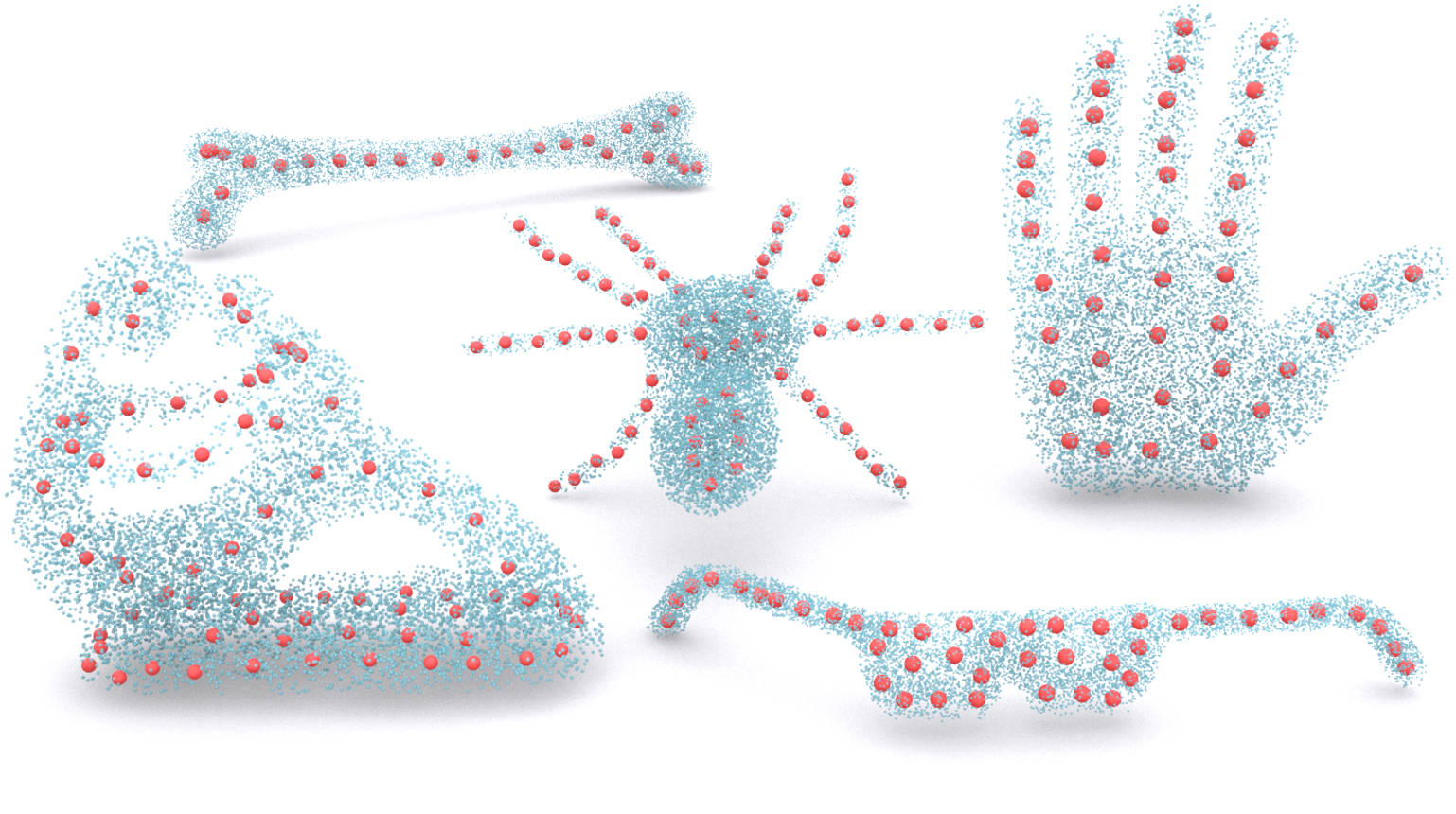}
       \vspace{-5mm}
    \caption{Skeletal point selection results from randomly sampled inner candidates. The blue points are candidate points generated by randomly sampling, while the orange points are selected points by our algorithm.}\label{fig:random_ablation}
           \vspace{-2mm}
\end{figure}
We first generate the point candidates by \textit{randomly} sampling inside the shape and then applying our point selection algorithm introduced in Sec~\ref{sec:point_selection}. The number of randomly generated inner points is $10000$, and the offset is set to $0.02$. A selection of results from these randomly generated points are shown in Figure~\ref{fig:random_ablation}. Thanks to the idea of optimal set coverage, it can be observed that the selected points are located at the centers of local geometry, even if the initial points are derived by random sampling. An evaluation of the influence of different randomly generated candidate inner point numbers $|P|$ for centrality is given in Figure~\ref{fig:centrality_ablation}. We find different numbers of candidate inner points do not show a large influence on the centrality.

\begin{figure}
    \centering
       \includegraphics[width=0.69\linewidth]{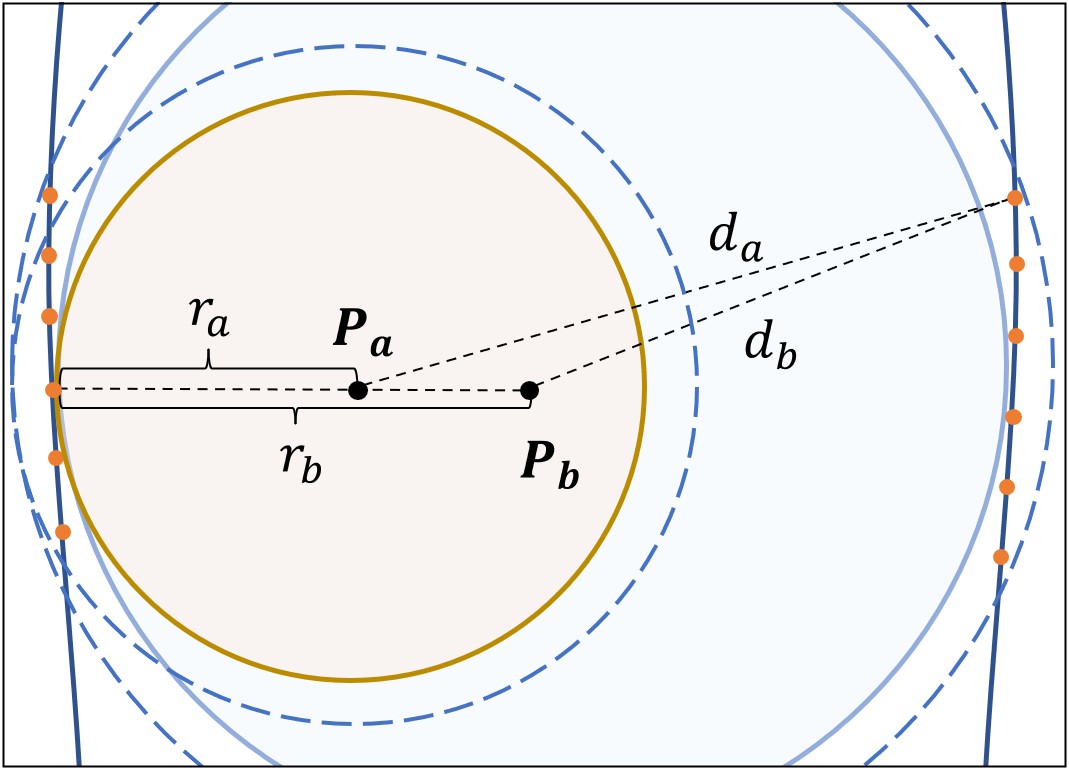}
       \vspace{-1mm}
    \caption{Illustration of the coverage of local geometry using medial balls. The orange points are surface samples. Balls centered at $P_a$ and $P_b$ are two candidate inner balls. The dilated balls are plotted using dashed circle.}
    \label{fig:proof}
    \vspace{-6mm}
\end{figure}

In the following, we give an analysis of the rationale of the centrality property of our algorithm. Recall in Eq.~\ref{eq:core_op}, the constraint of the optimization is that the union of dilated balls needs to cover the sampled surface points. As explained in Sec.~\ref{sec:amplification}, the dilated radius should not deviate too much from the original radius, which means $r' \approx r$. Consider a local shape shown in Figure~\ref{fig:proof}. The coverage goal of one candidate point $P_a$ is defined by its farthest distance to the local geometry, e.g., $d_a$, given that all the surface samples are covered only if the farthest sample is covered. According to the definition of inner balls, the coverage ability is determined by its radius $r_a$ (note that we have $r_a' \approx r_a$), which is actually the distance to the nearest surface samples of the local geometry. \\

The goal of finding the minimum number of dilated balls to cover all surface samples can be viewed as selecting those points with the most powerful coverage ability (cover all the target samples). That to say, \textit{its radius $r_a$ should be as equal as possible to the distance to farthest point $d_a$ so that we have $r_a' \ge d_a$ which satisfies the coverage constraint}. Those balls whose radii are close to the distance to the farthest point that needs to be covered will be preferred during optimization. Note that $r_a \approx d_a$ is consistent with the definition of the medial axis, which is intrinsically a set of centers of maximally inscribed spheres. In this case, the inner point $P_b$ is preferred for the local geometry coverage since $d_b - r_b < d_a - r_a$. Note that we always have $d_b \ge r_b$. Finally, as shown in Figure~\ref{fig:proof}, $P_b$ is a better approximation than $P_a$ of skeleton point and will be selected by our strategy.

\subsection{Robustness to Noise}
\label{sec:noise}

\begin{figure*}
    \centering
    \begin{overpic}
           [width=\linewidth]{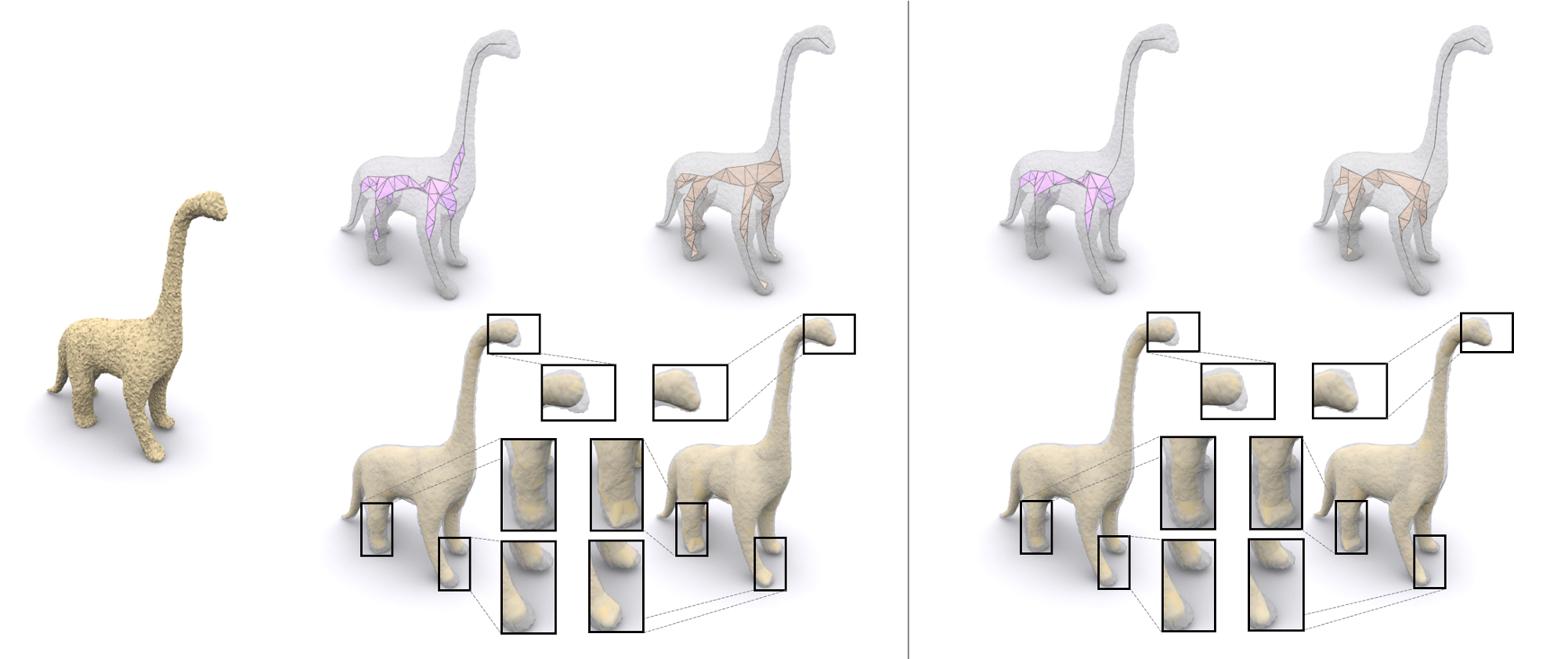}
           \put(4,-3){(a) Input model}
           \put(20,-3){(b) Q-MAT, $|V|=87$}
           \put(37.5,-3){(c) Coverage Axis, $|V|=87$}
           \put(60,-3){(d) Q-MAT, $|V|=61$}
           \put(77.5,-3){(e) Coverage Axis, $|V|=61$}
           \put(24,-5.5){$\overleftrightarrow{\epsilon} = 2.03\%$}
           \put(43.5,-5.5){$\overleftrightarrow{\epsilon} = 1.21\%$ }
           \put(64,-5.5){$\overleftrightarrow{\epsilon} = 2.08\%$ }
           \put(83,-5.5){$\overleftrightarrow{\epsilon} = 1.25\%$ }
           \put(44.1,-7.5){$\delta_r = 0.015$ }
           \put(83.6,-7.5){$\delta_r = 0.02 $}
    \end{overpic}
    \vspace{7mm}
    \caption{Comparison with Q-MAT on the robustness to surface noise. (a) Input model. (b)(c) and (d)(e) correspond to comparisons at different simplification levels.}
    \label{fig:noise}
\vspace{-7mm}
\end{figure*}
We evaluate the robustness of our method to surface noise. We compare with Q-MAT~\cite{li2015q}, which has the state-of-the-art performance in terms of approximation accuracy and goodness of structure (i.e., compactness and shape awareness).
The number of vertices of the skeletal points are set the same for both methods. 
As shown in Figure~\ref{fig:noise}, given noisy input, our method can still preserve the main features and achieve relatively high-precision shape approximation with a compact representation. The robustness mainly owes to the ball dilation strategy as well as the formulation based on global coverage. The dilated balls cover the insignificant details cause by local perturbations, while the overall coverage formulation gives a high-level optimization rather than focusing on local details. In contrast, Q-MAT~\cite{li2015q}, which uses step-by-step edge collapse based on local errors, cannot capture the fine details when the input surface has great noise. 

\subsection{Dilation Strategy}
\label{sec:diff_enlarge}
\begin{figure}
    \centering
       \includegraphics[width=0.78\linewidth]{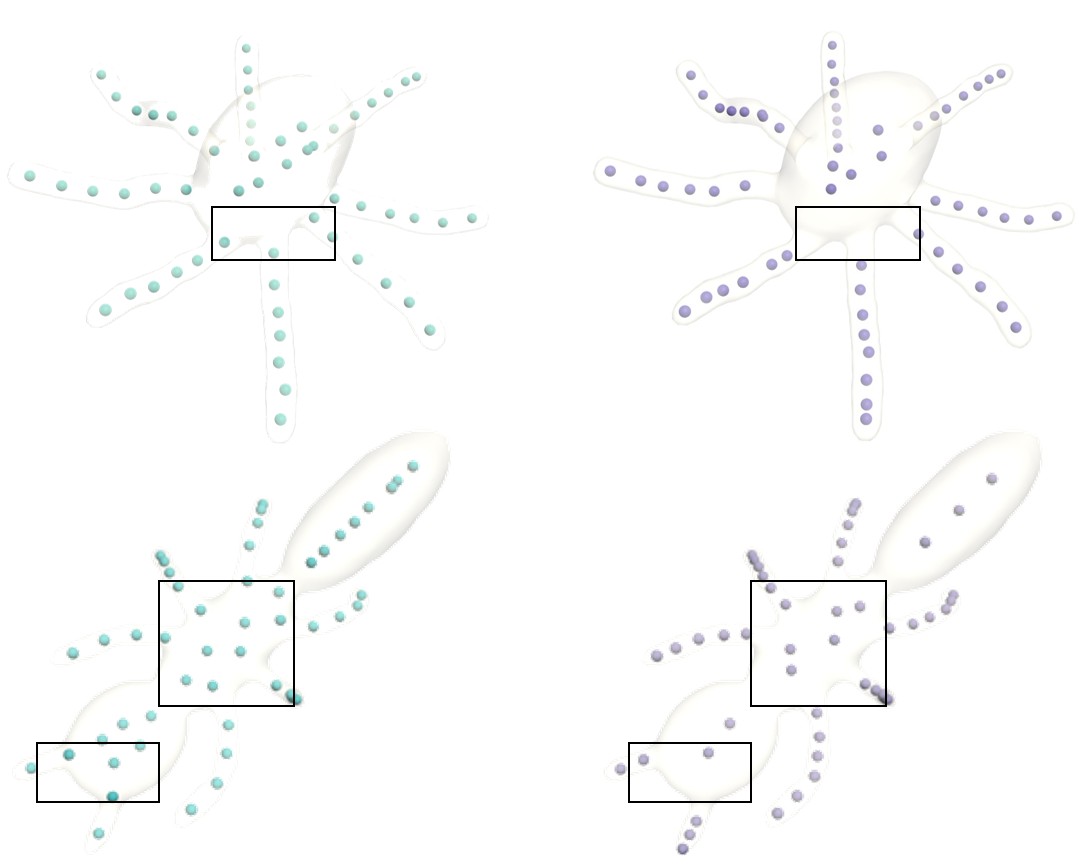}
       \vspace{-2mm}
          \makebox[0.01\linewidth][l]{ }
   \makebox[0.4\linewidth][c]{(a)}
        \makebox[0.48\linewidth][c]{(b)}\\
            \vspace{-1mm}
    \caption{Effect on different ball dilation strategies: (a) offset, (b) scaling.}\label{fig:enlarge_ablation}
    \vspace{-4mm}
\end{figure}
As aforementioned, the relationship between inner points and surface is built by the coverage, which is achieved by dilating the medial radius. There are two typical ways for inner ball dilation. One is by adding a specified value $\delta_r$ to the radius of each ball. The other is to scale the ball by multiplying the radius by a factor $\sigma_r$. We now discuss the effect of these two dilation strategies. To facilitate the discussion, we refer to the first way as \textit{offset} and the second as \textit{scaling}.

We take the model Octopus and Ant-3 as examples. Let $r$ and $r'$ denote the original medial radius and dilated medial radius respectively.
For the offset manner, we set $r' = r + \delta_r$ where $\delta_r = 0.02$, while for the scaling manner, we set $r' = r \times \sigma_r$ with $\sigma_r = 1.5$. 

As shown in Figure~\ref{fig:enlarge_ablation}, the offset manner tends to generate selected points that are distributed more evenly, while the scaling manner can distinguish parts that have different radii, which leads to relatively large gap around the joints of different parts (as shown in the box region). The reason behind this is, adding a constant value is independent of the original radius, but the scaling makes the larger ball dilate more and control more area, which results in part differentiation. 

\subsection{Parameter Analysis}
\label{sec:parameter_analysis}
\begin{figure}
    \centering
  \includegraphics[width=0.90\linewidth]{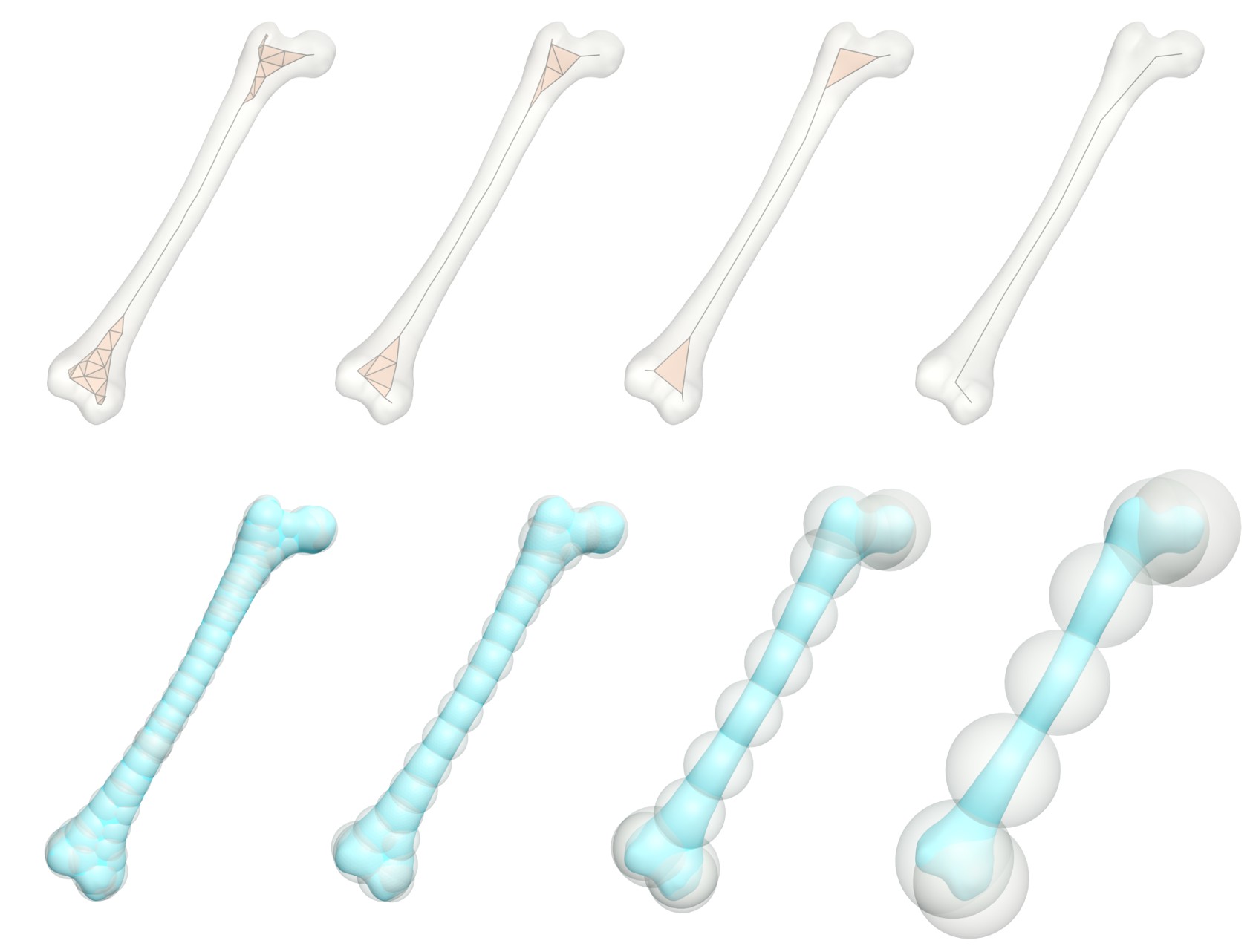}\\
  \vspace{-1mm}
 \makebox[0.20\linewidth][l]{(a) $|V|=47$}
 \makebox[0.20\linewidth][l]{(b) $|V|=24$}
 \makebox[0.20\linewidth][l]{(c) $|V|=13$}
 \makebox[0.20\linewidth][l]{(d) $|V|=7$}\\
 \makebox[0.05\linewidth][c]{}
 \makebox[0.2\linewidth][l]{$\delta_r=0.01$}
 \makebox[0.2\linewidth][l]{$\delta_r=0.02$}
 \makebox[0.2\linewidth][l]{$\delta_r=0.05$}
 \makebox[0.21\linewidth][l]{$\delta_r=0.1$}
 \vspace{-1mm}
    \caption{Simplification results using different dilation offsets. The first row and second row are simplification results and selected dilated coverage balls, respectively. We use $|V|$ to denote the number of vertices in the result and $\delta_r$ is the offset added to the original radius.}
    \label{fig:scale_factor}
    \vspace{-5mm}
\end{figure}

\paragraph{Dilation Factor} 
\label{sec:amplification}
The dilation factor $\delta_r$ is an important parameter that controls the simplicity of the output skeleton. In the following, we discuss the effect of different values of the parameter $\delta_r$. We use the Femur model as an example for demonstration, whose results are shown in Figure~\ref{fig:scale_factor}.

Obviously, a larger radius enables a ball to cover more surface points, which allows the algorithm to use fewer inner balls to cover the whole geometry. Therefore, a larger $\delta_r$ will lead to a higher simplified skeletal representation that captures less geometric details. Actually, the premise that we use surface information for simplification is that the envelope formed by dilated balls should not deviate significantly from the geometry of the input shape. On the one hand, we recommend not setting an excessively large $\delta_r$, as this leads to the over-dilated balls of which union are inconsistent with the original shape structure.  On the other hand,  $\delta_r$ being too small is also discouraged because it runs counter to the simplicity and compactness, which are related requirements for a skeletal representation. In this paper, we always set $\delta_r=0.02$ for our experiments. 

\paragraph{Surface points and inner points.} To investigate the influence of the number of surface points $|S|$ and candidate inner points $|P|$ (note it is determined by $|C|$, i.e., the number of samples to compute Voronoi diagram) on the algorithm performance, we conduct comprehensive experiments. We find that these two parameters do not show a significant influence on the selected point number $|V|$ and the quality of the resulting medial mesh. See more details in Appendix~C. However, it has more effect on the time efficiency, for which a detailed analysis is given in Sec.~\ref{sec:running_time}.

\subsection{Time Efficiency}
\label{sec:running_time}
We further discuss the time efficiency of our point selection strategy. The statistics is summarized in Table~\ref{tab:running_time_mesh}. 
We use Mixed-integer linear programming (MILP) solver in MATLAB (MathWorks 2021) on the platform explained in Sec~\ref{sec:exp_res}. 

As aforementioned, the experiments are conducted with mesh input using $1500$ surface point samples, candidate inner points generated by $4000$ surface samples. The dilation offset is set to $0.02$. It can be viewed from Table~\ref{tab:running_time_mesh} that the point selection only takes a few seconds for most shapes. For some cases, the complexity of the searching space of the Set Coverage problem may increase, which leads to one to two minutes of running time. The average running time for all the samples is $16.4$ s. 
\begin{table}
\small
\begin{center}
 \caption{Evaluation of the running time (s) of inner point selection.}\label{tab:running_time_mesh}
 \vspace{-1mm}
 \setlength{\tabcolsep}{1.1mm}{
\begin{tabular}{p{1.2cm}*{3}{c}||p{1.2cm}*{3}{c}}
\hline
{Model}   &  \makecell[c]{$|P|$} & \makecell[c]{$|V|$} & \makecell[c]{Time (s)} &{Model}   &   \makecell[c]{$|P|$} & \makecell[c]{$|V|$} & \makecell[c]{Time (s)}  \\
\hline
Ant-1 & $13490$ &$46$ & $2.8$ &   
Hand-1 & $15475$&$42$ &$6.7$ \\  
Armadillo  &$12348$  &$95$ & $5.4$ &
Hand-2  & $13665$ &  $60$& $48.8$ \\ 
Kitten  & $13718$  & $45$& $36.2$&
Human-1  &$13023$ &$45$&$0.9$\\
Bunny  &$14837$  & $106$ &$10.1$ &
Horse  &$13135$& $67$& $9.3$\\
Crab  & $13971$&$58$& $3.3$ & 
Octopus-1 &$13432$ & $49$ & $0.3$ \\ 
Camel  &$12807$ & $79$& $4.9$ & 
Octopus-2 &$13315$ & $68$& $0.4$\\ 
Dog  & $12623$ & $49$& $13.1$&
Pig & $13238$& $56$& $12.1$ \\
Dolphin  &$13998$& $43$& $55.4$& 
Pliers & $13097$& $31$& $0.9$ \\ 
Duck  &$17292$& $25$&$42.3$& 
Snake  &$14160$& $39$  & $0.2$\\ 
Elephant &$12503$ & $87$& $7.8$& 
Spider  &$13012$& $62$  & $15.5$\\ 
Eight  &$16393$& $28$& $16.1$& 
Spectacle  &$11235$ & $45$  & $19.2$\\
Fish  &$13279$ & $53$ & $42.1$ & 
Bear  &$13466$ & $32$  & $23.7$\\
Fertility &$13486$  & $79$ &  $34.5$ &
Vase  &$13174$ & $106$ &  $35.2$ \\
Femur & $14018$ & $24$ & $2.1$ &
Venus  &$14362$& $36$ &  $9.7$ \\
\hline
Average &  & &  &  &   &  & $16.4$\\
\hline
\end{tabular}}
\end{center}
\end{table}

\begin{table}
\small
\begin{center}
\caption{Running time~(with resulting vertex number in brackets) comparison (in seconds) with different methods using mesh and point cloud inputs.}
\vspace{-1mm}
\label{tab:comparison_running}
\begin{tabular}{p{0.66cm}|p{0.55cm}<{\centering}p{0.85cm}<{\centering}p{0.95cm}<{\centering}||p{0.82cm}p{0.75cm}<{\centering}p{0.85cm}<{\centering}}
\hline
\multirow{2}{*}{Model} & \multicolumn{3}{c||}{Mesh Input}                            & \multicolumn{3}{c}{Point Cloud Input}                            \\
& \multicolumn{1}{c}{\makecell[c]{Q-MAT}} & \multicolumn{1}{c}{SAT} & Ours & \multicolumn{1}{c}{DPC} & \multicolumn{1}{c}{P2S} & Ours \\ \hline
                  Ant-1 & $3.1~(46)$ & $38.2~(20\text{k})$ & $5.1~(46)$ & $6.1~(1094)$& $3.2~(100)$& $6.3~(46)$\\
                  Bunny & $24.6~(106)$ &  $63.7~(13\text{k})$ & $14.9~(106)$ & $7.2~(1094)$ & $3.3~(100)$ & $18.4~(106)$ \\
                  Horse	& $35.1~(67)$ & $52.8~(12\text{k})$ & $11.5~(67)$ & $8.2~(1094)$ & $3.4~(100)$ & $12.4~(67)$\\
                  Pig & $20.5~(56)$ & $67.1~(13\text{k})$ & $14.3~(56)$ &
                  $6.6~(1094)$ & $3.2~(100)$ &  $14.1~(56)$	\\
                  Venus & $12.7~(36)$ & $57.5~(8\text{k})$&	$13.7~(36)$ & $7.1~(1094)$	& $3.2~(100)$ & $15.1~(36)$	\\
                  \hline
\end{tabular}
\end{center}
\end{table}

Table~\ref{tab:running_time_mesh} also reveals the difference in running time for various shapes, where the difference is because the complexity of the solution space in SCP varies for different geometries. Generally, solving for the coverage of plate-like geometries is more complex than tube-like ones, since the plate-like models have more eligible candidates that satisfy the coverage constraints, which leads to a larger search space during optimization. Since many shapes consist of both tube-like and plate-like structures, there is a variation in running times.
We further report running time comparison in Table~\ref{tab:comparison_running}.

SCP is a known NP-Hard problem~\cite{hartmanis1982computers} and its running time complexity is $O(|P|^22^{|S|})$ where $|P|$ and $|S|$ are the number of candidate inner points and surface points, respectively. To investigate the influence of $|S|$ and $|P|$ on running time efficiency and the resulting number of selected skeletal points $|V|$, we conduct additional experiments of which results are summarized in Table~\ref{tab:ablation_surface} and Table~\ref{tab:ablation_inner} respectively. We adopt the Octopus-1 model for evaluation. We fix $|C| = 4000$ in Table~\ref{tab:ablation_surface} and $|S|=1500$ in Table~\ref{tab:ablation_inner}. From the results, both $|S|$ and $|P|$ do not show a significant influence on the selected vertex number. However, $|S|$ has a larger impact on the running time, while $|P|$ shows less effect, which is consistent with the aforementioned analysis of time complexity. 

\noindent
\begin{minipage}{\linewidth}
\begin{minipage}[t]{0.38\linewidth}
\makeatletter\def\@captype{table}
\small
\caption{Effect of different numbers of surface samples $|S|$ on selected point number and running time efficiency.}
\vspace{-1mm}
\label{tab:ablation_surface}
\begin{tabular}{p{0.7cm}<{\centering}|p{0.4cm}<{\centering}|p{0.8cm}<{\centering}}
\hline
 $|S|$ & $|V|$ & Time~(s) \\
\hline
$500$  & $47$ & $0.25$\\ 
$1500$ & $49$ & $0.29$\\ 
$2500$ & $49$ & $0.34$\\ 
$3500$ & $49$ & $0.73$\\ 
$4500$ & $49$ & $0.98$\\ 
$5500$ & $49$ & $1.33$\\ 
$6500$ & $49$ & $2.44$\\ 
$7500$ & $49$ & $3.24$\\ 
$8500$ & $49$ & $4.22$\\ 
\hline
\end{tabular}
\label{sample-table}
\end{minipage}
\quad
\begin{minipage}[t]{0.58\linewidth}
\makeatletter\def\@captype{table}
\small
\caption{Effect of different numbers of inner point candidates $|P|$ on selected point number and running time efficiency. $C$ denotes surface samples used for generating candidate inner points.}
\vspace{-1mm}
\label{tab:ablation_inner}
\begin{tabular}{p{0.7cm}<{\centering}|c|p{0.4cm}<{\centering}|p{0.8cm}<{\centering}}
\hline
 $|C|$ & $|P|$ &$|V|$ & Time~(s) \\
\hline
$1000$  & $2964$   & $48$ & $0.08$\\ 
$2000$  & $6894$   & $49$ & $0.18$\\ 
$3000$  & $10776$  & $49$ & $0.28$\\ 
$4000$  & $13432$  & $49$ & $0.31$\\ 
$5000$  & $16749$  & $49$ & $0.49$\\ 
$6000$  & $22301$  & $49$ & $0.82$\\ 
$7000$  & $26161$  & $49$ & $1.23$\\ 
$8000$  & $29859$  & $49$ & $1.43$\\ 
$9000$  & $33651$  & $49$ & $1.65$\\ 
\hline
\end{tabular}
\label{sample-table}
\end{minipage}
\end{minipage}

\begin{table*}
\small
\begin{center}
 \caption{Additional evaluations on time efficiency. We show the running time (s) of point selection from randomly generated candidates using different ball dilation strategies. }\label{tab:running_time_pc}
 \vspace{-1mm}
\setlength{\tabcolsep}{2mm}{
\begin{tabular}{c|cccc|cccc|cccc|cccc}
\hline
{Model} & \multicolumn{4}{c|}{Ant-2}  &  \multicolumn{4}{c|}{FEMUR}  & \multicolumn{4}{c|}{Hand-2}  & \multicolumn{4}{c}{Fertility}   \\
\hline
\makecell[c]{Offset} & $0.015$ & $0.02$& $0.025$& $0.03$ &  
$0.015$  & $0.02$& $0.025$& $0.03$ &  $0.015$  & $0.02$& $0.025$& $0.03$ &
$0.015$  & $0.02$& $0.025$& $0.03$ \\
\makecell[c]{Selected points}  &$98$& $59$ & $48$& $34$ &$39$ &$26$ &$21$ &$17$ &$86$ &$63$ &$57$ &$38$ &$153$&$82$ &$62$&$49$\\
\makecell[c]{Time (s)} & $3.2$& $2.1$  & $8.1$ &$7.5$  &$3.3$ &$2.2$ & $1.2$ &$0.3$  &$19.5$ &$15.8$ & $10.2$ &$9.8$& $21.0$&$17.4$ &
$9.3$&$10.7$\\
\hline
 \hline
 \makecell[c]{Scaling} & 
$1.5$ & $1.75$& $2$ & $2.25$ &
$1.5$& $1.75$& $2$ & $2.25$ &
$1.5$& $1.75$& $2$ & $2.25$ &
$1.5$& $1.75$& $2$ & $2.25$ \\
\makecell[c]{Selected points}  
&$98$& $77$ & $62$& $49$ 
&$30$ &$17$ &$12$ & $10$
&$65$ &$31$ &$22$ & $18$
&$80$&$64$&$40$ &$31$\\
\makecell[c]{Time (s)} 
&$3.1$ &$1.4$ & $0.9$ &$0.4$  
&$0.4$ & $0.8$ &$1.7$ &$0.3$
&$0.7$  & $3.3$ &$3.2$ &$2.5$ 
&$0.7$&$0.5$&$0.7$  &$1.6$ \\
 \hline
\end{tabular}}
\end{center}
\vspace{-4mm}
\end{table*}

To evaluate the time efficiency for different algorithm settings, we conduct an additional experiment where the initial candidate points are generated by random sampling. Here, we randomly sample $10000$ points inside a shape and test the time efficiency using two different dilation strategies and different dilation factors.
The results are given in Table ~\ref{tab:running_time_pc}. It can be observed the running time is stable in different configurations.

\section{Additional Features of Coverage Axis}
\label{sec:additional_feature}
In this section, we analyze some of the other interesting features of our method, including the ability to handle large models and the flexibility of user-specified point control.

\subsection{Divide and Conquer Strategy for Large Models}
\label{subsec:Divide_and_conquer}
Since the set coverage problem is NP-hard, the action space grows exponentially with respect to the number of surface points. Fortunately, an interesting feature of the proposed method is the flexibility in handling large complex models. The overall insights lie in the combination of divide and conquer strategy and our set coverage formulation. In this case, we take a mesh model with Voronoi initialization for explanation.

The first step is decomposing a mesh into smaller and meaningful sub-meshes with some existing methods, e.g.,~\cite{cgal:y-smsimpl-21b, hanocka2019meshcnn}. Then we have surface samples divided into several subsets $S_1, S_2, ..., S_n$ where $n$ is the number of surface components. According to the correspondence between the vertices on the inside Voronoi diagram and the surface samples, we partition the inside Voronoi diagram into  $P_1, P_2, ..., P_n$ accordingly. For each component $i$, with surface samples $S_i$ and inner point candidates $P_i$, we solve the SCP and obtain the selection result $P^+_i$ embedded on the sub-Voronoi diagram. This process reduces the size of the problem and speeds up our algorithm by divide and conquer. 

\begin{figure}
    \centering
    \begin{overpic}
[width=0.8\linewidth]{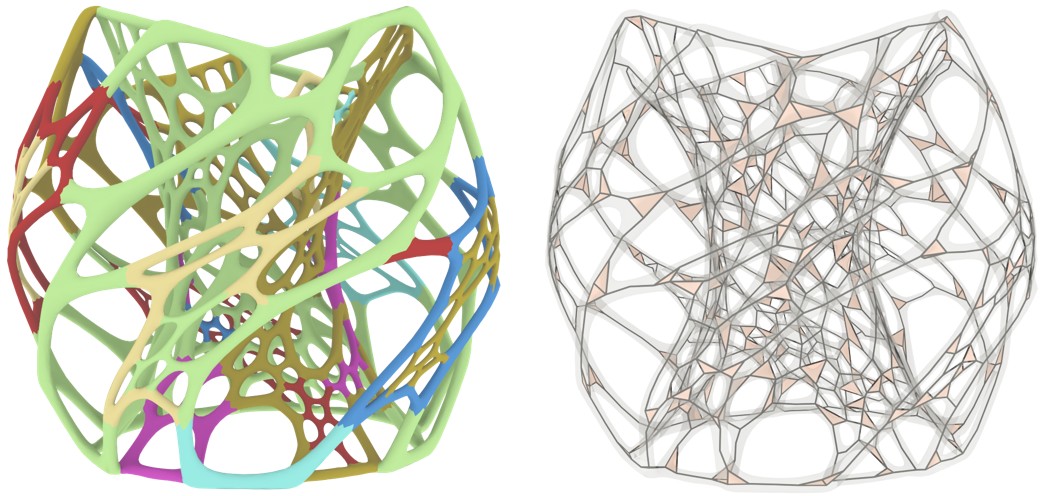}
\put(22,-5){(a)}
\put(74,-5){(b)}
\end{overpic}
\vspace{1mm}
    \caption{Coverage Axis with divide and conquer strategy. The model is divided into $29$ parts for solving the set coverage problem separately on each part. The average running time of inner point selection of each component is $1.04$~s. However, it takes more than $15$~min to solve for the entire model.}
    \label{fig:additional}
\end{figure}

Finally, we merge each Voronoi diagram as well as the selected points. Then, we have all the selected points embedded on the original inside Voronoi diagram so that the model's simplification can be achieved by removing redundant components as we described in~Sec.~\ref{subsec:connection}. We demonstrate the results in Figure~\ref{fig:additional} (a-b). Note that the complex model is with $30000$ surface samples which are further used for generating candidate inner points.

\subsection{User-specified Point Control}

Our method allows users to specify the ignorable surface parts that can be neglected. To indicate the ignorable surface points, the constraint in Eq~\ref{eq:core_op} are reformulated as $\mathbf{Dv} \ge \mathbf{B}$, where the corresponding values for the ignorable surface points in $\mathbf{B}$ are set to $0$ and the others to $1$. In this way, the optimization tends to remove the skeletal points in the specified parts to favor parsimony. Users can also directly indicate if a skeletal point should be preserved or removed, for which we just simply set the corresponding values in $\mathbf{v}$ to 0 or 1 after the point selection. 
\section{Limitations}
\begin{figure}
    \centering
  \includegraphics[width=0.9\linewidth]{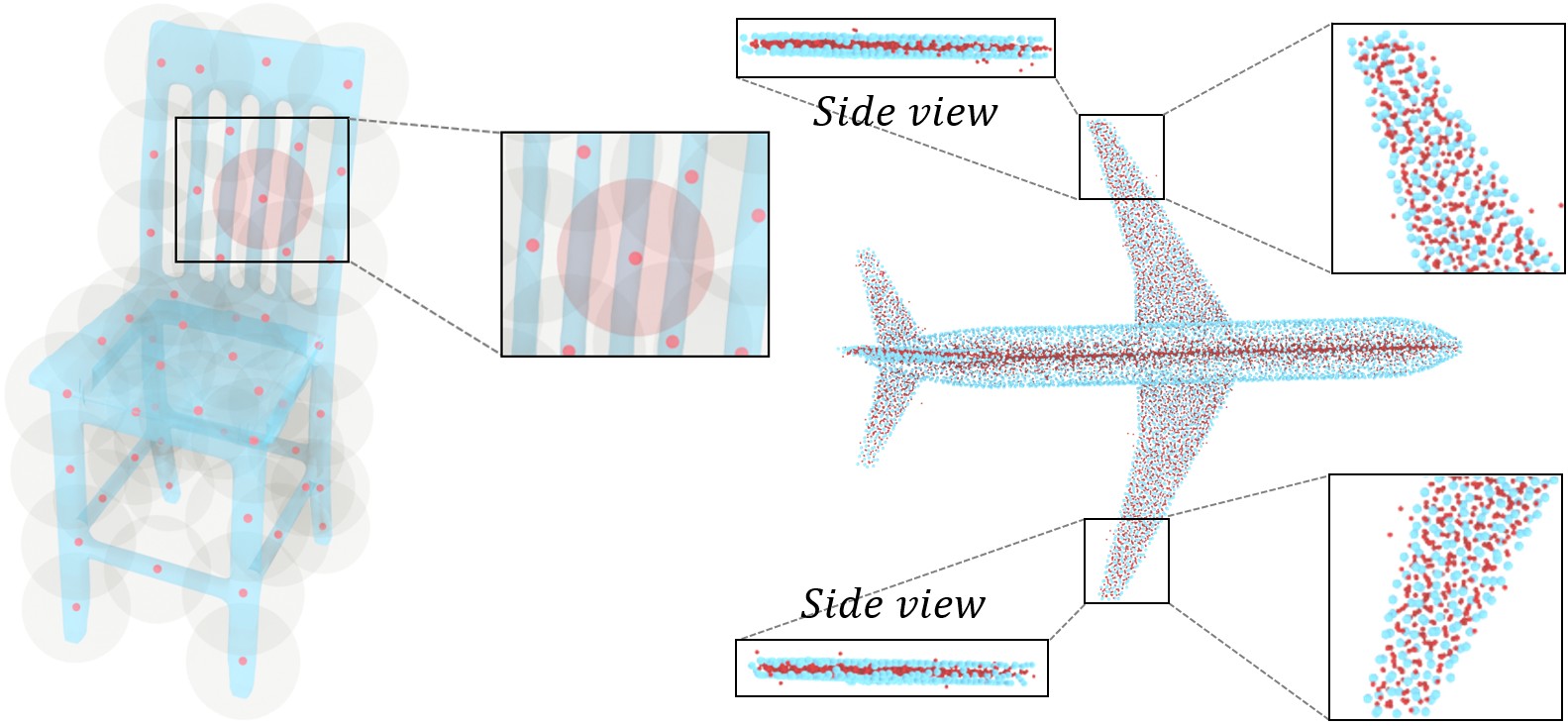}\\
          \makebox[0.01\linewidth][l]{ }
   \makebox[0.4\linewidth][c]{(a)}
        \makebox[0.48\linewidth][c]{(b)}\\ 
        \vspace{-2mm}
    \caption{Examples of failure case. (a) The over-dilated inner balls destroy the original shape topology. (b) Incorrect candidate initialization of thin flat model taken point clouds as the input. The red points are candidate points and the blue points are the input, where some candidates are located outside the shape.}
    \label{fig:limitation}
    \vspace{-2mm}
\end{figure}

\begin{figure}
    \centering
    \hspace{-2mm}
    \begin{overpic}
[width=0.97\linewidth]{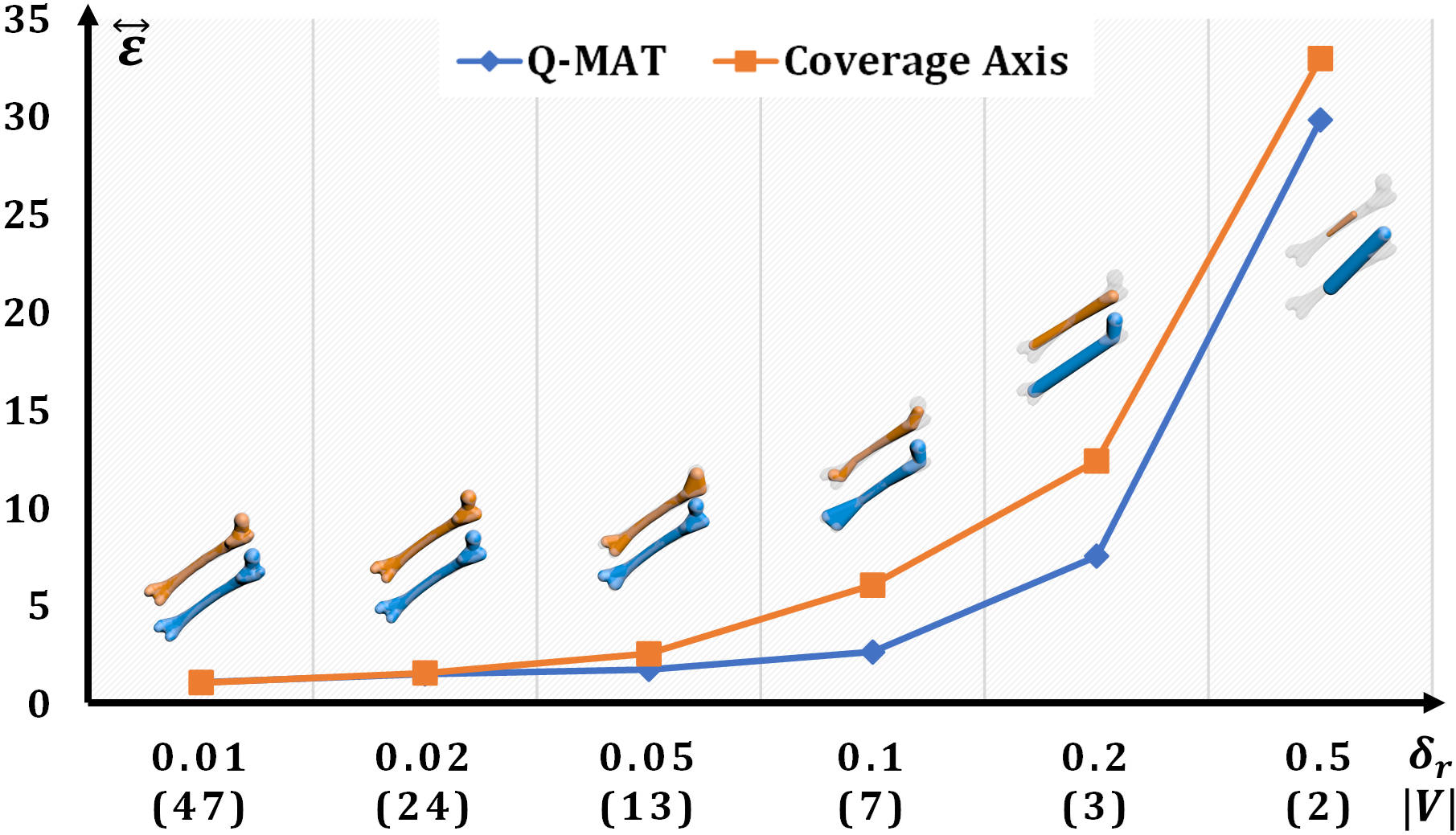}
\end{overpic}
\vspace{-2mm}
\caption{The change of the approximation error~(two-sided HD $\ora \epsilon$) w.r.t. the simplification level. Compared with Q-MAT, our method exhibits larger errors for highly decimated medial surfaces. We adopt the FEMUR model for evaluation.}
   \label{fig:extreme_decimated}
   \vspace{-6mm}
\end{figure}

In this section, we discuss the limitations of our method.
First, the number of selected points is related to the dilation parameter $\delta_r$ which leads to a difficulty in explicit control of the number of vertices of a skeleton. In particular, when inner balls are excessively dilated, the topology of the input surface tends to be destroyed by the coverage. See Figure~\ref{fig:limitation} (a). Another limitation lies in the difficulty in preserving approximation accuracy when computing extremely decimated representations (lower vertices number in the skeleton). As shown in Figure~\ref{fig:extreme_decimated}, Q-MAT~\cite{li2015q} shows a better control of the reconstruction error $\overleftrightarrow{\epsilon}$ than ours, as the number of target skeletal points $|V|$ decreases. This is because Coverage Axis relies on much larger dilation factors for computing highly decimated results, which brings difficulty in preserving the local geometry. Detailed statistics can be found in Appendix~D.

Our method usually suffers from a relatively high computational cost when given a large number of surface points to be covered. To alleviate this, we demonstrate a potential approach in designing a divide and conquer strategy based on geometric clues to achieve acceleration. In addition, for the point cloud input, especially the thin flat models, it remains an open problem to clearly distinguish between inner and outside volume. Therefore, for this kind of input, it is extremely challenging to generate legitimate candidate points that are located inside the shape, which can result in failure cases during point selection (Figure~\ref{fig:limitation} (b)). 

\section{Conclusion}
In this paper, we propose a novel, simple, yet effective formulation named Coverage Axis for 3D shape skeletonization of both meshes and point clouds. Inspired by the set cover problem (SCP), our goal is to cover all the surface points using as few inside medial balls as possible. Compared with previous methods, our approach shows a set of appealing properties including its simple formulation, high-level abstraction, robustness to noise, flexibility to handle various cases, and so on. Comprehensive experiments verify the aforementioned claims. In the future, we believe Coverage Axis, as a tool for encoding shapes, has a large potential value in various applications including shape abstraction, shape segmentation, simulation as well as animation, etc. Another point worth noting is the combination of Coverage Axis and learning techniques to improve the work of shape analysis and processing, such as 3D object detection, shape classification, automatic blending and so on. 

\printbibliography   
\clearpage

\appendix
\section{\textbf{Inner Points Labeling for Point Cloud with Normals.}}

\label{ap:pc_normal}
\setcounter{figure}{0}
\setcounter{table}{0}
\begin{figure}[H]
\vspace{-1.6mm}
    \centering
     \begin{overpic} [width=\linewidth]{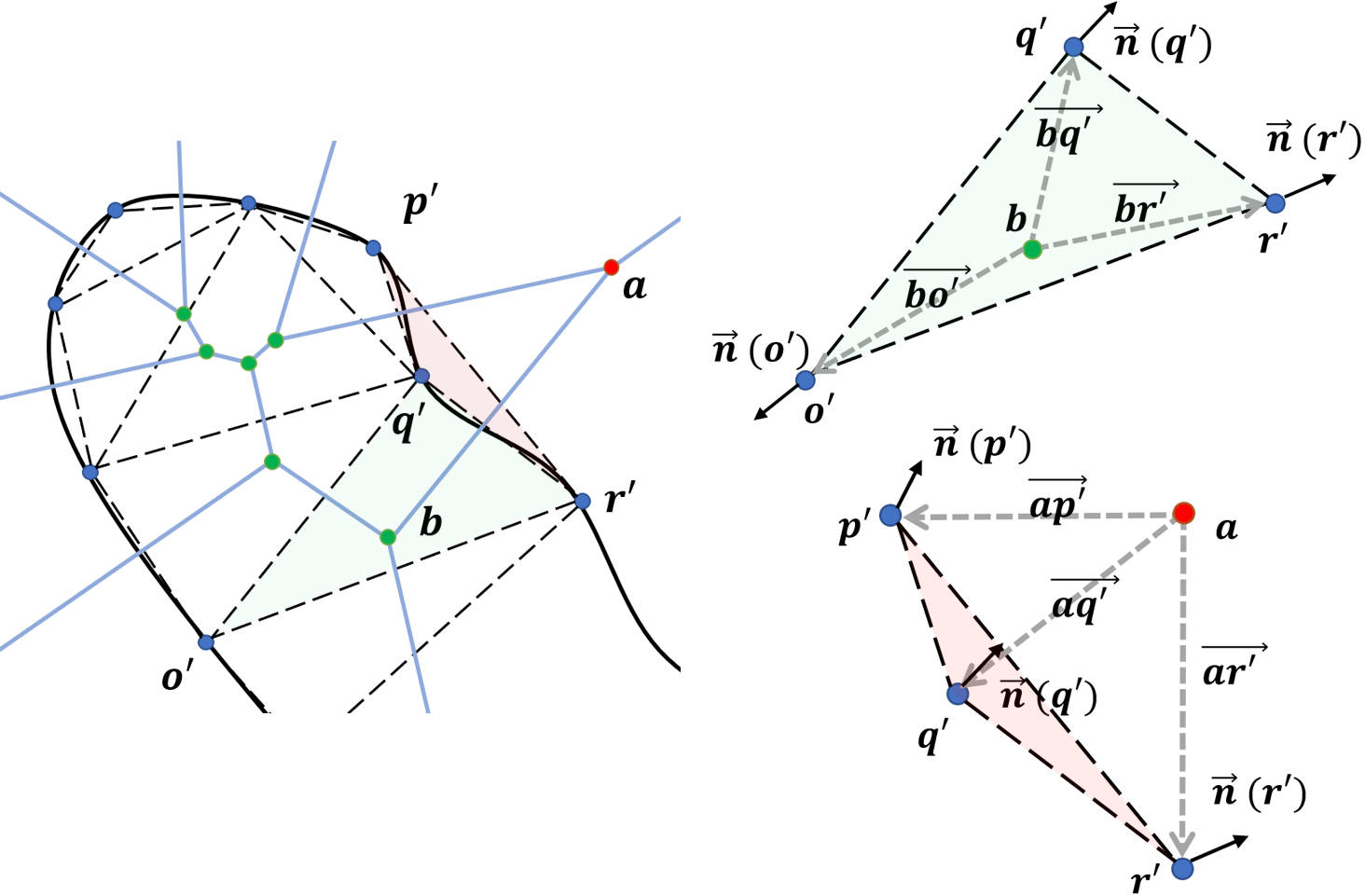}           \put(19,-3){(a)}
           \put(74,-3){(b)}
    \end{overpic}
    \vspace{0.5mm}
        \caption{Inner points labeling for an oriented point cloud in 2D.
    $b$ is labelled as an inner candidate since the dot products of $ \oraright{bo'} \cdot \oraright{n}(o')$, $\oraright{bq'} \cdot \oraright{n}(q')$ and $\oraright{br'} \cdot \oraright{n}(r')$ are all positive, while $a$ is labeled as an outside point because $\oraright{ap'} \cdot \oraright{n}(p'), \oraright{aq'} \cdot \oraright{n}(q')$ and $\oraright{ar'}\cdot \oraright{n}(r')$ are negative.  The Voronoi diagram and its dual graph Delaunay triangulation are denoted by blue lines and dashed lines respectively. Surface samples are in blue.}\label{fig:pc_normal}
    \vspace{-4mm}
\end{figure}
Our point selection strategy does not rely on mesh connection; it is able to handle more generalized inputs such as polygon soups, point clouds as long as candidate inner points can be identified inside the volume. Given a point cloud input, in this paper, we generate and label candidate inner points by utilizing normal vectors of the point cloud.  Recall that we first compute its Delaunay triangulation which is the dual of the Voronoi diagram w.r.t. the input point cloud. Consider a Voronoi vertex $p$ and the vertices of its dual tetrahedron $p'_0, p'_1, p'_2$ and $p'_3$. The candidate $p$ is considered as an inner point only if we have $\overrightarrow{pp'_i}\cdot \overrightarrow{n}(p'_i) >0$, $\forall i = 0,1,2,3$. Here $\overrightarrow{n}(p'_i)$ is the input normal of  $p'_i$ and $\cdot$ is dot product. A detailed 2D example is given in Figure~\ref{fig:pc_normal}.

\begin{figure}
    \centering
       \includegraphics[width=0.95\linewidth]{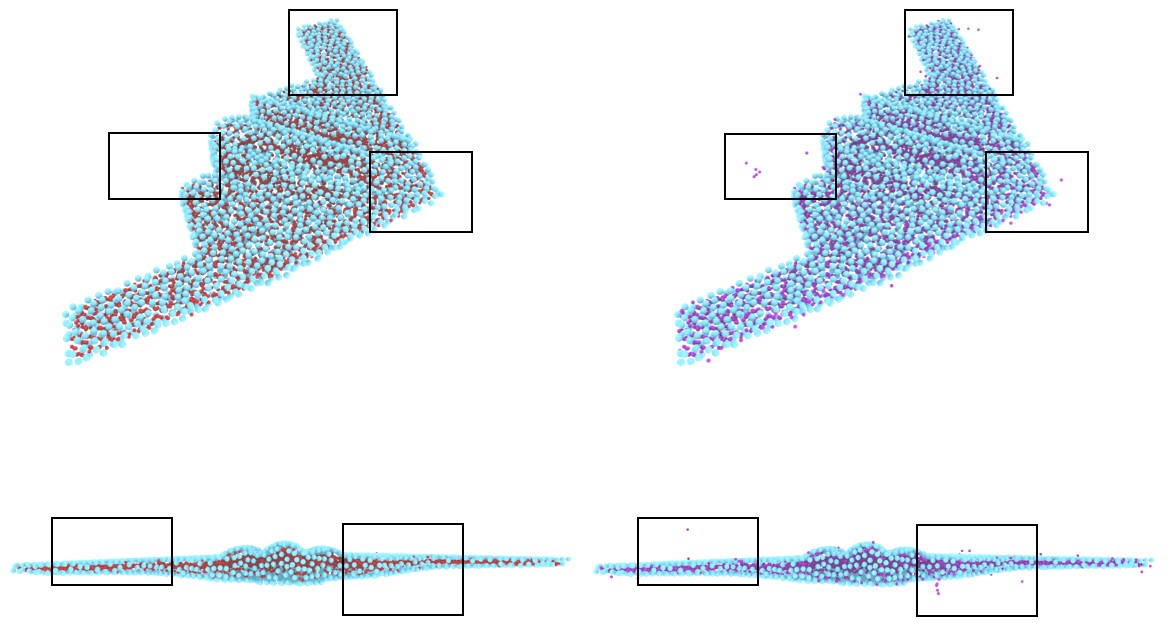}
          \makebox[0.48\linewidth][c]{(a)}
        \makebox[0.48\linewidth][c]{(b)}\\  
        \vspace{-2mm}
    \caption{Inner points labeling with filtering by clustering. (a) and (b) indicate the results after applying filtering by clustering (red points) and before applying filtering by clustering (purple points). The input point cloud is in blue. We provide both normal view (first row) and side view (second row). }\label{fig:filter_by_clustering}
    \vspace{-4mm}
\end{figure}

\begin{figure}
    \centering
       \includegraphics[width=0.95\linewidth]{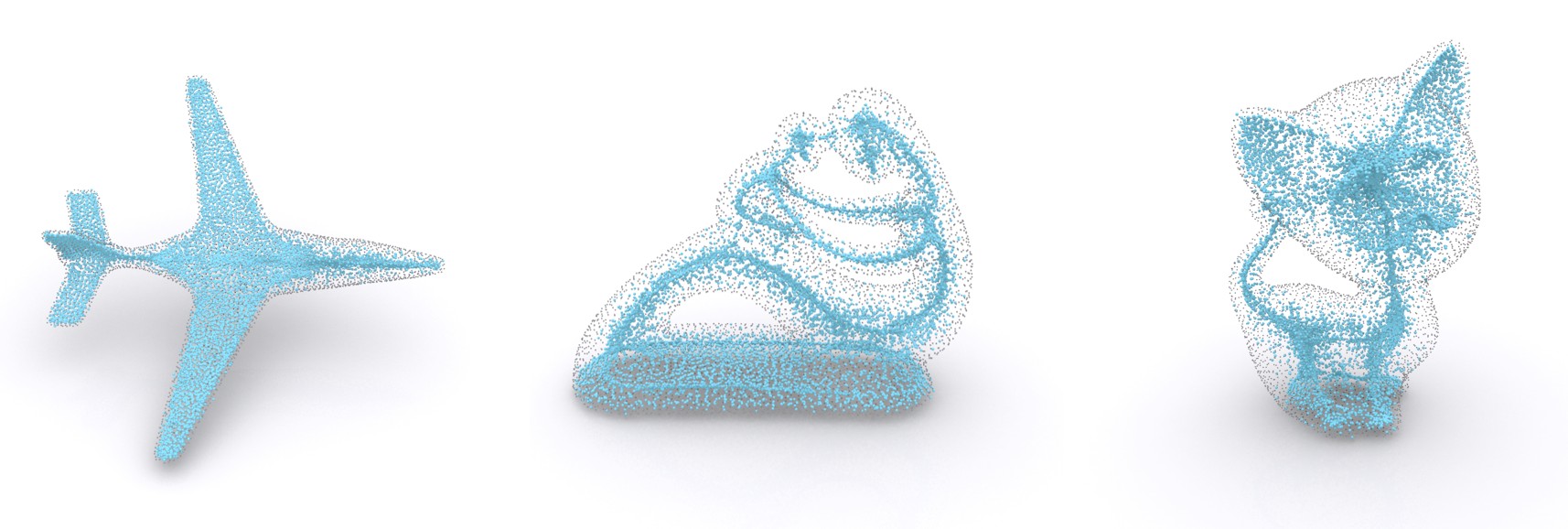}
       \vspace{-3mm}
    \caption{Inner point candidates generated from point clouds with normals.}\label{fig:pc_normal_result}
    \vspace{-8mm}
\end{figure}

After the initial labeling, we further apply filtering by clustering to those labeled inner points. For each inner point, we count the number of its neighbors within $0.02$ among the top $100$ nearest neighbors by K-neighbor searching~\cite{cgal:tf-ssd-21b}. If the result is less than $20$, the point is considered an outlier and is discarded. 
A result of the UFO model as an example is shown in Figure~\ref{fig:filter_by_clustering}.  Some labeling results are given in Figure~\ref{fig:pc_normal_result}.


\begin{wrapfigure}{r}{2.0cm}
\vspace{-3.5mm}
  \hspace*{-3.5mm}
  \centerline{
  \includegraphics[width=32mm]{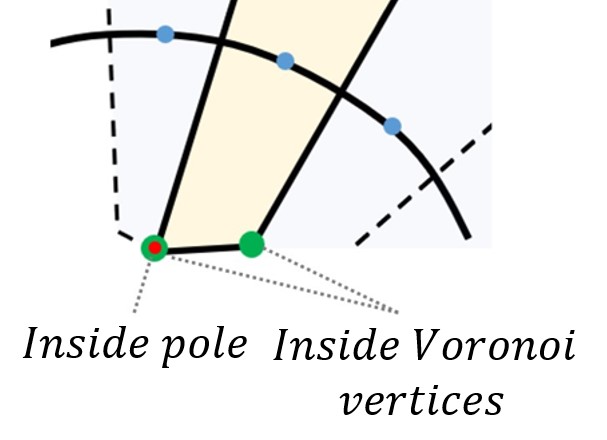}
  }
  \vspace*{-6mm}
\end{wrapfigure}
Besides, we also conduct an experiment taking inside poles~\cite{amenta2001power} (the inside Voronoi vertex with the furthest distance to the seed of the cell) instead of all Voronoi vertices enclosed by the surface as candidates.  We find there is no significant difference in point selection results.

\section{\textbf{Details of Connection Establishment.}}
\label{ap:details}
\subsection{Mesh Input}

Recall that for a mesh input, we take the whole subset of Voronoi structure inside the model generated by $4000$ surface samples ($|C| = 4000$) as the original inner point candidates $P$. After selecting inner points based on set coverage, we build up the connection structure.
For the Voronoi initialization style, all selected points are embedded on the Voronoi diagram. We further remove all redundant points and edges to achieve the simplification by edge collapse using quadric error metric (QEM) ~\cite{faraj2013progressive} same as~\cite{li2015q}. 

After that, we adopt LOP~\cite{de2010triangulations} algorithm to adjust the mesh tessellations as long as the increase of the local approximation error caused by the flipping operation is less than $5\%$. The local approximation error is measured by the Hausdorff distance from the covered surface to candidate local reconstructed surfaces (the envelope of two candidate triangulation.), a.k.a., one-sided Hausdorff distance. Note that we first unstitch each face patch based on non-manifold edges before performing LOP.\\

\subsection{Point Cloud Input} 
\begin{figure}[H]
    \centering
  \includegraphics[width=0.8\linewidth]{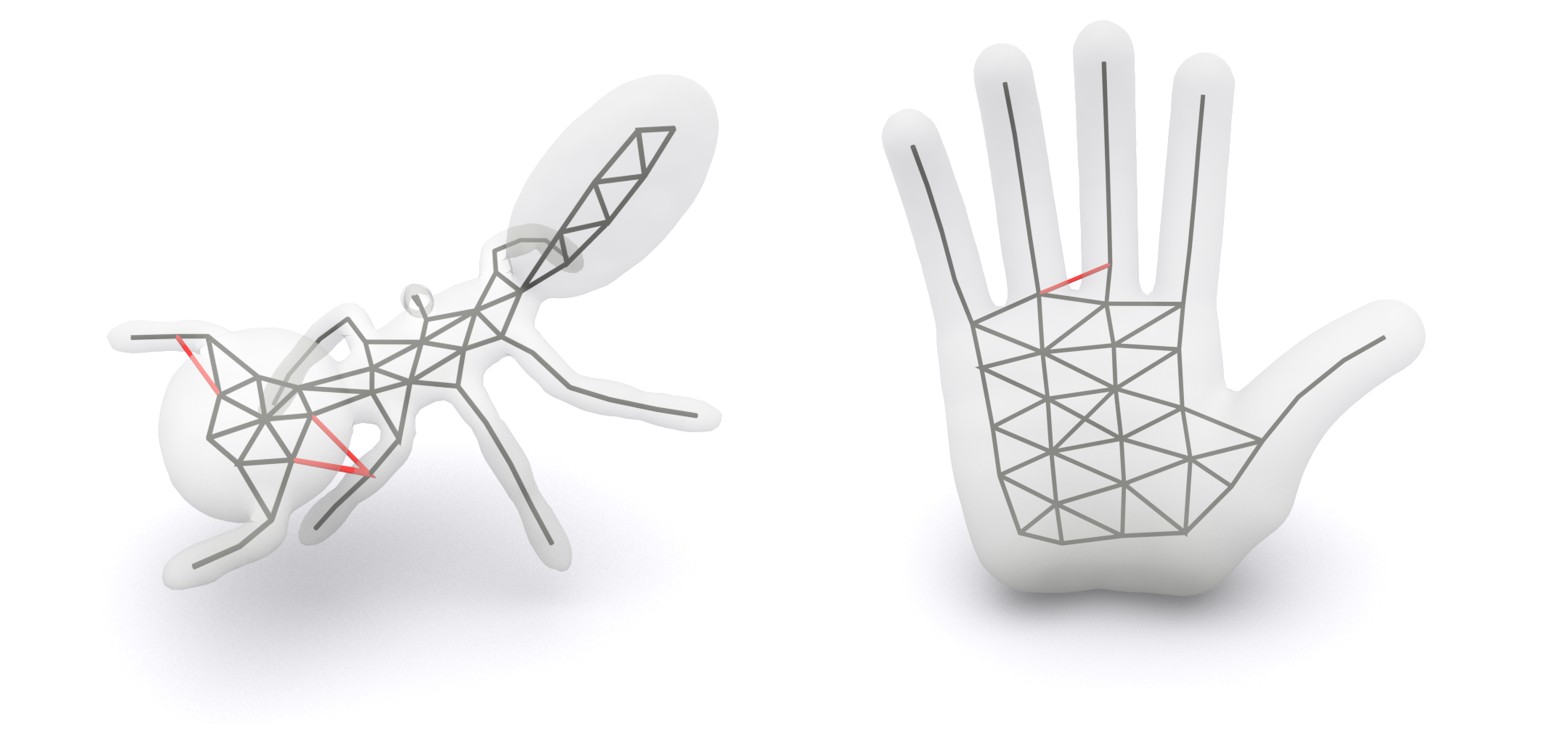}\\
    \caption{ Over-connection suppression by up-sampling. The over-connected structures (red edges) are suppressed after up-sampling. We visualize the original surface for illustration purposes.}
    \label{fig:upsampling}
    \vspace{-5mm}
\end{figure}

In order to suppress over-connection, we up-sample on the point cloud with normals following~\cite{huang2013edge}, leading to $30000$ surface samples. These points are treated as surface points in $\textbf{RT}$ during connection establishment. The effect of point cloud up-sampling is demonstrated in Figure~\ref{fig:upsampling}. 

\section{\textbf{\textbf{Parameter Analysis on Surface points and Candidate Inner Points.}}}
\label{ap:surface_inner_points}
We conduct comprehensive experiments on the number of surface points $|S|$ and candidate inner points $|P|$ (note it is determined by $|C|$, which is the number of samples to compute the Voronoi diagram) for the algorithm performance. Detailed results including approximation error are summarized in Table~\ref{tab:ablation_surface} and Table~\ref{tab:ablation_inner}, respectively. In Table~\ref{tab:ablation_surface}, we fix $|C| = 4000$ and in Table~\ref{tab:ablation_inner}, we always set $|S|=1500$. Same as the main paper, we use $|V|$ to denote the number of skeletal points, $\overrightarrow{\epsilon}$ to denote the HD from surface to reconstruction, $\overleftarrow{\epsilon}$ to denote the HD from reconstruction to surface, and $\overleftrightarrow{\epsilon}$ to represent two-sided HD. We find both $|S|$ and $|P|$ do not show a large influence on the selected point number as well as the approximation error. However, as we mentioned in the main paper, the two factors, especially the number of surface samples $|S|$, have more effect on the time efficiency.
\begin{table}
\small
\begin{center}
\vspace{-2mm}
\small
\centering
\caption{Effect of different surface point numbers $|S|$ on selected points and time efficiency.}
\label{tab:ablation_surface}
\begin{tabular}{p{0.7cm}<{\centering}|p{0.4cm}<{\centering}|p{0.8cm}<{\centering}|p{0.8cm}<{\centering}|p{0.8cm}<{\centering}|p{0.8cm}<{\centering}}
\hline
\rule{0pt}{11pt} $|S|$ & $|V|$ & Time~(s) & $\overrightarrow{\epsilon}$ &  $\overleftarrow{\epsilon}$ &  $\overleftrightarrow{\epsilon}$\\
\hline
$500$  & $47$ & $0.25$ &  $2.062 \%$ & $ 1.981 \%$ & $ 2.062\%$ \\
$1500$ & $49$ & $0.29$ & $ 1.843 \%$ & $ 1.884 \%$ & $ 1.884\%$ \\
$2500$ & $49$ & $0.34$ & $ 1.761 \%$ & $ 1.799 \%$ & $ 1.799\%$ \\
$3500$ & $49$ & $0.73$ & $ 2.031 \%$ & $ 1.937 \%$ & $ 2.031\%$ \\
$4500$ & $49$ & $0.98$ & $ 1.750 \%$ & $ 1.762 \%$ & $ 1.762\%$ \\
$5500$ & $49$ & $1.33$ & $ 1.883 \%$ & $ 1.862 \%$ & $ 1.883\%$ \\
$6500$ & $49$ & $2.44$ & $ 1.701 \%$ & $ 1.747 \%$ & $ 1.747\%$ \\
$7500$ & $49$ & $3.24$ & $ 1.810 \%$ & $ 1.894 \%$ & $ 1.894\%$\\
$8500$ & $49$ & $4.22$ & $ 1.624 \%$ & $ 1.670 \%$ & $ 1.670\%$ \\
\hline
\end{tabular}
\end{center}
\end{table}

\begin{table}
\small
\begin{center}
\vspace{-2mm}
\small
\centering
\caption{Effect of different numbers of inner point candidates $|P|$ on selected points and time efficiency. $C$ denotes surface samples used for generating candidate inner points.}
\vspace{-1mm}
\label{tab:ablation_inner}
\begin{tabular}{p{0.7cm}<{\centering}|c|p{0.4cm}<{\centering}|p{0.8cm}<{\centering}|p{0.8cm}<{\centering}|p{0.8cm}<{\centering}|p{0.8cm}<{\centering}}
\hline
\rule{0pt}{11pt} $|C|$ & $|P|$ &$|V|$ & Time~(s) &   $\overrightarrow{\epsilon}$ &  $\overleftarrow{\epsilon}$ &  $\overleftrightarrow{\epsilon}$\\
\hline
$1000$  & $2964$   & $48$ & $0.08$ & $ 1.982 \%$ & $ 2.196 \%$ & $ 2.196\%$ \\
$2000$  & $6894$   & $49$ & $0.18$ & $ 1.843 \%$ & $ 1.811 \%$ & $ 1.843\%$ \\
$3000$  & $10776$  & $49$ & $0.28$ & $ 1.905 \%$ & $ 1.952 \%$ & $ 1.952\%$ \\ 
$4000$  & $13432$  & $49$ & $0.31$ & $ 1.855 \%$ & $ 1.894 \%$ & $ 1.894\%$ \\ 
$5000$  & $16749$  & $49$ & $0.49$ & $ 1.674 \%$ & $ 1.744 \%$ & $ 1.744\%$ \\
$6000$  & $22301$  & $49$ & $0.82$ & $ 1.675 \%$ & $ 1.735 \%$ & $ 1.735\%$ \\ 
$7000$  & $26161$  & $49$ & $1.23$ & $ 1.765 \%$ & $ 1.807 \%$ & $ 1.807\%$ \\
$8000$  & $29859$  & $49$ & $1.43$ & $ 1.754 \%$ & $ 1.916 \%$ & $ 1.916\%$ \\
$9000$  & $33651$  & $49$ & $1.65$ & $ 1.936 \%$ & $ 1.948 \%$ & $ 1.948\%$ \\
\hline
\end{tabular}
\end{center}
\vspace{-4mm}
 \end{table}

\section{\textbf{Comparison with Q-MAT for Highly Decimated MAT Computation.}}
\label{ap:ablation_random}

\begin{table}
\small
\begin{center}
 \caption{Evaluations on approximation accuracy for highly decimated medial surfaces between Q-MAT and Coverage Axis. We adopt the FEMUR model for evaluation.}
 \label{apextreme_decimated}
\vspace{-2mm}
\setlength{\tabcolsep}{0.45mm}{
\begin{tabular}{p{1.19cm}<{\centering}|p{0.8cm}<{\centering}|ccc|ccc}
\hline
\multirow{2}{*}{Offset $\delta_r$} & \multirow{2}{*}{ \makecell[c]{$|V|$} } & \multicolumn{3}{c|}{Q-MAT} & \multicolumn{3}{c}{Coverage Axis}\\
  &  &      $\overrightarrow{\epsilon}$ &  $\overleftarrow{\epsilon}$ &  $\overleftrightarrow{\epsilon}$ &    $\overrightarrow{\epsilon}$ &  $\overleftarrow{\epsilon}$ &  $\overleftrightarrow{\epsilon}$ \\
  \hline
$0.01$ & $47$ & $1.015\%$ & $1.114\%$ & $ 1.114\%$ & $ 0.947 \%$ & $ 1.104 \%$ & $ \mathbf{1.104\%}$ \\
$0.02$ & $24$ & $1.545\%$ & $1.281\%$ & $ \mathbf{1.545\%}$ & $ 1.535 \%$ & $ 1.593 \%$ & $ 1.593\%$ \\
$0.05$ &$13$ & $1.779\%$ & ${1.783\%}$ & $ \mathbf{1.783\%}$ & $ 2.576 \%$ & $ 2.534 \%$ & $ 2.576\%$ \\
$0.1$&$7$  & $2.679\%$ & $2.480\%$ & $ \mathbf{2.679\%}$ & $ 6.086 \%$ & $ 4.218 \%$ & $ 6.086\%$ \\
$0.2$ & $3$  & $7.565\%$ & $4.099\%$ & $ \mathbf{7.565\%}$ & $ 12.434 \%$ & $ 4.271 \%$ & $ 12.434\%$ \\
$0.5$ &$2$  & $29.855\%$ & $5.743\%$ & $ \mathbf{29.855\%}$ & $ 32.993 \%$ & $ 2.919 \%$ & $ 32.993\%$ \\
 \hline
\end{tabular}}
\end{center}
 \end{table}
The quantitative comparison of surface reconstruction by highly decimated medial surfaces between Q-MAT and Coverage Axis is shown in Table~\ref{apextreme_decimated}.


\end{document}